\documentclass[11pt,onecolumn]{article}

\usepackage{fullpage}

\usepackage{epsfig}
\usepackage{color}
\usepackage{graphicx}
\usepackage{amssymb,latexsym,amsmath}
\usepackage[latin1]{inputenc}
\usepackage{enumerate}
\usepackage{url}
\usepackage{pstricks}

\usepackage{amsfonts,epsfig,graphicx}
\usepackage{amsmath,amssymb,amsthm}

\usepackage{multirow}
\usepackage{psfrag}

\makeatletter \long\def\@makecaption#1#2{
        \vskip 0.8ex
        \setbox\@tempboxa\hbox{\small {\bf #1:} #2}
        \parindent 1.5em  
        \dimen0=\hsize
        \advance\dimen0 by -3em
        \ifdim \wd\@tempboxa >\dimen0
                \hbox to \hsize{
                        \parindent 0em
                        \hfil
                        \parbox{\dimen0}{\def\baselinestretch{0.96}\small
                                {\bf #1.} #2
                                }
                        \hfil}
        \else \hbox to \hsize{\hfil \box\@tempboxa \hfil}
        \fi
        }
\makeatother

\newtheorem{theorem}{Theorem}
\newtheorem{lemma}{Lemma}

\newtheorem{proposition}{Proposition}

\theoremstyle{plain}

\theoremstyle{definition}
\newtheorem{exas}{Example}

\newcommand{\widgraph}[2]{\includegraphics[keepaspectratio,width=#1]{#2}}
\newcommand{\matsnorm}[2]{|\!|\!| #1 |\!|\!|_{{#2}}}
\newcommand{\vnorm}[2]{\ensuremath{\|#1\|_#2}}
\newcommand{\vecprod}[2]{\ensuremath{\inprod{#1}{#2}}}
\newcommand{\inprod}[2]{\ensuremath{\langle #1 , \, #2 \rangle}}

\newcommand{\var} {\operatorname{var}}
\newcommand{\diameter}{\operatorname{diam}}

\newcommand{\order}{\ensuremath{\mathcal{O}}}

\newcommand{\graph}{\mathcal{G}}
\newcommand{\node}{\mathcal{V}}
\newcommand{\edge}{\mathcal{E}}
\newcommand{\DirEdge}{\vec{\edge}}

\newcommand{\DirSet}{\mathcal{\vec{\edge}}}
\newcommand{\numnode}{\ensuremath{n}}
\newcommand{\dimn}{\ensuremath{d}}
\newcommand{\degr}{\ensuremath{\rho}}
\newcommand{\dmax}{\ensuremath{\rho_{\text{max}}}}
\newcommand{\cliq}{\ensuremath{I}}
\newcommand{\SetCliq}{\ensuremath{\mathcal{C}}}
\newcommand{\diam}{\ensuremath{r}}

\newcommand{\rv}{\ensuremath{X}}
\newcommand{\realize}{\ensuremath{x}}
\newcommand{\obrv}{\ensuremath{Y}}

\newcommand{\Alphabet}{\ensuremath{\mathcal{\rv}}}
\newcommand{\Dimn}{\ensuremath{D}}

\newcommand{\Fnode}{\ensuremath{u}}
\newcommand{\Snode}{\ensuremath{v}}
\newcommand{\Tnode}{\ensuremath{w}}
\newcommand{\Ftnode}{\ensuremath{s}}

\newcommand{\vari}{\ensuremath{x}}
\newcommand{\Fvar}{\ensuremath{\vari_\Fnode}}
\newcommand{\Svar}{\ensuremath{\vari_\Snode}}

\newcommand{\EPot}{\ensuremath{\psi_{\Fnode\Snode}}}
\newcommand{\NPot}{\ensuremath{\psi_{\Fnode}}}
\newcommand{\Pot}{\ensuremath{\psi}}

\newcommand{\pott}{\ensuremath{\gamma}}
\newcommand{\mean}{\ensuremath{\mu}}
\newcommand{\std}{\ensuremath{\sigma}}

\long\def\comment#1{}

\newcommand{\Prob}[1]{\mathbb{P} \big(#1\big)}
\newcommand{\mProb}{\mathbb{P}}
\newcommand{\Exp}[2]{\mathbb{E}_{#1}\big[ #2 \big]}
\newcommand{\Expt}[1]{\mathbb{E}\big[ #1 \big]}
\newcommand{\Exps}{\mathbb{E}}
\newcommand{\real}{\mathbb{R}} 

\newcommand{\bO}[1]{\ensuremath{\order \big(#1\big)}}

\newcommand{\Th}[1]{\ensuremath{\Theta \big(#1 \big)}}

\newcommand{\Mess}{\ensuremath{m}}
\newcommand{\Mesast}{\ensuremath{\Mess^{\ast}}}
\newcommand{\Messast}{\ensuremath{\Mess^{\ast}}}

\newcommand{\AgMess}{\ensuremath{M_{\Fnode\Snode}}}
\newcommand{\MessP}{\ensuremath{\Mess^{\prime}}}

\newcommand{\Mes}[3]{\ensuremath{\Mess^{#1}_{#2 #3}}}
\newcommand{\AgMes}[3]{\ensuremath{M^{#1}_{#2 #3}}}

\newcommand{\Neig}{\ensuremath{\mathcal{N}}}

\newcommand{\CompNeig}{\ensuremath{\Tnode \in \Neig(\Fnode)
    \backslash \{\Snode \} }}

\newcommand{\CompSize}{\ensuremath{2|\edge|\dimn}}

\newcommand{\Ball}{\ensuremath{\mathcal{S}}}

\newcommand{\Time}{\ensuremath{t}}
\newcommand{\step}{\ensuremath{\lambda^{\Time}}}
\newcommand{\UpFunc}{\ensuremath{F}}
\newcommand{\UpFun}{\ensuremath{\UpFunc_{\Fnode \Snode}}}

\newcommand{\tBPMat} {\ensuremath{\Gamma}}
\newcommand{\tBPMatrix}[1]{\ensuremath{{\Gamma}_{#1}}}

\newcommand{\ind}[2]{\ensuremath{J^{#1}_{#2}}}
\newcommand{\indx}{\ensuremath{J}}
\newcommand{\normVec}[1] {\ensuremath{\nu^{#1}}}

\newcommand{\tMatEntry}{\ensuremath{\tBPMat_{\Fnode\Snode } (i, j)}}
\newcommand{\ColSum}[1]{\ensuremath{\beta_{\Fnode \Snode} (#1)}}
\newcommand{\tMatCol}[1]{\ensuremath{\Gamma_{\Fnode \Snode} (: , #1)}}

\newcommand{\probmass}{\ensuremath{p}}
\newcommand{\pbmas}[2]{\ensuremath{\probmass}_{#1} (#2)}

\newcommand{\Leigvec}[1]{\ensuremath{z_{#1}}}

\newcommand{\Lips}{\ensuremath{L}}

\newcommand{\LBound}[3]{\ensuremath{\underline{B}_{#1} ^{#3} (#2)}}

\newcommand{\LBnd}[2]{\ensuremath{\underline{B}_{#1} (#2)}}
\newcommand{\UBound}[3]{\ensuremath{\overline{B}_{#1}^{#3} (#2) }}
\newcommand{\UBnd}[2]{\ensuremath{\overline{B}_{#1} (#2) }}

\newcommand{\LocalJacob}[2]{\ensuremath {\frac{ \ro q_{#1} (\Mess) }
    {\ro \Mess_{#2}}}}

\newcommand{\TermPhi}[2]{\ensuremath {\phi_{#1, #2}}}
\newcommand{\TermChi}[2]{\ensuremath {\chi_{#1, #2}}}
\newcommand{\Term}{\ensuremath{G}}
\newcommand{\secterm}{\ensuremath{T}}

\newcommand{\marg}{\ensuremath{\tau}}

\newcommand{\JacMat}{\ensuremath{\grad q}}
\newcommand{\SpMat}{\ensuremath{A}}
\newcommand{\SpMatEntry}[2]{\ensuremath{\SpMat (#1, #2)}}
\newcommand{\ro}{\ensuremath{\partial}}
\newcommand{\slack}{\ensuremath{\mu}}

\newcommand{\errvec}[1]{\ensuremath{e^{#1}}}
\newcommand{\Error}[1]{\ensuremath{\vnorm{e^{#1}}{2}^2}}
\newcommand{\MSE}[1]{\ensuremath{\mathbb{E}[\Error{#1}]}}

\newcommand{\vecone}{\ensuremath{y}}
\newcommand{\vectwo}{\ensuremath{z}}

\newcommand{\VF}{\ensuremath{H}}
\newcommand{\MVF}{\ensuremath{h}}

\newcommand{\MarDif}[1] {\ensuremath{Y^{#1}}}

\newcommand{\Filt}[1]{\ensuremath{\mathcal{F}^{#1}}}

\newcommand{\defn}{\ensuremath{: = }}

\newcommand{\grad}{\ensuremath{\nabla}}
\newcommand{\const}{\ensuremath{c}}
\newcommand{\Coef}{\ensuremath{\alpha}}

\newcommand{\nilpot}{\ensuremath{A}}
\newcommand{\diredge}[2]{\ensuremath{#1 \to #2}}

\newcommand{\parderv}{\ensuremath{\frac{\partial \UpFunc_{\Fnode
        \Snode}}{\partial \Mess_{\Tnode \Ftnode}}}}

\newcommand{\blkMat}[2]{\ensuremath{A_{#1,#2}}} 

\newcommand{\blkInd}{\ensuremath{B}}

\newcommand{\smallTerm}[1] {\ensuremath{Z^{#1}}}
\newcommand{\zerovec}{\ensuremath{\vec{0}}}
\newcommand{\onevec}{\ensuremath{\vec{1}}}

\newcommand{\mprob}{\ensuremath{\mathbb{P}}}

\newcommand{\conegeq}{\ensuremath{\succeq}}
\newcommand{\coneleq}{\ensuremath{\preceq}}

\newcommand{\PHIFUN}{\ensuremath{\Phi_1}}
\newcommand{\PHIFUNTWO} {\ensuremath{\Phi_2}}

\newcommand{\Exs}{\ensuremath{\mathbb{E}}}

\newcommand{\prefac}{\ensuremath{K}}

\newcommand{\SPECCON}{4}

\newcommand{\newtime}{\ensuremath{\tau}} 

\newcommand{\blkdiag}{\ensuremath{\operatorname{blkdiag}}}

\newcommand{\Xspace}{\ensuremath{\Alphabet}}
\newcommand{\PotTil}{\ensuremath{\widetilde{\Pot}}}

\begin{document}

\begin{center}

{\bf{\LARGE{ Stochastic Belief Propagation: 
\\ A Low-Complexity Alternative to the Sum-Product Algorithm}}}

\vspace*{.2in}

{\large{
\begin{tabular}{ccc}
Nima Noorshams$^1$ & & Martin J. Wainwright$^{1,2}$ \\
{\texttt{nshams@eecs.berkeley.edu}} & & {\texttt{wainwrig@eecs.berkeley.edu}}
\end{tabular}
}}

\vspace*{.1in}

\begin{center}
Department of Statistics$^2$ and \\
Department of Electrical Engineering $\&$ Computer Science$^{1}$\\
University of California Berkeley 
\end{center}



\vspace*{.1in}

April 2012

\vspace*{.2in}

\begin{abstract}
The sum-product or belief propagation (BP) algorithm is a widely-used
message-passing algorithm for computing marginal distributions in
graphical models with discrete variables.  At the core of the BP
message updates, when applied to a graphical model with pairwise
interactions, lies a matrix-vector product with complexity that is
quadratic in the state dimension $\dimn$, and requires transmission of
a $(\dimn-1)$-dimensional vector of real numbers (messages) to its
neighbors.  Since various applications involve very large state
dimensions, such computation and communication complexities can be
prohibitively complex. In this paper, we propose a low-complexity
variant of BP, referred to as stochastic belief propagation (SBP). As
suggested by the name, it is an adaptively randomized version of the
BP message updates in which each node passes randomly chosen
information to each of its neighbors. The SBP message updates reduce
the computational complexity (per iteration) from quadratic to linear
in $\dimn$, without assuming any particular structure of the potentials,
and also reduce the communication complexity significantly, requiring
only $\log{\dimn}$ bits transmission per edge.  Moreover, we establish a
number of theoretical guarantees for the performance of SBP, showing
that it converges almost surely to the BP fixed point for any
tree-structured graph, and for graphs with cycles satisfying a
contractivity condition.  In addition, for these graphical models, we
provide non-asymptotic upper bounds on the convergence rate, showing
that the $\ell_{\infty}$ norm of the error vector decays no
slower than $\bO{1/\sqrt{\Time}}$ with the number of iterations
$\Time$ on trees and the mean square error decays as $\bO{1/\Time}$
for general graphs.  These analysis show that SBP can provably yield
reductions in computational and communication complexities for various
classes of graphical models.\footnote{Portions of the results given
  here were initially reported at the Allerton Conference on
  Communications, Control, and Computing (September 2011).}
\end{abstract}
\end{center}



\section{Introduction}
\label{SecIntro}

Graphical models provide a general framework for describing
statistical interactions among large collections of random
variables. A broad range of fields---among them statistical signal
processing, computer vision, coding and information theory, and
bioinformatics---involve problems that can be fruitfully tackled using
the formalism of graphical models.  A computational problem central to
such applications is that of \emph{marginalization}, meaning the
problem of computing marginal distributions over a subset of random
variables.  Naively approached, these marginalization problems have
exponential complexity, and hence are computationally
intractable. Therefore, graphical models are only useful when combined
with efficient algorithms. For graphs without cycles, the
marginalization problem can be solved exactly and efficiently via an
algorithm known as the sum-product or belief propagation (BP)
algorithm.  It is a distributed algorithm, in which each node performs
a set of local computations, and then relays the results to its graph
neighbors in the form of so-called messages.  For graphs with cycles,
BP is no longer an exact method, but nonetheless is widely used and
known to be extremely effective in many settings.  For a more detailed
discussion of the role of the marginalization problem and the use of
sum-product, we refer the reader to various overview papers
(e.g.,~\cite{KasEtal01,Loeliger04,WaiJorBook08, McEliece02}).

In many applications of BP, the messages themselves are
high-dimensional in nature, either due to discrete random variables
with a very large number of possible realizations $\dimn$, which will
be reffered to as the number of states, factor nodes with high degree,
or continuous random variables that are discretized. Examples of such
problems include disparity estimation in computer vision, tracking
problems in sensor networks, and error-control decoding.  For such
problems, it may be expensive to compute and/or store the messages,
and as a consequence, BP may run slowly, and be limited to small-scale
instances.  Motivated by this challenge, researchers have studied a
variety of techniques to reduce complexity of BP in different
applications (e.g., see the papers~\cite{FlezHutt06, SuddEtal03,
  McauCaet11, IsaEtal09, KersEtal09, CouShe07, SongEtal11} and
references therein).  At the core of sum-product message-passing is a
matrix-vector multiplication, with complexity scaling quadratically in
the number of states $\dimn$.  Certain graphical models have special
structure that can be exploited so as to reduce this complexity.  For
instance, in application to the decoding of low-density parity check
codes in channel coding (e.g.,~\cite{Gallager63,KasEtal01}), the
complexity of message-passing, if performed naively, would scale
exponentially in the factor degrees.  However, a clever use of the
fast Fourier transform over $\mbox{GF}(2)$ reduces this complexity to
linear in the factor degrees~\cite{SonCru03}. Other problems arising
in computer vision involve pairwise factors with a circulant structure
for which the fast Fourier transform can also reduce
complexity~\cite{FlezHutt06}. Similarly, computation can be
accelerated by exploiting symmetry in factors~\cite{KersEtal09}, or
additional factorization properties of the
distribution~\cite{McauCaet11}.  In the absence of structure to
exploit, other researchers have proposed different types of
quantization strategies for BP message
updates~\cite{CouShe07,IsaEtal09}, as well as stochastic methods based
on particle filtering or non-parametric belief propagation
(e.g.,~\cite{AruEtal02,SuddEtal03,DouFreGor}) that approximate
continuous messages by finite numbers of particles. For certain
classes of these methods, it is possible to establish consistency as
the number of particles tends to infinity~\cite{DouFreGor} or
establish finite-length results inversely proportional to the square
root of the number of particles~\cite{IhlMcA09}. As the number of
particles diverges, the approximation error becomes negligible, a
property that underlies such consistency proofs. Researchers have also
proposed stochastic techniques to improve the decoding efficiency of
binary error-correcting codes~\cite{TehEtal06, RapEtal03}. These
techniques, which are based on encoding messages with sequences of
Bernoulli random variables, lead to efficient decoding hardware
architectures.

In this paper, we focus on the problem of implementing BP in
high-dimensional discrete spaces, and propose a novel low-complexity
algorithm, which we refer to as \emph{stochastic belief propagation}
(SBP).  As suggested by its name, it is an adaptively randomized
version of the BP algorithm, where each node only passes randomly
selected partial information to its neighbors at each round. The SBP
algorithm has two features that makes it practically appealing.
First, it reduces the computational cost of BP by an order of
magnitude; in concrete terms, for arbitrary pairwise potentials over
$\dimn$ states, it reduces the per iteration computational complexity
from quadratic to linear---that is, from $\Th{\dimn^2}$ to
$\Th{\dimn}$. Second, it significantly reduces the
message/communication complexity, requiring transmission of only
$\log{\dimn}$ bits per edge as opposed to $(\dimn-1)$ real numbers in
the case of BP.

Even though SBP is based on low-complexity updates, we are able to
establish conditions under which it converges (in a stochastic sense)
to the exact BP fixed point, and moreover, to establish quantitative
bounds on this rate of convergence.  These bounds show that SBP can
yield provable reductions in the complexity of computing a BP fixed
point to a tolerance $\delta > 0$.  In more precise terms, we first
show that SBP is strongly consistent on any tree-structured graph,
meaning that it converges almost surely to the unique BP fixed point;
in addition, we provide non-asymptotic upper bounds on the
$\ell_\infty$ norm (maximum value) of the error vector as a function
of iteration number (Theorem~\ref{ThmTree}).  For general graphs with
cycles, we show that when the ordinary BP message updates satisfy a
type of contraction condition, then the SBP message updates are
strongly consistent, and converge in mean-squared error at the rate
$\order(1/t)$ to the unique BP fixed point, where $t$ is the number of
iterations. We also show that the typical performance is sharply
concentrated around its mean (Theorem~\ref{ThmMain}).  These
theoretical results are supported by simulation studies, showing the
convergence of the algorithm on various graphs, and the associated
reduction in computational complexity that is possible.

The remainder of the paper is organized as follows.  We begin in
Section~\ref{SecProbState} with background on graphical models as well
as the BP algorithm.  In Section~\ref{SecSBP}, we provide a precise
description of the SBP, before turning in Section~\ref{SecMainResult}
to statements of our main theoretical results, as well as discussion
of some of their consequences.  Section~\ref{SecProof} is devoted to
the proofs of our results, with more technical aspects of the proofs
deferred to the Appendices.  In Section~\ref{SecSimulations}, we
demonstrate the correspondence between our theoretical predictions and
the algorithm's practical behavior.


\section{Background}
\label{SecProbState}
In this section, we provide some background on graphical models as
well as the sum-product or belief propagation algorithm.


\subsection{Graphical Models} 
\label{SubSecGraphModel}

Consider a random vector $\rv \defn \{ \rv_1, \rv_2, \ldots,
\rv_\numnode \}$, where for each $u = 1, 2, \ldots, \numnode$, the
variable $\rv_\Fnode$ takes values in some discrete space $\Xspace
\defn \{1, 2, \ldots, \dimn \}$ with cardinality $\dimn$.  An
undirected graphical model, also known as a Markov random field,
defines a family of joint probability distributions over this random
vector by associating the index set $\{1, 2, \ldots, \numnode \}$ with
the vertex set $\node$ of an undirected graph $\graph = (\node,
\edge)$.  In addition to the vertex set, the graph consists of a
collection of edges $\edge \subset \node \times \node$, where a pair
$(\Fnode, \Snode) \in \edge$ if and only if nodes $\Fnode$ and
$\Snode$ are connected by an edge.  The structure of the graph
describes the statistical dependencies among the different random
variables---in particular, via the cliques\footnote{ A clique $\cliq$
  of a graph is a subset of vertices that are all joined by edges, and
  so form a fully connected subgraph.}  of the graph.  For each clique
$\cliq$ of the graph, let $\Pot_\cliq: \Xspace^{|\cliq|} \rightarrow
(0, \infty)$ be a function of the sub-vector $\rv_\cliq \defn \{
\rv_\Fnode, \; \Fnode \in \cliq \}$ of random variables indexed by the
clique, and then consider the set of all distributions over $\rv$ that
factorize as
\begin{align}
\label{EqnGeneralFact}
\mprob(\realize_1, \ldots, \realize_\numnode) \;  & \propto
\prod_{\cliq \in \SetCliq} \Pot_\cliq(\realize_\cliq),
\end{align}
where $\SetCliq$ is the set of all cliques in the graph.

\begin{figure}[h]
\begin{center}
\begin{tabular}{ccc}
\psfrag{*cedge*}{$\Pot_{\Fnode \Snode}$}
\psfrag{*cvar1*}{$\Pot_{\Fnode}$}
\psfrag{*cvar2*}{$\qquad \quad \Pot_{\Snode}$}
\widgraph{.45\textwidth}{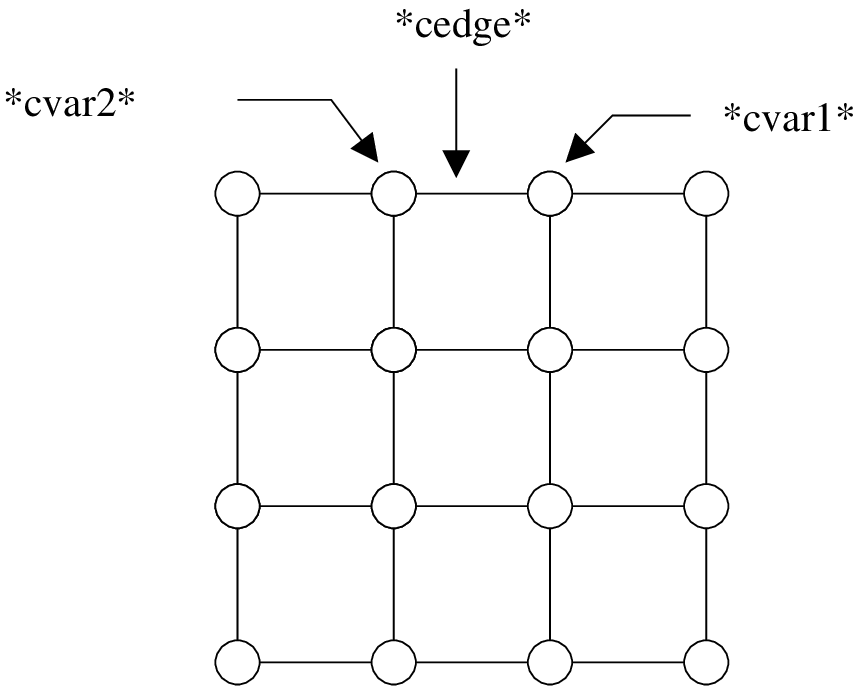} & & 
\psfrag{#x1#}{$x_1$}
\psfrag{#x2#}{$x_2$}
\psfrag{#x3#}{$x_3$}
\psfrag{#x4#}{$x_4$}
\psfrag{#x5#}{$x_5$}
\psfrag{#y1#}{$y_1$}
\psfrag{#y2#}{$y_2$}
\psfrag{#y3#}{$y_3$}
\psfrag{#y4#}{$y_4$}
\psfrag{#y5#}{$y_5$}
\raisebox{.2in}{\widgraph{.45\textwidth}{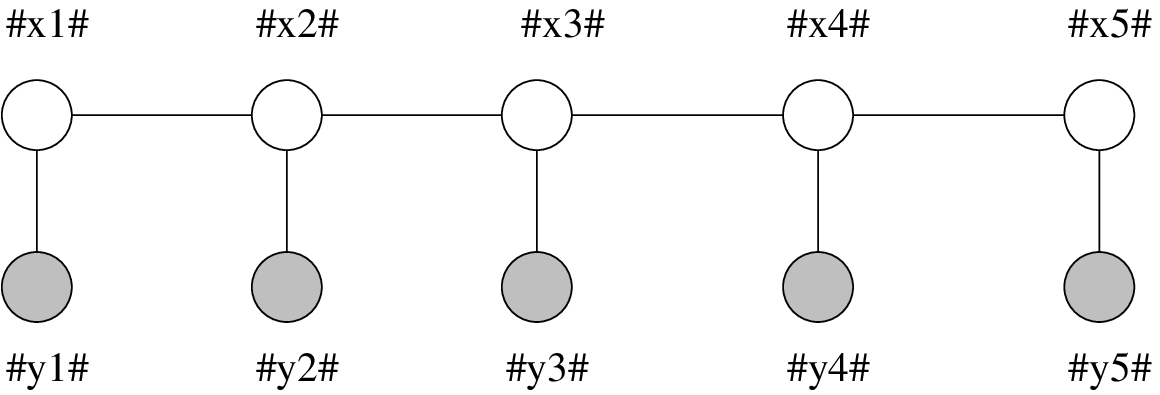}} \\
(a) & & (b)
\end{tabular}
\end{center}
\caption{Examples of pairwise Markov random fields. (a) A
  two-dimensional grid: potential functions $\Pot_\Fnode$ and
  $\Pot_\Snode$ are associated with nodes $\Fnode$ and $\Snode$
  respectively, whereas potential function $\Pot_{\Fnode \Snode}$ is
  associated with edge $(\Fnode, \Snode)$.  (b) Markov chain model
  including both hidden variables $(x_1, \ldots, x_5)$, represented as
  white nodes, and observed variables $(y_1, \ldots, y_5)$ represented
  as shaded nodes.}
\label{FigGraphicalModels}
\end{figure}

As a concrete example, consider the two-dimensional grid shown in
Figure~\ref{FigGraphicalModels}(a).  Since its cliques consist of the
set of all vertices $\node$ together with the set of all edges
$\edge$, the general factorization~\eqref{EqnGeneralFact} takes the
special form
\begin{align}
\label{EqnPairwiseFact}
\mprob(\realize_1, \ldots, \realize_\numnode) \; & \propto \; \prod_{\Fnode
  \in \node} \Pot_\Fnode(\realize_\Fnode) \prod_{(\Fnode, \Snode)
  \in \edge} \Pot_{\Fnode \Snode}(\realize_\Fnode, \realize_\Snode),
\end{align}
where $\Pot_\Fnode: \Xspace \rightarrow (0,\infty)$ is the node
potential function for node $\Fnode$, and $\Pot_{\Fnode
  \Snode}: \Xspace \times \Xspace \rightarrow (0,\infty)$ is the edge
potential function for the edge $(\Fnode,\Snode)$.  A factorization of
this form~\eqref{EqnPairwiseFact} is known as a \emph{pairwise Markov
  random field}.  It is important to note that there is no loss of
generality in assuming a pairwise factorization of this form; indeed,
any graphical model with discrete random variables can be converted
into a pairwise form by suitably augmenting the state space (e.g., see
Yedidia et al.~\cite{Yedidia05} or Wainwright and
Jordan~\cite{WaiJorBook08}, Appendix E.3).  Moreover, the sum-product
message updates can be easily translated from the original graph to
the pairwise graph, and vice versa.  Accordingly, for the remainder of
this paper, we focus on the case of a pairwise MRF.

In various application contexts, the random vector $(\rv_1, \ldots,
\rv_\numnode)$ is an unobserved or ``hidden'' quantity, and the goal
is to draw inferences on the basis of a collection of observations
$(\obrv_1, \ldots, \obrv_\numnode)$.  The link between the observed
and hidden variables is specified in terms of a conditional
probability distribution, which in many cases can be written in the 
product form $\mprob(y \mid x) = \prod_{\Fnode = 1}^\numnode
\mprob(y_\Fnode \mid \realize_\Fnode)$.  For instance, in
error-control coding using a low-density parity check code, the vector
$X$ takes values in a linear subspace of $\{0,1\}^\numnode$,
corresponding to valid codewords, and the observation vector $Y$ is
obtained from some form of memoryless channel (e.g., binary symmetric,
additive white Gaussian noise, etc.).  In image denoising applications, the
vector $X$ represents a rasterized form of the image, and the
observation $Y$ corresponds to a corrupted form of the image.

In terms of drawing conclusions about the hidden variables based on
the observations, the central object is the posterior distribution
$\mprob(\realize \mid y)$.  From the definition of conditional
probability and the form of the prior and likelihoods, this posterior
can also be factorized in pairwise form
\begin{align}
\mprob(x \mid y) & \propto \; \mprob(x_1, \ldots, x_\numnode) \;
\prod_{\Fnode =1}^\numnode \mprob(y_\Fnode \mid x_\Fnode) \; = \;
\prod_{\Fnode \in \node} \PotTil_\Fnode(x_\Fnode) \prod_{(\Fnode,
  \Snode) \in \edge} \Pot_{\Fnode \Snode}(x_\Fnode, x_\Snode),
\end{align}
where $\PotTil_\Fnode(x_\Fnode) \defn \Pot_\Fnode(x_\Fnode)
\mprob(y_\Fnode \mid x_\Fnode)$ is the new node compatibility
function.  (Since the observation $y_\Fnode$ is fixed, there is no
need to track its functional dependence.)  Thus, the problem of
computing marginals for a posterior distribution can be
cast\footnote{For illustrative purposes, we have assumed here that the
  distribution $\mprob(y \mid x)$ has a product form, but a somewhat
  more involved reduction also applies to a general observation
  model.} as an instance of computing marginals for a pairwise Markov
random field~\eqref{EqnPairwiseFact}.

Our focus in this paper is the \emph{marginalization problem}, meaning
the computation of the single-node marginal distributions
\begin{align}
\label{EqnDefnSingleNode}
\mprob(\realize_\Fnode) \;\;  & \defn \! \sum_{ \{x' \, \mid \, x'_\Fnode =
  x_\Fnode \}} \mprob \left(\realize'_1, \ldots, \realize'_\numnode \right) \quad
\mbox{for each $\Fnode \in \node$,}
\end{align}
and more generally, higher-order marginal distributions on edges and
cliques. Note that to calculate this summation, brute force is not
tractable and requires $\numnode \dimn ^{\numnode - 1}$
computations.  For any graph without cycles---known as a tree---this
computation can be carried far more efficiently in only
$\order(\numnode \dimn^2)$ operations using an algorithm known as
the beilef propagation algorithm, to which we now turn.


\subsection{Sum-product Algorithm} 
\label{SubSecBP}

Belief propagation, also known as the sum-product algorithm, is an
iterative algorithm consisting of a set of local message-passing
rounds, for computing either exact or approximate marginal
distributions. For tree-structured (cycle-free) graphs, it is known
that BP message updates converge to the exact marginals in a finite
number of iterations. However, the same message-passing updates can
also be applied to more general graphs, and are known to be effective
for computing approximate marginals in numerous applications.  Here we
provide a very brief treatment, referring the reader to various
standard sources~\cite{KasEtal01,McEliece02,Yedidia05,WaiJorBook08}
for further background.

In order to define the message-passing updates, we require some
further notation.  For each node $\Fnode \in \node$, let
$\Neig(\Fnode) \defn \{ \Tnode \; | \; (\Tnode, \Fnode) \in \edge \}$
denote its set of neighbors, and let \mbox{$\DirSet(\Fnode) \defn
  \{(\Fnode \to \Snode) \; | \; \Snode \in \Neig (\Fnode) \}$} denote
the set of all directed edges emanating from $\Fnode$.  Finally, we
define $\DirSet \defn \cup_{\Fnode \in \node} \DirSet(\Fnode)$, the
set of \emph{all directed edges} in the graph; note that $\DirSet$ has
cardinality $2 |\edge|$.  In the BP algorithm, one message
$\Mess_{\Fnode \Snode} \in \real^{\dimn}$ is assigned to every
directed edge $(\Fnode \to \Snode) \in \DirSet$.  By concatenating all
of these $\dimn$-vectors, one for each of the $2 |\edge|$ members of
$\DirEdge$, we obtain a $\Dimn$-dimensional vector of messages $\Mess
= \{\Mes{}{\Fnode}{\Snode} \} _ {(\Fnode \to \Snode) \in \DirSet}$,
where $\Dimn \defn \CompSize$.

\begin{figure}[h]
\begin{center}
\begin{tabular}{cc}
\psfrag{#1#}{$s_2$}
\psfrag{#6#}{$s_1$} 
\psfrag{#2#}{$v$} 
\psfrag{#3#}{$u$}
\psfrag{#4#}{$w_2$} \psfrag{#5#}{$w_1$} 
\psfrag{#m43#}{$\Mess_{w_2 u}$}
\psfrag{#m53#}{$\Mess_{w_1 u}$} 
\psfrag{#m32#}{$\Mess_{uv}$}
\widgraph{0.45\textwidth}{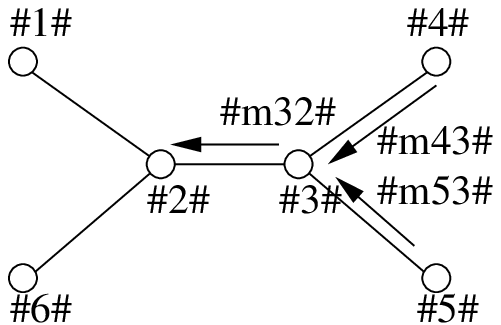} &
\psfrag{#1#}{$s_2$}
\psfrag{#6#}{$s_1$} 
\psfrag{#2#}{$v$} 
\psfrag{#3#}{$u$}
\psfrag{#4#}{$w_2$} \psfrag{#5#}{$w_1$} 
\psfrag{#m43#}{$\Mess_{s_2 v}$}
\psfrag{#m53#}{$\Mess_{s_1 v}$} 
\psfrag{#m32#}{$\Mess_{uv}$}
\widgraph{0.45\textwidth}{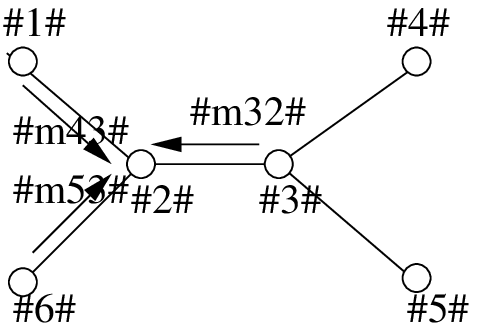} \\
(a) & (b)
\end{tabular}
\end{center}
\caption{Graphical representation of message-passing algorithms. (a)
  Node $\Fnode$ transmits the message $\Mess_{\Fnode\Snode} =
  \UpFunc_{\Fnode \Snode} (\Mess)$, derived from
  equation~\eqref{EqnBPUpdateEntry}, to its neighbor $\Snode$. (b)
  Upon receiving all the messages, node $\Snode$ updates its marginal
  estimate according to~\eqref{EqnUpdateMarg}.}
\label{FigMessPass}
\end{figure}

At each round $\Time = 1, 2, \ldots$, every node $\Fnode \in \node$
calculates a message $\Mes{\Time+1}{\Fnode}{\Snode} \in \real^\dimn$
to be sent to its neighbor $\Snode \in \Neig(\Fnode)$. In mathematical
terms, this operation can be represented as an update of the form
$\Mes{\Time+1}{\Fnode}{\Snode} = \UpFun(\Mess^{\Time})$ where $\UpFun
: \real^{\Dimn} \to \real ^{\dimn}$ is the local update function of
the directed edge $(\Fnode \to \Snode)$.  In more detail, for each
$\Svar \in \Xspace$, we have\footnote{It is worth mentioning that
  $\Mess^{\Time + 1}_{\Fnode \Snode}$ is only a function of the
  messages $\Mess^{\Time }_{\Tnode \Fnode}$ for
  $\CompNeig$. Therefore, we have $\UpFun :
  \real^{(\degr_{\Fnode}-1)\dimn} \to \real ^{\dimn}$, where
  $\degr_{\Fnode}$ is the degree of the node $\Fnode$. Since it is
  clear from the context and for the purpose of reducing the notation
  overhead, we say $\Mes{\Time+1}{\Fnode}{\Snode} =
  \UpFun(\Mess^{\Time})$ instead of $\Mes{\Time+1}{\Fnode}{\Snode} =
  \UpFun(\{\Mess^{\Time }_{\Tnode \Fnode}\}_{\CompNeig})$.}
\begin{align}
\label{EqnBPUpdateEntry}
\Mess^{\Time + 1}_{\Fnode \Snode} (\Svar) \, = & \, [\UpFun
  (\Mess^{\Time})] (\Svar) \; = \; \kappa \sum_{\Fvar \in \Alphabet}
\bigg( \Pot_{\Fnode \Snode}( \Fvar, \Svar) \NPot (\Fvar) \!\!
\prod_{\CompNeig} \Mes{\Time}{\Tnode}{\Fnode} (\Fvar) \bigg),
\end{align}
where $\kappa$ is a normalization constant chosen to ensure that
$\sum_{\Svar} \Mess^{\Time+1}_{\Fnode \Snode}(\Svar) =
1$. Figure~\ref{FigMessPass}(a) provides a graphical representation of
the flow of information in this local update.

Equation~\eqref{EqnBPUpdateEntry} is basically an iterative way of
solving a set of fixed-point equations in $\real^\Dimn$. More
precisely, by concatenating the local
updates~\eqref{EqnBPUpdateEntry}, we obtain a global update function
$\UpFunc: \real^{\Dimn} \to \real^{\Dimn}$ of the form
\begin{align}
\label{EqnGlobalUpFun}
\UpFunc (\Mess) \, = \, \{\UpFunc_{\Fnode \Snode} (\Mess)\}
_{(\diredge{\Fnode}{\Snode}) \in \DirSet}.
\end{align}
Typically, the goal of message-passing is to obtain a \emph{fixed
  point}, meaning a vector $\Messast \in \real^\Dimn$ such that
$\UpFunc (\Messast) = \Messast$.  For any tree-structured graph, it is
known that the update~\eqref{EqnGlobalUpFun} has a unique fixed point.
For a general graph (with some mild conditions on the potentials; see
Yedidia et al.~\cite{Yedidia05} for details), it is known that the
global update~\eqref{EqnGlobalUpFun} has at least one fixed point, but
it is no longer unique in general.  However, there are various types
of contraction conditions that can be used to guarantee uniqueness on
a general graph (e.g.,~\cite{Tatikonda02,Ihler05, MooijKapp07,
  RoostaEtal08}).

Given a fixed point $\Messast$, node $\Snode$ computes its marginal
(approximation) $\marg^{\ast}_\Snode$ by combining the local potential
function $\Pot_\Snode$ with a product of all incoming messages as
\begin{align}
\label{EqnUpdateMarg}
\marg^{\ast}_\Snode (\Svar)\;  = \; \kappa \; \Pot_{\Snode}(\Svar) \!
\prod_{\Fnode \in \Neig(\Snode)} \Messast_{\Fnode \Snode} (\Svar),
\end{align}
where $\kappa$ is a normalization constant chosen so that $\sum_{\Svar
  \in \Alphabet} \marg^{\ast}_\Snode(\Svar) = 1$.  See
Figure~\ref{FigMessPass}(b) for an illustration of this computation.
For any tree-structured graph, the quantity $\marg^\ast_\Snode(\Svar)$
is equal to the single-node marginal $\mprob(\Svar)$, as previously
defined~\eqref{EqnDefnSingleNode}.  For a graph with cycles, the
vector $\marg^\ast_\Snode$ represents an approximation to the
single-node marginal, and is known to be a useful approximation for
many classes of graphical models.


\section{Algorithm and Main Results}
\label{SecSBP}

We now turn to a description of the SBP algorithm
(Section~\ref{SecDesc}), as well as the statement of our main
theoretical guarantees on its behavior (Section~\ref{SecMainResult}).

\subsection{Stochastic Belief Propagation}
\label{SecDesc}

When applied to a pairwise graphical model with random variables
taking $\dimn$ states, the number of summations and multiplications
required by original BP algorithm is $\Th{\dimn^2}$ per iteration, as
can be seen by inspection of the message update
equation~\eqref{EqnBPUpdateEntry}.  This quadratic complexity---which
is incurred on a per iteration, per edge basis---is prohibitive in
many applications, where the state dimension may be on the order of
thousands.  As discussed earlier in Section~\ref{SecIntro}, although
certain graphical models have particular structures that can be
exploited to reduce complexity of the updates, not all problems have
such special structures, so that a general-purpose approach is of
interest.  In addition to computational cost, a standard BP message
update can also be expensive in terms of communication cost, since
each update requires transmitting $(\dimn-1)$ real numbers along each
edge.  For applications that involve power limitations, such as sensor
networks, reducing this communication cost is also of interest. \\

Stochastic belief propagation is an adaptively randomized form of the
usual BP message updates that yields savings in both
computational and communication cost.  It is motivated by a simple
observation---namely, that the message-passing update along the
directed edge $(\Fnode \to \Snode)$ can be formulated as an
expectation over suitably normalized columns of the compatibility
matrix.  Here the probability distribution in question depends on the
incoming messages, and changes from iteration to iteration.  This
perspective leads naturally to an \emph{adaptively randomized variant}
of BP: instead of computing and transmitting the full expectation at
each round---which incurs $\Theta(\dimn^2)$ computational cost and
requires sending $\Theta(\dimn)$ real numbers---the SBP algorithm
simply picks a single normalized column with the appropriate
(message-dependent) probability, and performs a randomized update.  As
we show, each such operation can be performed in $\order(\dimn)$ time
and requires transmitting only $\log \dimn$ bits, so that the SBP
message updates are less costly by an order of magnitude.  \\

With this intuition in hand, we are now ready for a precise
description of the SBP algorithm.  Let us view the edge potential
function $\Pot_{\Fnode \Snode}$ as a matrix of numbers $\Pot_{\Fnode \Snode}(i,j)$, 
for $i,j = 1, \ldots, \dimn$. For the directed edge
$(\Fnode \rightarrow \Snode)$, define the collection of column vectors
\begin{align}
\label{EqnNormalizedCol}
\tMatCol{j} \: \defn \: \frac{ \Pot_{\Fnode \Snode}(:, \: j) \;
  \Pot_\Fnode(j)}{\ColSum {j} }, \quad \text{for $j = 1, 2, \ldots,
  \dimn$},
\end{align}
where $\ColSum{j} \defn \sum_{i=1}^\dimn \Pot_{\Fnode \Snode}(i, j) \:
\Pot_\Fnode(j)$.  We assume that the column vectors $\tMatCol{j}$ and
normalization constants $\ColSum{j}$ have been pre-computed and
stored, which can be done in an off-line manner. In addition, the
algorithm makes use of a positive sequence of step sizes
$\{\step\}_{t=0}^\infty$.  In terms of these quantities, the SBP
algorithm consists of the steps shown in Figure~\ref{FigAlgorithm}. \\
\begin{figure}[h]
\begin{center}
\framebox[1\textwidth]{\parbox{.97\textwidth}{\textbf{Stochastic
      Belief Propagation Algorithm:}
\begin{enumerate}
\item[(I)] Initialize the message vector $\Mes{0}{}{} \in \real^\Dimn$.
%
\item[(II)] For iterations $\Time = 0, 1, 2, 3, \ldots$, and for each
  directed edge $(\Fnode \to \Snode) \in \DirEdge$:
\begin{enumerate}
\item[(a)] Compute the product of incoming messages:
\begin{align}\label{EqnProdIncMess}
\AgMes{\Time}{\Fnode}{\Snode} (i) \; \; \; & = \! \!\prod_{\CompNeig}
\Mes{\Time}{\Tnode}{\Fnode} (i) \quad \mbox{for $i \in \{1, \ldots, \dimn \}$.}
\end{align}
\item[(b)] Pick a random index $\ind{\Time + 1}{\Fnode \Snode} \in
  \{1, 2, \ldots, \dimn \}$ according to the probability distribution
\begin{align}
\label{EqnProbMass}
\probmass^{\Time} _{\Fnode\Snode} (j) \; \propto \; \AgMes{\Time}{\Fnode}{\Snode}
(j) \: \ColSum{j} \quad \mbox{for $j \in \{ 1, \ldots, \dimn \}$.}
\end{align}

\item[(c)] For a given step size $\step \in (0, 1)$, update the
  message $\Mes{\Time+1}{\Fnode}{\Snode} \in \real^\dimn$ via
\begin{align}
\label{EqnSBPUpdate}
\Mes{\Time+1}{\Fnode}{\Snode} \, & = \, (1 - \step) \:
\Mes{\Time}{\Fnode}{\Snode} \, + \, \step \:
\tMatCol{\ind{\Time+1}{\Fnode\Snode}}.
\end{align}
\end{enumerate}
\end{enumerate}
}}
\vspace{.1in}
\caption{Specification of stochastic belief propagation.}
\label{FigAlgorithm}
\end{center}
\end{figure}

The per iteration computational complexity of the SBP algorithm lies
in calculating the probability mass function $\probmass_{\Fnode
  \Snode}$, defined in equation~\eqref{EqnProbMass}; generating a
random index $\indx_{\Fnode\Snode}$ according to the mass
function~\eqref{EqnProbMass}, and performing the weighted
update~\eqref{EqnSBPUpdate}.  Denoting the maximum degree of the graph
by $\dmax$, we require at most $(\dmax - 1) \dimn$ multiplications to
compute $\AgMess$.  Moreover, an additional $3\dimn$ operations are
needed to compute the probability mass function $\probmass_{\Fnode
  \Snode}$. On the other hand, generating a random index
$\indx_{\Fnode\Snode}$, can be done with less than $\dimn$ operations
by picking a number $U$ uniformly at random from $[0, 1]$ and
setting\footnote{It is known that for any distribution function
  $G(\cdot)$, the random variable $G^{-1}(U)$ has the distribution
  $G(\cdot)$.}  $\indx_{\Fnode\Snode} \: \defn \: \inf \big\{ j:
\sum_{i = 1} ^{j} \probmass_{\Fnode \Snode} (i) > U \big\}$.  Finally
the update~\eqref{EqnSBPUpdate} needs $3 \dimn + 3$ operations. Adding
up these contributions, we find that the SBP algorithm requires at
most $(\dmax + 6) \dimn + 3$ multiplications and/or summations per
iteration per edge to update the messages. As can be seen from
equation~\eqref{EqnBPUpdateEntry}, the regular BP complexity is
$\Th{\dimn ^2}$. Therefore, for graphs with bounded degree (of most
interest in practical applications), the SBP message updates have
reduced the per iteration computational complexity by a factor of
$\dimn$. In addition to computational efficiency, SBP provides us with
a significant gain in message/communication complexity over BP. This
can be observed from the fact that the normalized compatibility matrix
$\tBPMat_{\Fnode\Snode}$ is only a function of edge potentials
$\EPot$, hence known to the node $\Snode$. Therefore, node $\Fnode$
has to transmit the random column index $\indx_{\Fnode\Snode}$ to node
$\Snode$, which can be done with only $\log{\dimn}$ bits. This is a
significant gain over BP that requires transmitting a
$(\dimn-1)$-dimensional vector of real numbers per edge at every
round.  Here we summarize the features of our algorithm that make it
appealing for practical purposes.
\begin{itemize}
\item \emph{Computational complexity}: SBP reduces the per iteration
  complexity by an order of magnitude from $\Theta(\dimn^2)$ to
  $\Theta(\dimn)$.
\item \emph{Communication complexity}: SBP requires transmitting only
  $\log{\dimn}$ bits per edge in contrast to transmitting a
  $(\dimn-1)$-dimensional vector of real numbers in the case of BP.
\end{itemize}

The remainder of the paper is devoted to understanding when, and if
so, how quickly the SBP message updates converge to a BP fixed point.
Let us provide some intuition as to why such a behavior might be
expected.  Recall that the update~\eqref{EqnSBPUpdate} is random,
depending on the choice of index $J$ chosen in step II(b).  Suppose
that we take expectations of the update~\eqref{EqnSBPUpdate} only over
the distribution~\eqref{EqnProbMass}, in effect conditioning on all
past randomness in the algorithm.  (We make this idea precise via the
notion of $\sigma$-fields in our analysis.)  Doing so yields that the
the expectation of the update~\eqref{EqnSBPUpdate} is given by
\begin{align*}
\Exs \big[\Mes{\Time+1}{\Fnode}{\Snode} \mid \Mes{\Time}{\Fnode}{\Snode}\big] \; & = \; (1 - \step) \:
\Mes{\Time}{\Fnode}{\Snode} \, + \, \step \: \sum_{j=1}^\dimn \probmass ^{\Time}_{\Fnode\Snode} (j) \: \tBPMat_{\Fnode \Snode}(:, j).
%
\end{align*}
Recalling the definitions~\eqref{EqnNormalizedCol}
and~\eqref{EqnProbMass} of the matrix $\tBPMat$ and mass function
$\probmass$, respectively, and performing some algebra, we see that,
in an average sense, the SBP message update is equivalent to (a damped
version of the) usual BP message update.  The technical difficulties lie in
showing that despite the fluctuations around this average behavior,
the SBP updates still converge to the BP fixed point when the stepsize
or damping parameter $\step$ is suitably chosen.  We now turn to
precisely this task.


\subsection{Main Theoretical Results}
\label{SecMainResult}

Thus far, we have proposed a stochastic variant of the usual belief
propagation (BP) algorithm.  In contrast to the usual deterministic
updates, this algorithm generates a random sequence
$\{\Mess^\Time\}_{\Time=0}^\infty$ of message vectors.  This
randomness raises two natural questions:
\begin{itemize}
\item Is the SBP algorithm \emph{strongly consistent}?  More
  precisely, assuming that the ordinary BP algorithm has a unique
  fixed point $\Mesast$, under what conditions do we have
  $\Mess^{\Time} \to \Mesast$ almost surely as $\Time \to \infty$?
\item When convergence occurs, \emph{how fast} does it take place?
  The computational complexity per iteration is significantly reduced,
  but what are the trade-offs incurred by the number of iterations
  required?
\end{itemize} 

The goal of this section is to provide some precise answers to these
questions, ones which show that under certain conditions, there are
provable gains to be achieved by the SBP algorithm. We begin with the
case of trees, for which the ordinary BP message updates are known to
have a unique fixed point for any choice of potential functions.  For
any tree-structured problem, the upcoming Theorem~\ref{ThmTree}
guarantees that the SBP message updates are strongly consistent, and
moreover that in terms of the elementwise $\ell_\infty$ norm they
converge in expectation at least as quickly as
$\order(1/\sqrt{\Time})$, where $\Time$ is the number of iterations.
We then turn to the case of general graphs.  Although the BP fixed
point need not be unique in general, a number of contractivity
conditions that guarantee uniqueness and convergence of ordinary BP
have been developed (e.g.,~\cite{Tatikonda02,Ihler05, MooijKapp07,
  RoostaEtal08}).  Working under such conditions, we show in
Theorem~\ref{ThmMain} that the SBP algorithm is strongly consistent,
and we show that the mesn square error decays at least as quickly as
$\order(1/\Time)$.  In addition, we provide high probability bounds on
the error at each iteration, showing that the typical performance is
highly concentrated around its average. Finally, in
Section~\ref{SecContLipschitz}, we provide a new set of sufficient
conditions for contractivity in terms of node/edge potentials and the
graph structure. As we discuss, our theoretical analysis shows not
only that SBP is provably correct, but also that in various regimes,
substantial gains in overall computational complexity can be obtained
relative to the ordinary BP.


\subsubsection{Guarantees for Tree-structured Graphs}
\label{SecMainResultTree}

We begin with the case of a tree-structured graph, meaning a graph
$\graph$ that contains no cycles.  As a special case, the Markov chain
shown in Figure~\ref{FigGraphicalModels}(b) is an instance of such a
tree-structured graph.  Recall that for some integer $r \geq 1$, a
square matrix $\nilpot$ is said to be nilpotent of degree $r$ if
$\nilpot^r = 0$.  (We refer the reader to Horn and
Johnson~\cite{Horn85} for further background on nilpotent matrices and
their properties.)  Also recall the definition of the diameter of a
graph $\graph$, denoted by $\diameter(\graph)$, as the length (number
of edges) of the longest path between any pair of nodes in the graph.
For a tree, this diameter can be at most $\numnode -1$, a bound
achieved by the chain graph.  In stating Theorem~\ref{ThmTree}, we
make use of the following definition: for vectors $x, y \in
\real^{\Dimn}$, we write $x \coneleq y$ if and only if $x(i) \le y(i)$
for all $i = 1, 2, \ldots, \Dimn$. Moreover, for an arbitrary $x \in
\real^{\Dimn}$, let $|x|$ denote the vector obtained from taking the
absolute value of its elements. With this notation in hand, we are now
ready to state our first result.

\begin{theorem}[Tree-structured graphs]
\label{ThmTree} 
For any tree-structured Markov random field, the sequence of messages
$\{\Mess^{\Time}\}_{\Time = 0} ^{\infty}$ generated by the SBP
algorithm with step size $\step = 1 / (\Time + 1)$, has the following
properties:
\begin{enumerate}[(a)]
\item The message sequence $\{\Mess^\Time\}_{\Time = 0}^\infty$
  converges almost surely to the unique BP fixed point $\Mesast$ as
  $\Time \to \infty$.
\item 
There exist a nilpotent matrix $\nilpot \in \real^{\Dimn \times
  \Dimn}$ of degree at most $r = \diameter(\graph)$ such that
the \mbox{$\Dimn$-dimensional} error vector $\Mess^\Time - \Mesast$
satisfies the elementwise inequality
\begin{align}
\label{EqnErrvecTreeElementwise}
\Expt{|\Mess^\Time - \Mesast|} & \coneleq \; 4 \; (I - 2\nilpot)^{-1}
\: \frac{\onevec}{\sqrt{\Time}} \qquad \mbox{for all iterations $\Time
  = 1, 2, \ldots$.}
\end{align}
\end{enumerate}
\end{theorem}

\paragraph{Remarks:} 
The proof of this result is given in Section~\ref{SubSecProofTree}.
Part (a) shows that the SBP algorithm is guaranteed to converge almost
surely to the unique BP fixed point, regardless of the choice of
node/edge potentials and the initial message vector.  Part (b) refines
this claim by providing a quantitative upper bound on the rate of
convergence: in expectation, the $\ell_\infty$ norm of the error
vector is guaranteed to decay at the rate $\order(1/\sqrt{\Time})$.
As noted by a helpful reviewer, the upper bound in part (b) is likely
to be conservative at times, since the inverse matrix $(1-2A)^{-1}$
may have elements that grow exponentially in the graph diameter $r$.
As shown by our experimental results, the theory is overly
conservative in this way, as SBP still behaves well on trees with
large diameters (such as chain).  Indeed, in the following section, we
provide results for general graphs under contractive conditions that
are less conservative.


\subsubsection{Guarantees for General Graphs}
\label{SecMainResultGeneral}

Our next theorem addresses the case of general graphs. In contrast to
the case of tree-structured graphs, depending on the choice of
potential functions, the BP message updates may have multiple fixed
points, and need not converge in general.  A sufficient condition for
both uniqueness and convergence of the ordinary BP message updates, which we
assume in our analysis of SBP, is that the update function $\UpFunc$,
defined in~\eqref{EqnGlobalUpFun}, is \emph{contractive}.  In
particular, it suffices that there exist some $0 < \slack < 2$ such
that
\begin{align}
\label{EqnLipschitz} \|\UpFunc(\Mess) - \UpFunc(\Mess')\|_2 & \leq
\big(1 - \frac{\slack}{2} \big) \, \|\Mess - \Mess'\|_2.
\end{align}
Past work has established contractivity conditions of this form when
the BP updates are formulated in terms of log
messages~\cite{Tatikonda02,Ihler05, MooijKapp07, RoostaEtal08}.  In
Section~\ref{SecContLipschitz}, we use related techniques to establish
sufficient conditions for contractivity for the BP message update
$\UpFunc$ that involves the messages (as opposed to log messages).

Recalling the normalized compatibility matrix with columns
$\tMatCol{j} \defn \EPot (:, j) \NPot(j) / \ColSum{j}$, we define its
minimum and maximum values per row as follows:\footnote{As will be
  discussed later, we can obtain a sequence of more refined (tighter)
  lower $\{\LBound{\Fnode \Snode}{i}{\ell}\}_{\ell=0}^{\infty}$, and
  upper $\{\UBound{\Fnode \Snode}{i}{\ell}\}_{\ell=0}^{\infty}$
  bounds by confining the space of feasible messages.}
\begin{align}
\label{EqnZeroBound}
\LBound{\Fnode \Snode}{i}{0} \defn \min_{j \in \Alphabet}
\tBPMatrix{\Fnode \Snode}(i, j) \, > \, 0 , \quad \text{and} \quad
\UBound{\Fnode \Snode}{i}{0} \defn \max_{j \in \Alphabet}
\tBPMatrix{\Fnode \Snode}(i, j) \, < \,1.
\end{align}
The pre-factor in our bounds involves the constant
\begin{align}
\label{EqnDefnprefac}
\prefac(\Pot) & \defn \; 4 \, \frac {\sum _{(\Fnode \to \Snode) \in
    \DirSet} \big( \max_{i\in \Alphabet} \UBound{\Fnode \Snode} {i}
  {0} \big)} {\sum _{(\Fnode \to \Snode) \in \DirSet} \big( \min_{i\in
    \Alphabet} \LBound{\Fnode \Snode} {i} {0} \big)}.
\end{align}
With this notation, we have the following result:

\begin{theorem}[General graphs]
\label{ThmMain}
Suppose that the BP update function $\UpFunc: \real^\Dimn \rightarrow
\real^\Dimn$ satisfies the contraction condition~\eqref{EqnLipschitz}.
\begin{enumerate}[(a)]
\item Then BP has a unique fixed point $\Mesast$, and the SBP message
  sequence $\{\Mess^{\Time}\}_{\Time=0}^\infty$, generated with the
  step size $\step = \order(1/\Time)$, converges almost surely to
  $\Mesast$ as $t \rightarrow \infty$.
\item With the step size $\step = \Coef / (\slack \, (\Time + 2))$ for
  some fixed $1 < \Coef < 2$, we have
\begin{align}
\label{EqnMSEBound}
\frac {\Expt{\vnorm {\Mess^{\Time} - \Mesast} {2} ^2 }} {\vnorm
  {\Mesast}{2} ^2} \; \le \; \frac {3^\Coef \: \prefac(\Pot) \:
  \Coef^2 } {2^{\Coef} \: \slack ^2 (\Coef - 1)} \: \bigg(\frac{1}
      {\Time}\bigg) \, + \, \frac {\vnorm {\Mess^{0} - \Mesast} {2}
          ^2}{\vnorm {\Mesast}{2} ^2} \: \bigg( \frac {2} {\Time}
      \bigg)^{\Coef}
\end{align}
for all iterations $\Time = 1, 2,\ldots.$
\item With the step size \mbox{$\step =
  1 / (\slack \,(\Time + 1))$}, we have
\begin{align}
\label{EqnMSEBoundPartc}
\frac {\Expt{\vnorm {\Mess^{\Time} - \Mesast} {2} ^2 }} {\vnorm
  {\Mesast}{2} ^2} \; \le \; \frac{\prefac(\Pot)}{\slack^2} \: \bigg(
\frac{1 + \log{\Time}}{\Time} \bigg);
\end{align}
also for every $0< \epsilon <1$ and $\Time \ge 2$, we have
\begin{align}
\label{EqnFiniteSampleBound}
\frac {\vnorm {\Mess^{\Time} - \Mesast} {2} ^2 } {\vnorm
  {\Mesast}{2} ^2} \; & \leq \; \frac{\prefac(\Pot)}{\slack^2} \: \bigg(1
 + \frac{8}{\sqrt{\epsilon}}\bigg) \bigg( \frac{1 + \log{\Time}}{\Time} \bigg)
\end{align}
with probability at least $1 - \epsilon$.

\end{enumerate}
\end{theorem}

\paragraph{Remarks:} 
The proof of Theorem~\ref{ThmMain} is given in
Section~\ref{SubSecProofMain}.  Here we discuss some of the various
guarantees that it provides.  First, part (a) of the theorem shows
that the SBP algorithm is strongly consistent, in that it converges
almost surely to the unique BP fixed point.  This claim is analogous
to the almost sure convergence established in Theorem~\ref{ThmTree}(a)
for trees.  Second, the bound~\eqref{EqnMSEBound} in
Theorem~\ref{ThmMain}(b) provides a non-asymptotic bound on the
normalized mean-squared error $\Exs[\|\Mess^{\Time} -
  \Mesast\|_2^2]/\|\Mesast\|_2^2]$.  For the specified choice of
  step-size \mbox{($1 < \Coef < 2$),} the first component of the
  bound~\eqref{EqnMSEBound} is dominant, hence the expected error (in
  squared $\ell_2$-norm) is of the order\footnote{At least
    superficially, this rate might appear faster than the
    $1/\sqrt{\Time}$ rate established for trees in
    Theorem~\ref{ThmTree}(b); however, the reader should be careful to
    note that Theorem~\ref{ThmTree} involves the elementwise
    $\ell_\infty$-norm, which is not squared, as opposed to the
    squared $\ell_2$-norm studied in Theorem~\ref{ThmMain}.}
  $1/\Time$. Therefore, after $\Time = \Theta(1/\delta)$ iterations,
  the SBP algorithm returns a solution with MSE at most
  $\order(\delta)$. Finally, part (c) provides bounds, both in
  expectation and with high probability, for a slightly different step
  size choice.  On one hand, the bound in
  expectation~\eqref{EqnMSEBoundPartc} is of the order $\order(\log
  \Time /\Time)$, and so includes an additional logarithmic factor not
  present in the bounds from part (b).  However, as shown in the high
  probability bound~\eqref{EqnFiniteSampleBound}, the squared error is
  also guaranteed to satisfy a sample-wise version of the same bound
  with high probability.  This theoretical claim is consistent with
  our later experimental results, showing that the error exhibits
  tight concentration around its expected behavior.\\

Let us now compare the guarantees of SBP to those of BP.  Under the
contraction condition of Theorem~\ref{ThmMain}, the ordinary BP
message updates are guaranteed to converge geometrically quickly,
meaning that $\Theta(\log(1/\delta))$ iterations are sufficient to
obtain $\delta$-accurate solution.  In contrast, under the same
conditions, the SBP algorithm requires $\Theta(1/\delta)$ iterations
to return a solution with MSE at most $\delta$, so that its iteration
complexity is larger.  However, as noted earlier, the BP message
updates require $\Theta(\dimn^2)$ operations for each edge and
iteration, whereas the SBP message updates require only
$\Theta(\dimn)$ operations.  Putting the pieces together, we conclude
that:
\begin{itemize}
\item on one hand, ordinary BP requires $\Theta \big(|\edge| \,
  \dimn^2 \, \log(1/\delta) \big)$ operations to compute the fixed
  point to \mbox{ $\delta$-accuracy};
\item in comparison, SBP requires $\Theta \big( |\edge| \, \dimn \, (1
  /\delta) \big)$ operations to compute the fixed point to expected accuracy
  $\delta$.
\end{itemize}
Consequently, we see that as long the desired tolerance is not too
small---in particular, if $\delta \geq 1/\dimn$---then SBP leads to
computational savings.  In many practical applications, the state
dimension is on the order of $10^3$ to $10^5$, so that the precision
$\delta$ can be of the order $10^{-3}$ to $10^{-5}$ before the
complexity of SBP becomes of comparable order to that of BP. Given
that most graphical models represent approximations to reality, it is
likely that larger tolerances $\delta$ are often of interest.


\subsubsection{Sufficient Conditions for Contractivity}
\label{SecContLipschitz}

Theorem~\ref{ThmMain} is based on the assumption that the update
function is contractive, meaning that its Lipschitz constant $\Lips$
is less than one.  In past work, various authors have developed
contractivity conditions, based on analyzing the log messages, that
guarantee uniqueness and convergence of ordinary BP
(e.g.,~\cite{Tatikonda02,Ihler05, MooijKapp07, RoostaEtal08}).  Our
theorem requires contractivity on the messages (as opposed to log
messages), which requires a related but slightly different argument.
In this section, we show how to control $\Lips$ and thereby provide
sufficient conditions for Theorem~\ref{ThmMain} to be applicable.

Our contractivity result applies when the messages under consideration
belong to a set of the form
\begin{align}
\label{EqnDefnBall}
\Ball \defn \bigg\{\Mess \in \real ^{\Dimn} \; | \; \sum_{i \in
  \Alphabet} \Mess_{\Fnode\Snode} (i) = 1, \; \LBnd {\Fnode \Snode}
      {i} \le \Mess_{\Fnode\Snode} (i) \le \UBnd {\Fnode \Snode} {i}
      \quad \forall (\Fnode \to \Snode) \in \DirSet, \; \forall i \in
      \Alphabet \bigg\},
\end{align}
for some choice of the upper and lower bounds---namely, $\UBnd{ \Fnode
  \Snode}{i}$ and $\LBnd{ \Fnode \Snode}{i}$ respectively.  For
instance, for all iterations $\Time = 0, 1, \ldots$, the messages
always belong to a set of this form\footnote{It turns out that the BP
  update function on the directed edge ($\Fnode \to \Snode$) is a
  convex combination of normalized columns $\tMatCol{j}$ for $j = 1,
  \ldots, \dimn$.  Therefore, we have $\LBound {\Fnode \Snode} {i} {0}
  \le \Mess_{\Fnode \Snode} (i) \le \UBound {\Fnode \Snode} {i} {0}$,
  for all $i = 1, \ldots, \dimn$.}  with $\LBnd{ \Fnode \Snode}{i} =
\LBound{\Fnode \Snode}{i}{0}$ and $\UBnd{\Fnode \Snode}{i} =
\UBound{\Fnode \Snode}{i}{0}$, as previously
defined~\eqref{EqnZeroBound}.  Since the bounds $(\LBound{\Fnode
  \Snode }{i}{0}, \UBound{\Fnode \Snode}{i}{0})$ do not involve the
node potentials, one suspects that they might be tightened at
subsequent iterations, and indeed, there is a progressive refinement
of upper and lower bounds of this form.  Indeed, assuming that the
messages belong to a set $\Ball$ at an initial iteration, then for any
subsequent iterations, we are guaranteed the inclusion
\begin{align}
\label{EqnRefined}
\Mess \in \UpFunc (\Ball) \defn \big\{ \UpFunc (\MessP) \in
\real ^{\Dimn} \; \mid \; \MessP \in \Ball \big\},
\end{align}
which then leads to the refined upper and lower bounds
\begin{align*}
\LBound {\Fnode \Snode} {i} {1} & \defn \inf_{\Mess \in \Ball} \;
\bigg\{ \sum _{j = 1} ^{\dimn} \: \tMatEntry \: \frac { \ColSum{j} \:
  \AgMes{}{\Fnode}{\Snode}(j)} {\sum_{\ell = 1} ^{\dimn} \ColSum{\ell}
  \; \AgMes{}{\Fnode}{\Snode}(\ell) } \bigg\}, \quad \text{and}
\\ \UBound {\Fnode \Snode} {i} {1} & \defn \sup_{\Mess \in \Ball} \;
\bigg\{ \sum _{j = 1} ^{\dimn} \: \tMatEntry \: \frac {\ColSum{j} \;
  \AgMes{}{\Fnode}{\Snode} (j)} {\sum_{\ell = 1} ^{\dimn}
  \ColSum{\ell} \; \AgMes{}{\Fnode}{\Snode} (\ell)} \bigg\},
\end{align*}
where we recall the quantity $\AgMes{}{\Fnode}{\Snode} =
\prod_{\CompNeig} \Mes{}{\Tnode}{\Fnode}$ previously
defined~\eqref{EqnProdIncMess}. While such refinements are possible,
in order to streamline our presentation, we focus primarily on the
zero'th order bounds $\LBnd{\Fnode \Snode}{i} = \LBound{\Fnode
  \Snode}{i}{0}$, and $\UBnd{\Fnode \Snode}{i} = \UBound{\Fnode
  \Snode}{i}{0}$.

Given a set $\Ball$ of the form~\eqref{EqnDefnBall}, we associate with
the directed edge $(\Fnode \to \Snode)$ and $(\Tnode \to \Fnode)$
(where $\CompNeig$) the non-negative numbers
\begin{subequations}
\begin{align}
\label{EqnDefnBigPhi}
\PHIFUN(\Fnode, \Snode) & \defn  \sum_{\CompNeig} \big( \TermPhi{\Fnode
  \Snode}{\Tnode \Fnode} \: (\TermPhi{\Fnode \Snode} {\Tnode \Fnode} +
\TermChi{\Fnode \Snode} {\Tnode \Fnode}) \big) ^{\frac{1}{2}}, \quad \mbox{and} \\
\label{EqnDefnBigPhiTwo}
\PHIFUNTWO(\Tnode, \Fnode) & \defn \sum_{\Snode \in \Neig (\Fnode) \setminus \{\Tnode\}} \big( \TermPhi{\Fnode
  \Snode}{\Tnode \Fnode} \: (\TermPhi{\Fnode \Snode} {\Tnode \Fnode} +
\TermChi{\Fnode \Snode} {\Tnode \Fnode}) \big) ^{\frac{1}{2}},
\end{align}
\end{subequations}
where
\begin{subequations}
\begin{align}
\label{EqnDefnPhi}
\TermPhi{\Fnode \Snode} {\Tnode \Fnode} & \defn \max_{j \in \Alphabet} \: \sup_{\Mess \in \Ball}
\bigg\{ \frac{\ColSum {j} \: \AgMess (j) } {\sum _{k
    =1} ^{\dimn} \ColSum {k} \: \AgMess (k) } \: \frac {1}
   {\Mess_{\Tnode \Fnode} (j)} \bigg\}, \quad \mbox{and} \\
\label{EqnDefnChi}
\TermChi {\Fnode \Snode} {\Tnode \Fnode} & \defn \max_{j \in
  \Alphabet} \: \sup _{\Mess \in \Ball}
\bigg\{ \frac {\ColSum {i} \: \AgMess
     (i)} {\big(\sum_{k=1} ^{\dimn} \ColSum {k} \: \AgMess (k) \big)
     ^{2}} \: \sum_{j=1} ^{\dimn} \frac{\ColSum{j} \: \AgMess(j)}
   {\Mess_{\Tnode \Fnode (j)}} \bigg\}.
\end{align}
\end{subequations}

Recall the normalized compatibility matrix $\tBPMat_{\Fnode \Snode}
\in \real ^{\dimn \times \dimn}$ on the directed edge $(\Fnode \to
\Snode)$, as previously defined in equation~\eqref{EqnNormalizedCol}.
Since $\tBPMatrix{\Fnode \Snode} ^T$ has positive entries, the
Perron-Frobenius theorem~\cite{Horn85} guarantees that the maximal
eigenvalue is equal to one, and is associated with a pair of left and
right eigenvectors (unique up to scaling) with positive entries.
Since $\tBPMatrix{\Fnode \Snode}^T$ is row-stochastic, any multiple of
the all-one vector $\onevec$ can be chosen as the right eigenvector.
Letting $\Leigvec{\Fnode\Snode} \in \real^{\dimn}$ denote the left
eigenvector with positive entries, we are guaranteed that $\onevec^T
\Leigvec{\Fnode\Snode} > 0$, and hence we may define the matrix
$\tBPMatrix{\Fnode \Snode}^T - \onevec \Leigvec{\Fnode\Snode}^T /
(\onevec^T \Leigvec{\Fnode\Snode})$.  By construction, this matrix has
all of its eigenvalues strictly less than $1$ in absolute value (Lemma
8.2.7,~\cite{Horn85}).

\begin{proposition}
\label{PropUpFunLipschitz}
The global update function $\UpFunc: \real ^{\Dimn} \to \real^{\Dimn}$
defined in equation~\eqref{EqnGlobalUpFun} is Lipschitz with constant
at most
\begin{align}
\label{EqnUpFunLipsConst}
\Lips \; & \defn \; 2 \max_{(\Fnode \to \Snode) \in \DirSet}
\matsnorm{\tBPMatrix{\Fnode \Snode} - \frac{ \Leigvec{\Fnode\Snode}
    \onevec^T}{\onevec^T \Leigvec{\Fnode\Snode} } } {2} \; \max_{
  (\Fnode \to \Snode) \in \DirSet}\PHIFUN(\Fnode, \Snode) \; \max_{
  (\Tnode \to \Fnode) \in \DirSet} \PHIFUNTWO(\Tnode, \Fnode),
\end{align}
where $\matsnorm{\cdot}{2}$ denotes the maximum singular value of a
matrix.
\end{proposition}

\noindent In order to provide some intuition for
Proposition~\ref{PropUpFunLipschitz}, let us consider a simple but
illuminating example.

\begin{exas}[Potts model]
\label{ExaPotts}
The \emph{Potts model}~\cite{FlezHutt06, SunEtal03,KlaEtal06} is often
used for denoising, segmentation, and stereo computation in image
processing and computer vision.  It is a pairwise Markov random field
that is based on edge potentials of the form
\begin{align*}
\EPot (i, j) & = \begin{cases} 1 & \mbox{if $i = j$, and} \\ \pott &
  \mbox{if $i \neq j$.}
\end{cases},
\end{align*}
for all edges $(\Fnode, \Snode) \in \edge$ and $i,j \in \{1,2, \ldots,
\dimn\}$.  The parameter $\pott \in (0,1]$ can be tuned to enforce
  different degrees of smoothness: at one extreme, setting $\pott = 1$
  enforces no smoothness, whereas a choice close to zero enforces a
  very strong type of smoothness.  (To be clear, the special structure
  of the Potts model can be exploited to compute the BP message
  updates quickly; our motivation in considering it here is only to
  provide a simple illustration of our contractivity condition.)
 
For the Potts model, we have $\ColSum{j} \: = \: \NPot (j) \: (1 +
(\dimn - 1) \pott)$, and hence $\tBPMat_{\Fnode\Snode}$ is a symmetric
matrix with
\begin{align*}
\tMatEntry \; = \; 
\begin{cases}
\frac{1}{1 + (\dimn - 1) \pott} & \mbox{if $i = j$}
\\ \frac{\pott}{1 + (\dimn - 1) \pott} & \mbox{if $i \neq j$.}
\end{cases}
\end{align*}
Some straightforward algebra shows that the second largest singular
value of $\tBPMat_{\Fnode \Snode}$ is given by $(1 - \pott) / (1 +
(\dimn - 1) \pott)$, whence
\begin{align*}
\max_{(\Fnode \to \Snode) \in \DirSet} \matsnorm {\tBPMatrix{\Fnode
    \Snode} - \frac{ \Leigvec{\Fnode\Snode} \onevec^T}{\onevec^T
    \Leigvec{\Fnode\Snode} }}{2} \; = \; \frac{1 - \pott}{ 1 + (\dimn
  - 1) \pott}.
\end{align*}

The next step is to find upper bounds on the terms $\PHIFUN(\Fnode,
\Snode)$ and $\PHIFUNTWO (\Tnode, \Fnode)$, in particular by upper
bounding the quantities $\TermPhi{\Fnode \Snode}{\Tnode \Fnode}$ and
$\TermChi{\Fnode \Snode}{\Tnode \Fnode}$, as defined in
equations~\eqref{EqnDefnPhi} and~\eqref{EqnDefnChi} respectively.  In
Appendix~\ref{AppPotts}, we show that the Lipschitz function of $\UpFun$ is
upper bounded as
\begin{align*}
\Lips & \leq 4 \: (1 - \pott) (1 + (\dimn - 1)\pott) \: \max_{\Fnode
  \in \node} \bigg\{ \frac {(\degr_{\Fnode} - 1)^2} {\pott ^{2
    \degr_{\Fnode}} } \, \max_{j \in \Alphabet} \bigg\{ \frac
      {\NPot(j)} {\sum_{\ell=1}^{\dimn} \NPot(\ell)} \bigg\} ^2 \bigg\},
\end{align*}
where $\degr_{\Fnode}$ is the degree of node $\Fnode$.  Therefore, a
sufficient condition for contractivity in the case of the Potts model
is
\begin{align}
\label{EqnSuffPotts}
\max_{\Fnode \in \node} \bigg\{ \frac {(\degr_{\Fnode} - 1)}
    {\pott^{\degr_{\Fnode}} } \, \max_{j \in \Alphabet} \bigg\{ \frac
    {\NPot(j)} {\sum_{\ell=1}^\dimn \NPot(\ell)} \bigg\} \bigg\} \; <
    \; \left(\frac{1} {4 \: (1 - \pott) \: (1 + (\dimn -
      1)\pott)}\right)^{\frac{1}{2}}.
\end{align}
To gain intuition, consider the special case in which the node
potentials are uniform, so that $\NPot(j) / (\sum_{\ell=1}^\dimn
  \NPot(\ell)) = 1/\dimn$.  In this case, for any graph with bounded
node degrees, the bound~\eqref{EqnSuffPotts} guarantees contraction
for all $\pott$ in an interval $[\epsilon, 1]$.  For non-uniform node
potentials, the inequality~\eqref{EqnSuffPotts} is weaker, but it can
be improved via the refined sets~\eqref{EqnRefined} discussed
previously.
\end{exas}


\section{Proofs}
\label{SecProof}

We now turn to the proofs of our two main results, namely Theorems
\ref{ThmTree} and~\ref{ThmMain}, as well as the auxiliary result,
Proposition~\ref{PropUpFunLipschitz}, on contractivity of the BP
message updates.  For our purposes, it is convenient to note that the
ordinary BP update can be written as an expectation of the form
\begin{align}
\label{EqnBPExpUpdate}
\UpFun (\Mess^{\Time}) = \Exp{\ind{\Time+1}{\Fnode \Snode} \sim \probmass
  _{\Fnode\Snode}^{\Time}} {\tMatCol{\ind{\Time+1}{\Fnode\Snode}} },
\end{align}
for all $\Time = 0, 1, \ldots$. Here the index $\ind{\Time+1}{\Fnode \Snode}$
is chosen randomly according to the probability mass
function~\eqref{EqnProbMass}.


\subsection{Proof of Theorem~\ref{ThmTree}}
\label{SubSecProofTree}

We begin by stating a lemma that plays a central role in the proof of
Theorem~\ref{ThmTree}.  
\begin{lemma}
\label{LemNilpotent}
For any tree-structured Markov random field, there exist a nilpotent
matrix \mbox{$\nilpot \in \real^{\Dimn \times \Dimn}$} of degree at
most $r = \diameter(\graph)$ such that
\begin{align}
\label{EqnLemNilpot}
|\UpFunc (\Mess) - \UpFunc (\MessP)| \; \coneleq \; \nilpot \: |\Mess
- \MessP|,
\end{align}
for all $\Mess, \MessP \in \Ball$.
\end{lemma}
\noindent 
The proof of this lemma is somewhat technical, so that we defer it to
Appendix~\ref{AppLemNilpotent}.  In interpreting this result, the
reader should recall that for vectors $x, y \in \real^{\Dimn}$, the
notation $x \coneleq y$ denotes inequality in an elementwise
sense---i.e., $x(i) \leq y(i)$ for $i = 1, \ldots, \Dimn$. \\

An immediate corollary of this lemma is the existence and uniqueness
of the BP fixed point. Since we may iterate
inequality~\eqref{EqnLemNilpot}, we find that
\begin{align*}
|\UpFunc^{(\ell)} (\Mess) - \UpFunc^{(\ell)} (\MessP)| & \coneleq \;
\nilpot^{\ell} \: |\Mess - \MessP|, 
\end{align*}
for all iterations $\ell = 1, 2, \ldots$, and arbitrary messages
$\Mess$, $\MessP$, where $\UpFunc^{(\ell)}$ denotes the composition of
$\UpFunc$ with itself $\ell$ times.  The nilpotence of $\nilpot$
ensures that $\nilpot^\diam = 0$, and hence $\UpFunc^{(\diam)}(\Mess)
= \UpFunc^{(\diam)}(\MessP)$ for all messages $\Mess$, and
$\MessP$. Let $\Mesast = \UpFunc^{(\diam)}(\Mess)$ denote the common
value. The claim is that $\Mesast$ is the unique fixed point of the BP
update function $\UpFunc$. This can be shown as follows: from
Lemma~\ref{LemNilpotent} we have
\begin{align*}
|\UpFunc(\Mesast) \, - \, \Mesast| \; = \; |\UpFunc^{(\diam+1)}(\Mess)
\, - \, \UpFunc^{(\diam)}(\Mess)| \; \coneleq \; \nilpot \:
|\UpFunc^{(\diam)}(\Mess) \, - \, \UpFunc^{(\diam-1)}(\Mess)|.
\end{align*}
Iterating the last inequality for the total of $\diam$ times, we obtain
\begin{align*}
|\UpFunc(\Mesast) \, - \, \Mesast| \; \coneleq \; \nilpot^{\diam} \:
|\UpFunc(\Mess) \, - \, \Mess| \; = \; 0,
\end{align*}
and hence $\UpFunc(\Mesast) = \Mesast$. On the other hand, the
uniqueness of the BP fixed point is a direct consequence of the facts
that for any fixed point $\Mesast$ we have $\UpFunc^{(\diam)}(\Mesast)
= \Mesast$, and for all arbitrary messages $\Mess$, $\MessP$ we have
$\UpFunc^{(\diam)} (\Mess) = \UpFunc^{(\diam)} (\MessP)$. Accordingly,
we see that Lemma~\ref{LemNilpotent} provides an alternative proof of
the well-known fact that BP converges to a unique fixed point on trees
after at most $\diam = \diameter(\graph)$ iterations.\\

\noindent We now show how Lemma~\ref{LemNilpotent} can be used to
establish the two claims of Theorem~\ref{ThmTree}.

\subsubsection{Part (a): Almost Sure Consistency}

We begin with the almost sure consistency claim of part (a).  By
combining all the local updates, we form the global update rule
\begin{align}
\label{EqnGlobalUpdate}
\Mess^{\Time + 1} \, = \, (1 - \step) \: \Mess ^{\Time} + \step \:
\normVec{\Time + 1} \quad \mbox{for iterations $\Time = 0, 1, 2,
  \ldots$,}
\end{align}
where \mbox{$\normVec{\Time + 1} \defn \{
  \tMatCol{\ind{\Time+1}{\Fnode \Snode}} \}_{(\Fnode \to \Snode) \in
    \DirSet}$} is the $\Dimn$-dimensional vector obtained from
stacking up all the normalized columns $\tMatCol{\ind{\Time+1}{\Fnode
    \Snode}}$. Defining the vector $\MarDif {\Time + 1} \defn \normVec
{\Time +1} - \UpFunc (\Mess ^{\Time}) \in \real ^{\Dimn}$, we can
rewrite the update~\eqref{EqnGlobalUpdate} as
\begin{align}
\label{EqnGlobalUpdateTwo}
\Mess ^{\Time + 1} \, = \, (1 - \step) \: \Mess ^{\Time} + \step \:
\UpFunc (\Mess ^{\Time}) + \step \: \MarDif {\Time + 1} \quad
\mbox{for $\Time = 0, 1, 2, \ldots$.}
\end{align}
With our step size choice $\step = 1 / (\Time + 1)$, unwrapping the
recursion~\eqref{EqnGlobalUpdateTwo} yields the representation
\begin{align*}
\Mess^{\Time} \, = \, \frac {1}{\Time} \: \sum _{\ell = 0}
  ^{\Time - 1} \UpFunc (\Mess ^{\ell}) \, + \, \frac{1}{\Time} \:
  \sum_{\ell = 1} ^{\Time} \MarDif{\ell}.
\end{align*}
Subtracting the unique fixed point $\Mesast$ from both sides then
leads to
\begin{align}
\label{EqnSunrise}
\Mess^{\Time} - \Mesast \, = \, \frac {1}{\Time} \: \sum _{\ell = 1}
^{\Time - 1} ( \UpFunc (\Mess ^{\ell}) - \UpFunc (\Mesast) ) \, + \,
\underbrace{\frac{1}{\Time} \, \sum_{\ell = 1} ^{\Time} \MarDif{\ell}
  + \frac{1}{\Time} \, ( \UpFunc (\Mess ^{0}) - \UpFunc
  (\Mesast))}_{\smallTerm{\Time}},
\end{align}
where we have introduced the convenient shorthand $\smallTerm{\Time}$.
We may apply triangle inequality to each element of this vector
equation; doing so and using Lemma~\ref{LemNilpotent} to upper bound
the terms $|\UpFunc (\Mess ^{\ell}) - \UpFunc (\Mesast)|$, we obtain
the element-wise inequality
\begin{align*}
|\Mess^{\Time} - \Mesast| & \coneleq \frac {1}{\Time} \sum_{\ell = 1}
^{\Time - 1} \nilpot \: |\Mess^{\ell} - \Mesast| \, + \,
|\smallTerm{\Time}| \quad \mbox{for $\Time = 1, 2, \ldots$.}
\end{align*}
Since $\nilpot^{\diam}$ is the all-zero matrix, unwrapping the last
inequality $\diam = \diameter (\graph)$ times yields the element-wise
upper bound
\begin{align}
\label{EqnUnwrapElemwisIneq}
|\Mess ^{\Time} - \Mesast| & \coneleq \; \Term_0^\Time \, + \, \nilpot
\Term_1^{\Time} + \nilpot^2 \Term_2^\Time + \cdots + \nilpot^{\diam-1}
\Term_{\diam-1}^\Time,
\end{align}
where the terms $\Term_\ell^\Time$ are defined via the recursion
$\Term_{\ell}^\Time \defn \frac{1}{t} \sum_{j=1}^{t-1}
\Term_{\ell-1}^{j}$ \mbox{for $\ell = 1, \ldots, \diam -1$,} with
initial conditions $\Term_0^\Time \defn |\smallTerm{\Time}|$. \\

It remains to control the sequences $\{\Term_\ell^{\Time}\}_{\Time=1}
^{\infty}$ for $\ell = 0, 1, \ldots, \diam - 1$.  In order to do so,
we first establish a martingale difference property for the variables
$\MarDif{\Time}$ defined prior to equation~\eqref{EqnGlobalUpdateTwo}.
For each $\Time = 0, 1, 2, \ldots$, define the $\sigma$-field
$\Filt{\Time} \defn \sigma (\Mess^{0}, \Mess ^{1}, \ldots,
\Mess^{\Time})$, as generated by the randomness in the messages up to
time $\Time$.  Based on the representation~\eqref{EqnBPExpUpdate}, we
see that $\Expt {\MarDif {\Time + 1} | \Filt{\Time}} = \zerovec$,
showing that $\{\MarDif{\Time + 1}\}_{\Time = 0} ^{\infty}$ forms
martingale difference sequence with respect to the filtration
$\{\Filt{\Time}\}_{\Time=0}^\infty$.  From the definition, it can be
seen that the entries of $\MarDif{\Time + 1}$ are bounded; more
precisely, we have $|\MarDif{\Time + 1} (i)| \le 1$ for all iterations
$\Time = 0, 1, 2, \ldots$, and all states $i = 1, 2, \ldots \Dimn$.
Consequently, the sequence $\{\MarDif{\ell}\} _{\ell = 1} ^{\infty}$
is a bounded martingale difference sequence. \\

We begin with the term $\Term_0^\Time$. Since $\MarDif{\ell}$ is a
bounded martingale difference, standard convergence
results~\cite{Durrett95} guarantee that $|\sum _{\ell = 1} ^{\Time}
\MarDif{\ell} | / \Time \to \zerovec$ almost surely.  Moreover, we
have the bound $|\UpFunc(\Mess^0) - \UpFunc(\Mesast)|/\Time \coneleq
\onevec / \Time$.  Recalling the definition of $\smallTerm{\Time}$
from equation~\eqref{EqnSunrise}, we conclude that $\Term_0^{\Time} =
|\smallTerm{\Time}|$ converges to the all-zero vector almost surely as
$\Time \to \infty$.  In order to extend our argument to the terms
$\Term_\ell^\Time$ for $\ell = 1, \ldots, \diam -1$, we make use of
the following fact: for any sequence of real numbers
$\{x^\Time\}_{\Time=0} ^{\infty}$ such that $x^{\Time} \to 0$, then we
also have $(\sum_{\ell = 0}^{\Time -1} x^{\ell}) / \Time \to 0$ (e.g.,
see Royden~\cite{Royden}). Consequently, for any realization $\omega$
such that the deterministic sequence $\{\Term_0^\Time(\omega)\}_{\Time
  = 0}^\infty$ converges to zero, we are also guaranteed that the
sequence $\{\Term_1^\Time(\omega)\}_{\Time = 0}^\infty$, with elements
$\Term_1^\Time(\omega) = (\sum_{j = 1}^{\Time-1} \Term_0 ^j (\omega))
/ \Time$, converges to zero.  Since we have shown that $\Term_0^\Time
\stackrel{\mathrm{a.s.}}{\rightarrow} 0$, we conclude that $\Term_1^\Time
\stackrel{\mathrm{a.s.}}{\rightarrow} 0$ as well.  This argument can be
iterated, thereby establishing almost sure convergence for all of the
terms $\Term_\ell^\Time$.  Putting the pieces together, we conclude
that the vector $|\Mess ^{\Time} - \Mesast|$ converges almost surely
to the all-zero vector as $\Time \to \infty$, thereby completing the
proof of part (a).


\subsubsection{Part (b): Bounds on Expected Absolute Error}

We now turn to part (b) of Theorem~\ref{ThmTree}, which provides upper
bounds on the expected absolute error.  We establish this claim by
exploiting some martingale concentration inequalities~\cite{ChuLu06}.
From part (a), we know that $\{\MarDif{\Time}\}_{\Time = 1} ^{\infty}$
is a bounded martingale difference sequence, in particular with
$|\MarDif{\Time}(i)| \leq 1$.  Applying the Azuma-Hoeffding
inequality~\cite{ChuLu06} yields the tail bound
\begin{align*}
\mprob\bigg( \frac{1}{\Time} \, \mid \sum _{\ell = 1} ^{\Time}
  \MarDif{\ell}(i)| \, > \, \gamma \bigg) \;  \leq \; 2 \exp\bigg(
-\frac{\Time \: \gamma^2}{2} \bigg),
\end{align*}
for all $\gamma>0$, and $i = 1, 2, \ldots, \Dimn$.  By integrating this
tail bound, we can upper bound the mean: in particular, we have
\begin{align*}
\Exs \bigg[\frac{1}{\Time} \, | \sum _{\ell = 1} ^{\Time}
  \MarDif{\ell}(i)|\bigg] \; = \; \int_{0} ^{\infty} \mprob\bigg(\frac{1}{\Time}
  \, | \sum _{\ell = 1} ^{\Time} \MarDif{\ell}(i)| \: > \: \gamma\bigg) \:
d\gamma \; \le \; \sqrt {\frac {2\pi} {\Time }}, 
\end{align*}
and hence
\begin{align}
\label{EqnExpElemwiseIneq}
\Expt{\Term_0^\Time} \; = \; \Expt {|\smallTerm{\Time}|} \; \coneleq
\; \sqrt {\frac {2\pi} {\Time }} \: \onevec \, + \,
\frac{\onevec}{\Time} \; \coneleq \; \frac{\SPECCON} {\sqrt{\Time}} \:
\onevec.
\end{align}
Turning to the term $\Term_1^\Time$, we have
\begin{align*}
\Exs[\Term_1^\Time] & \; = \frac{1}{\Time} \, \sum_{\ell=1}^{\Time-1}
\Exs[\Term_0^\ell] \; \; \stackrel{\mathrm{(i)}}{\coneleq} \; \;
\frac{1}{\Time} \, \sum_{\ell=1}^{\Time-1} \frac{\SPECCON} {\sqrt{\ell}}
\: \onevec \; \stackrel{\mathrm{(ii)}}{\coneleq} \; \frac{2 \, \cdot \,
  4}{\sqrt{\Time}} \: \onevec,
\end{align*}
where step (i) uses the inequality~\eqref{EqnExpElemwiseIneq}, and
step (ii) is based on the elementary upper bound \mbox{$\sum_{\ell =
    1} ^{\Time - 1} 1 / \sqrt{\ell} \, \leq \, 1 \, + \,
  \int_{1} ^{\Time - 1} 1 / \sqrt{x} \: dx \, < \, 2
  \sqrt{\Time}$.}  By repeating this same argument in a recursive
manner, we conclude that \mbox{$\Expt{\Term_{\ell} ^{\Time}} \coneleq
  (2 ^{\ell} \, \cdot 4 / \sqrt{\Time}) \: \onevec$} for $\ell = 2, 3,
\ldots, \diam - 1$. Taking the expectation on both sides of the the
inequality~\eqref{EqnUnwrapElemwisIneq} and substituting these upper
bounds, we obtain
\begin{align*}
\Expt {|\Mess^{\Time} - \Mesast|} \; \coneleq \; 4 \;
\bigg(\sum_{\ell=0}^{\diam - 1} 2^\ell \nilpot^\ell \bigg) \:
\frac{\onevec}{\sqrt{\Time}} \; = \; 4 \, (I - 2 \nilpot) ^{-1} \:
\frac{\onevec}{\sqrt{\Time}},
\end{align*}
where we have used the fact that $\nilpot^{\diam} = 0$.


\subsection{Proof of Theorem~\ref{ThmMain}}
\label{SubSecProofMain}

We now turn to the proof of Theorem~\ref{ThmMain}. Note that since the
update function is contractive, the existence and uniqueness of the BP
fixed point is an immediate consequence of the Banach fixed-point
theorem~\cite{AgarwalBook04}.


\subsubsection{Part (a): Almost Sure Consistency}
\label{PargPartA}

We establish part (a) by applying the Robbins-Monro theorem, a
classical result from stochastic approximation theory
(e.g.,~\cite{RobbinsMonro51,Benveniste90}). In order to do so, we
begin by writing the update~\eqref{EqnSBPUpdate} in the form
\begin{align*}
\Mes{\Time+1}{\Fnode}{\Snode} & = \Mes{\Time}{\Fnode}{\Snode} - \step
\underbrace{ \big\{ \Mes{\Time}{\Fnode}{\Snode}
  -\tMatCol{\ind{\Time+1}{\Fnode\Snode}} \big \}}_{\VF_{\Fnode \Snode}
  (\Mes{\Time}{\Fnode}{\Snode} , \ind{\Time+1}{\Fnode \Snode})},
\end{align*}
where for any realization $\bar{J}_{\Fnode \Snode} \in \{1, 2, \ldots,
\dimn\}$, the mapping $\Mes{}{\Fnode}{\Snode} \mapsto \VF_{\Fnode
  \Snode}(\Mes{}{\Fnode}{\Snode}, \bar{J}_{\Fnode \Snode})$ should
be understood as a function from $\real^\dimn$ to $\real^\dimn$.  By
concatenating together all of these mappings, one for each directed
edge $(\Fnode \to \Snode)$, we obtain a family of mappings $\VF(\cdot , 
\bar{J})$ from $\real^{\Dimn}$ to $\real^{\Dimn}$, one for each
realization $\bar{J} \in \{1, 2, \ldots, \dimn\}^{2|\DirEdge|}$ of
column indices.  

With this notation, we can write the message update of the SBP
algorithm in the compact form
\begin{align}
\label{EqnStochApprox}
\Mess^{\Time + 1} \, = \, \Mess^{\Time} \, - \, \step \: \VF
(\Mess^{\Time}, \indx^{\Time + 1}), \quad \mbox{valid for for $\Time =
  1, 2, \ldots$,}
\end{align}
suitable for application of the Robbins-Monro theorem.\footnote{The
  theorem states that if the vector field function $\VF (\Mess,
  \cdot)$ has a bounded second moment---that is $\Exp{}{\vnorm{\VF
      (\Mess , \indx)}{2}^2} \, \le \, \const (1 +
  \vnorm{\Mess}{2}^2)$ for some constant $\const$, the conditional
  distribution of the random vector $\indx ^{\Time + 1}$ knowing the
  past depends only on $\Mess ^{\Time}$---that is $\Prob {\indx
    ^{\Time +1} | \indx ^{\Time}, \indx ^{\Time - 1}, \cdots, \Mess
    ^{\Time}, \Mess ^{\Time - 1}, \cdots} \, = \, \Prob {\indx ^{\Time
      +1} | \Mess ^{\Time}}$, denoting the expected vector field
  function $\MVF(\Mess) \defn \Expt{\VF (\Mess, \indx)|\Mess}$, there
  exist a vector $\Mesast$ such that
\begin{align*}
\inf_{\Mess \in \Ball \backslash \{\Messast \}} \; \vecprod{ \Mess -
  \Mesast} {\MVF (\Mess)} \, > \, 0,
\end{align*}
and finally the step sizes satisfy the conditions
$\sum_{\Time=0}^{\infty} \step = \infty$, and $\sum_{\Time=0}^{\infty}
(\step)^2 < \infty$, then the sequence
$\{\Mess^{\Time}\}_{\Time=0}^{\infty}$ converges almost surely to
$\Mesast$.}  In order to apply this result, we need to verify its
hypotheses.  First of all, it is easy to see that we have a bound of
the form
\begin{align*}
\Exp{}{\vnorm{\VF (\Mess , \indx)}{2}^2} \, \le \, \const (1 +
\vnorm{\Mess}{2}^2),
\end{align*}
for some constant $\const$.  Moreover, the conditional distribution of
the vector $\indx ^{\Time + 1}$, given the past, depends only on
$\Mess ^{\Time}$; more precisely we have
\begin{align*}
\Prob {\indx ^{\Time +1} | \indx ^{\Time}, \indx ^{\Time - 1},
\ldots, \Mess ^{\Time}, \Mess ^{\Time - 1}, \ldots} \, = \, \Prob
{\indx ^{\Time +1} | \Mess ^{\Time}}.
\end{align*}
Lastly, defining the averaged function $\MVF(\Mess) \defn \Expt{\VF
  (\Mess, \indx)|\Mess} \, = \, \Mess - \UpFunc(\Mess)$, the final
requirement is to verify that the fixed point $\Mesast$ satisfies the
stability condition 
\begin{align}
\inf_{\Mess \in \Ball \backslash \{\Messast \}} \; \vecprod{ \Mess -
  \Mesast} {\MVF (\Mess)} \, > \, 0,
\end{align}
where $\vecprod{\cdot}{\cdot}$ denotes the Euclidean inner product,
and $\Ball$ denotes the compact set in which the messages lie.  Using
the Cauchy-Schwartz inequality and the fact that $\UpFunc$ is
Lipschitz with constant $\Lips = 1 - \slack / 2$, we obtain
\begin{align}
\nonumber \vecprod{\Mess - \Mesast} {\MVF (\Mess) - \MVF (\Mesast)}
\; = &\; \vnorm {\Mess - \Mesast} {2} ^2 \, - \, \vecprod{\Mess -
\Mesast} {\UpFunc (\Mess) -
  \UpFunc (\Mesast)} \\
\label{EqnLowBndCurv} \;  \ge & \; \frac{\slack}{2} \, \vnorm {\Mess
- \Mesast} {2} ^2 \; > \; 0,
\end{align}
where the strict inequality holds for all $\Mess \neq \Mesast$.  Since
$\Mesast$ is a fixed point, we must have \mbox{$\MVF(\Mesast) =
  \Mesast - \UpFunc(\Mesast) = 0$,} which concludes the proof.


\subsubsection{Part (b): Non-asymptotic Bounds on Mean-squared Error}
\label{PargPartB}

Let $\errvec{\Time} \defn (\Mess^{\Time} - \Messast) / \vnorm
{\Mesast}{2}$ denote the re-normalized error vector.  In order to
upper bound $\Expt{\Error{\Time}}$ for all $\Time = 1, 2, \ldots$,
we first control the quantity \mbox{$\Error{\Time + 1} -
  \Error{\Time}$,} corresponding to the increment in the squared
error. Doing some simple algebra yields
\begin{align*}
\Error{\Time + 1} - \Error{\Time} \; & = \; \frac{1}
{\vnorm{\Mesast}{2} ^2} \, \big( \vnorm{\Mess^{\Time + 1} -
\Mesast}{2} ^2 -
\vnorm{\Mess^{\Time} - \Mesast}{2} ^2 \big) \\
\; &= \; \frac{1} {\vnorm{\Mesast}{2} ^2}  \;  \vecprod
{\Mess^{\Time +1} - \Mess^{\Time}} {\Mess^{\Time +1} + \Mess^{\Time}
- 2 \Mesast}.
\end{align*}
Recalling the update equation~\eqref{EqnStochApprox}, we obtain
\begin{align}
\Error {\Time + 1} - \Error {\Time} \; & = \; \frac{1} {\vnorm{\Mesast}{2}
  ^2} \;   \vecprod {-\step \VF (\Mess^{\Time} , \indx ^ {\Time+
      1} )} {-\step \VF (\Mess^{\Time} , \indx ^{\Time + 1} ) + 2
    (\Mess^{\Time} - \Mesast) } \nonumber\\
\label{EqnSquErrDiff}
\; & = \; \frac{(\step) ^2} {\vnorm{\Mesast}{2} ^2} \, \vnorm {\VF
  (\Mess^{\Time}, \indx^{\Time + 1})} {2}^2 \, - \, \frac{2 \step}
      {\vnorm{\Mesast}{2} ^2} \; \vecprod {\VF (\Mess^{\Time},
        \indx^{\Time + 1}) } {\Mess^{\Time} - \Mesast}.
\end{align}
Now taking the expectation from both sides of the
equation~\eqref{EqnSquErrDiff} yields
\begin{align}
\MSE{\Time+1} - \MSE{\Time} \; & = \; \frac{(\step) ^2}
    {\vnorm{\Mesast}{2} ^2} \, \Expt{\vnorm {\VF (\Mess^{\Time},
        \indx^{\Time + 1})} {2}^2} \, - \, \frac{2 \step}
    {\vnorm{\Mesast}{2} ^2} \, \Expt{\Expt{\vecprod {\VF
          (\Mess^{\Time}, \indx^{\Time + 1}) } {\Mess^{\Time} -
          \Mesast}|\Filt{\Time}}} \nonumber\\
\label{EqnSimple}
\; & = \; \frac{(\step) ^2} {\vnorm{\Mesast}{2} ^2} \, \Expt{\vnorm
  {\VF (\Mess^{\Time}, \indx^{\Time + 1})} {2}^2} \, - \, \frac{2
  \step} {\vnorm{\Mesast}{2} ^2} \, \Expt{\vecprod {\MVF
      (\Mess^{\Time}) - \MVF(\Mesast) } {\Mess^{\Time} -
      \Mesast}},
\end{align}
where we used the facts that $\Exps[\VF (\Mess^{\Time}, \indx^{\Time +
    1})| \Filt{\Time}] = \MVF(\Mess^{\Time})$ and $\MVF(\Mesast) =
0$. We continue by upper bounding the term \mbox{$\Term _1 = \vnorm
  {\VF (\Mess^{\Time}, \indx^{\Time + 1})} {2}^2 / \vnorm{\Mesast}{2}
  ^2$} and lower bounding the term \mbox{$\Term_2 = \vecprod {\MVF
    (\Mess^{\Time}) - \MVF (\Mesast) } {\Mess^{\Time} - \Mesast} /
  \vnorm{\Mesast}{2} ^2$}.

\paragraph{Lower bound on $\Term_2$:}
Recalling~\eqref{EqnLowBndCurv} from our proof of part (a),
we see that
\begin{align}
\label{EqnLowerG}
\Term_2 \; \ge \; \frac {\slack} {2} \, \Error{\Time}.
\end{align} 

\paragraph{Upper bound on $\Term_1$:}
From the definition of the update function, we have
\begin{align*}
\vnorm {\VF (\Mess^{\Time}, \indx ^{\Time + 1})} {2} ^2 \;\;  & = \! \sum
_{(\Fnode \to \Snode) \in \DirSet} \vnorm { \Mes{\Time}{\Fnode}
  {\Snode} - \tMatCol{ \ind {\Time} {\Fnode\Snode}}}{2}^2 \;\; \leq \; 2
\!\! \sum _{(\Fnode \to \Snode) \in \DirSet} \big( \vnorm { \Mes{\Time}
  {\Fnode} {\Snode}} {2} ^2 \, + \, \vnorm{ \tMatCol{ \ind{\Time}
    {\Fnode\Snode}} } {2} ^2 \big).
\end{align*}
Recalling the bounds~\eqref{EqnZeroBound} and using the fact that
vectors $\Mes {\Time} {\Fnode} {\Snode}$ and $ \tMatCol{ \ind {\Time}
  {\Fnode\Snode}}$ sum to one, we obtain
\begin{align*}
\vnorm {\VF (\Mess ^{\Time}, \indx ^{\Time + 1})} {2} ^2 \; \; & \le \; 2
\sum _{(\Fnode \to \Snode) \in \DirSet} \big( \max_{i\in \Alphabet}
\UBound{\Fnode \Snode} {i} {0} \big) \, \big( \vnorm { \Mes {\Time}
  {\Fnode} {\Snode}} {1} \, + \, \vnorm{ \tMatCol{ \ind {\Time}
    {\Fnode\Snode}} } {1} \big) \\
\;\; & = \; 4  \sum _{(\Fnode \to \Snode) \in \DirSet} \big(
\max_{i\in \Alphabet} \UBound{\Fnode \Snode} {i} {0} \big).
\end{align*}
On the other hand, we also have
\begin{align*}
\vnorm {\Mesast} {2} ^{2} \;\; & \geq \sum _{(\Fnode \to \Snode) \in
  \DirSet} \big( \min _{i\in \Alphabet} \LBound{\Fnode \Snode} {i} {0}
\big) \vnorm {\Mesast_{\Fnode \Snode}} {1} \; = \; \sum _{(\Fnode \to
  \Snode) \in \DirSet} \big( \min _{i \in \Alphabet} \LBound{\Fnode
  \Snode} {i} {0} \big).
\end{align*}
Combining the pieces, we conclude that the term $\Term_1$ is upper
bounded as
\begin{align}
\label{EqnUpperG}
\Term_1 \; \le \; \prefac(\Pot) \; \defn \; 4 \, \frac {\sum _{(\Fnode
    \to \Snode) \in \DirSet} \big( \max_{i\in \Alphabet}
  \UBound{\Fnode \Snode} {i} {0} \big)} {\sum _{(\Fnode \to \Snode)
    \in \DirSet} \big( \min_{i\in \Alphabet} \LBound{\Fnode \Snode}
  {i} {0} \big)}.
\end{align}

Since both $\Term_1$ and $\Term_2$ are non-negative, the
bounds~\eqref{EqnUpperG} and~\eqref{EqnLowerG} also hold in
expectation.  Combining these bounds with the
representation~\eqref{EqnSimple}, we obtain the upper bound
\mbox{$\MSE {\Time + 1} - \MSE {\Time} \leq \; \prefac(\Pot) \:
  (\step)^2 - \step \slack \: \MSE {\Time}$,} or equivalently
\begin{align*}
\MSE {\Time + 1} \, \le \, \prefac(\Pot) \: (\step)^2 \: + \: (1 -
\step \slack) \: \MSE {\Time}.
\end{align*}
Setting $\step = \Coef / (\slack (\Time + 2))$ and unwrapping this
recursion yields
\begin{align}
\label{EqnUnwrapBound}
\MSE{\Time + 1} \; \le \; \frac{\prefac(\Pot) \: \Coef ^2} {\slack^2}
\, \sum_{i = 2} ^ {\Time + 2} \, \bigg( \frac {1} {i^2} \prod_{\ell =
  i+1} ^{\Time + 2} \left( 1 - \frac {\Coef} {\ell} \right) \bigg) \; + \;
\prod _{\ell = 2} ^{\Time + 2} \, \left( 1 - \frac {\Coef} {\ell} \right)
\, \MSE {0},
\end{align}
where we have adopted the convention that the inside product is equal
to one for $i = \Time + 2$. The following lemma, proved in
Appendix~\ref{AppLemAux}, provides a useful upper bound on the
products arising in this expression:
\begin{lemma}
\label{LemAux}
For all $i \in \{1, 2, \ldots, \Time + 1\}$, we have 
\begin{align*}
\prod_{\ell = i + 1} ^{\Time + 2} \left(1 - \frac {\Coef} {\ell}
\right) \; \leq \; \left( \frac {i + 1} {\Time + 3} \right) ^{\Coef}.
\end{align*}
\end{lemma}

\noindent 
Substituting this upper bound into the
inequality~\eqref{EqnUnwrapBound} yields
\begin{align*}
\MSE {\Time + 1} \, & \le \, \frac{\prefac(\Pot) \: \Coef ^2} {\slack
  ^2 (\Time + 3) ^{\Coef}} \: \sum _{i = 2} ^{\Time + 2} \: \frac{(i +
  1) ^{\Coef}}{i ^2}\, + \, \left( \frac{2}{\Time + 3} \right) ^{\Coef}
\MSE {0} \\
 & \le \, \frac{\prefac(\Pot) \: \Coef ^2} {\slack ^2 (\Time + 3)
  ^{\Coef}} \big( \frac{3}{2} \big) ^\Coef \: \sum _{i = 2} ^{\Time +
  2} \: \frac{1} {i ^{2 - \Coef}} \, + \, \left( \frac{2}{\Time + 3}
\right) ^{\Coef} \MSE {0}.
\end{align*}
It remains to upper bound the term $\sum_{i = 2}^{\Time + 2} 1 / i^{2
  - \Coef}$.  Since the function $1 / x^{2 - \Coef}$ is decreasing in
$x$ for $\Coef < 2$, we have the integral upper bound $\sum_{i = 2}
^{\Time + 2} \: 1 / i ^{2 - \Coef} \: \le \: \int_{1}^{\Time +2} 1 /
x^{2-\Coef} \: dx$, which yields
\begin{align*}
\MSE {\Time + 1} & \leq \begin{cases} \big( \frac{3}{2} \big) ^{\Coef}
     \frac {\prefac(\Pot) \: \Coef ^2 } {\slack ^2 (1 - \Coef)} \: \frac{1}
     {(\Time + 3) ^\Coef} \; + \; \big( \frac {2} {\Time + 3} \big)
     ^{\Coef} \MSE {0} & \mbox{if $ 0 < \Coef < 1$} \\
\frac {3}{2} \: \frac {\prefac(\Pot)} {\slack ^2} \: \frac { \log (\Time + 2)
} {\Time + 3} \; + \; \frac {2} {\Time + 3} \; \MSE {0} & \mbox{if
  $\Coef = 1$} \\
\big( \frac{3}{2} \big) ^{\Coef} \frac {\prefac(\Pot) \: \Coef ^2}
    {\slack ^2 (\Coef - 1)} \: \frac{ (\Time + 2 ) ^{\Coef - 1}}
    {(\Time + 3) ^{\Coef}} \; + \; \big( \frac {2} {\Time + 3} \big)
    ^{\Coef} \MSE {0} & \mbox{if $1 < \Coef < 2$}
\end{cases}.
\end{align*}
If we now focus on the range of $\Coef \in (1,2)$, which
yields the fastest convergence rate, some simple algebra yields the
form of the claim given in the theorem statement.


\subsubsection{High Probability Bounds}
\label{PargPartC}


Recall the algebra in the beginning of the Section~\ref{PargPartB}.
Subtracting the conditional mean of the second term of the
equation~\eqref{EqnSquErrDiff} yields
\begin{align*}
\Error {\Time +1} - \Error {\Time} \; & = \; \frac{(\step) ^2}
       {\vnorm{\Mesast}{2} ^2} \, \vnorm {\VF (\Mess^{\Time},
         \indx^{\Time + 1})} {2}^2 \, - \, \frac{2 \step}
       {\vnorm{\Mesast}{2} ^2} \; \vecprod {\MVF (\Mess^{\Time}) }
       {\Mess^{\Time} - \Mesast} + \, 2 \step\, \vecprod{\MarDif{\Time
           + 1}}{\errvec{\Time}},
\end{align*}
where we have denoted the term
\begin{align*}
\MarDif{\Time + 1} \; \defn \; \frac{\MVF (\Mess^{\Time}) - \VF
  (\Mess^{\Time}, \indx^{\Time + 1})}{\vnorm{\Mesast}{2}}.
\end{align*}
Recalling the bounds on $\Term_1 = \vnorm {\VF (\Mess^{\Time},
  \indx^{\Time + 1})} {2}^2 \: / \: \vnorm{\Mesast}{2} ^2$ and
$\Term_2 = \vecprod {\MVF (\Mess^{\Time}) } {\Mess^{\Time} -
  \Mesast}\: / \: \vnorm{\Mesast}{2} ^2$ from part (b), we have
\begin{align*}
\Error {\Time +1} - \Error {\Time} \; \le \; \prefac(\Pot) \: (\step)
^2 \, - \, \slack \step \Error {\Time} \, + \, 2 \step\,
\vecprod{\MarDif{\Time + 1}}{\errvec{\Time}},
\end{align*}
or equivalently
\begin{align*}
\Error {\Time +1} \; & \leq \; \prefac(\Pot) \, (\step) ^2 \, + \, (1
- \slack \step) \Error {\Time} \, + \, 2 \step\,
\vecprod{\MarDif{\Time + 1}}{\errvec{\Time}}.
\end{align*}
Substituting the step size choice $\step = 1 / (\slack (\Time + 1))$ 
and then unwrapping this recursion yields
\begin{align}
 \Error{\Time + 1} \; & \leq \; \frac {\prefac(\Pot)}{\slack^2 (\Time
   + 1)} \sum_{\newtime=1}^{\Time+1} \frac{1}{\newtime} \, + \,
 \frac{2}{\slack \: (\Time+1)} \sum_{\newtime = 0}^{\Time}
 \: \vecprod{\MarDif{\newtime + 1}}{\errvec{\newtime}} \nonumber \\
\label{EqnBoundActualError} 
\; & \leq \; \frac{\prefac(\Pot)}{\slack^2} \frac{1 + \log(\Time +
1)}{\Time + 1} \, + \, \frac{2}{\slack \: (\Time+1)} \sum_{\newtime =
    0}^{\Time} \: \vecprod{\MarDif{\newtime + 1}}{\errvec{\newtime}}.
\end{align}
Note that by construction, the sequence
$\{\MarDif{\newtime}\}_{\newtime=1}^{\infty}$ is a martingale
difference sequence with respect to the filtration
\mbox{$\Filt{\newtime} = \sigma (\Mess^{0}, \Mess^{1}, \ldots,
  \Mess^{\newtime})$} that is $\Expt{\MarDif{\newtime+1} \mid
  \Filt{\newtime}} = \zerovec$ and accordingly
$\Expt{\MarDif{\newtime+1}} = 0$ for $\newtime = 0, 1, 2, \ldots$.  We
continue by controlling the stochastic term $(\sum_{\newtime =
  0}^{\Time} \vecprod{\MarDif{\newtime + 1}}{\errvec{\newtime}})
/(\Time+1)$---namely its variance,
\begin{align*}
\var \bigg(\frac{1}{\Time+1} \sum_{\newtime = 0}^{\Time} \:
\vecprod{\MarDif{\newtime + 1}}{\errvec{\newtime}}\bigg) \; = & \;
\frac{1}{(\Time+1)^2} \, \Exps\bigg[\big(\sum_{\newtime = 0}^{\Time }
  \vecprod{\MarDif{\newtime + 1}}{\errvec{\newtime}}\big)^2\bigg]
\\ \; =& \; \underbrace{\frac{1}{(\Time+1)^2} \, \sum_{\newtime =
    0}^{\Time} \Exps\big[ \vecprod{\MarDif{\newtime +
        1}}{\errvec{\newtime}}^2 \big]}_{\secterm_1} \\ \, & + \,
\underbrace{\frac{2}{(\Time+1)^2} \, \sum_{0 \le \newtime_2 <
    \newtime_1\le \Time} \Exps\big[\vecprod{\MarDif{\newtime_1 +
        1}}{\errvec{\newtime_1}}\vecprod{\MarDif{\newtime_2 +
        1}}{\errvec{\newtime_2}} \big]}_{\secterm_2}.
\end{align*}
Since we have
\begin{align*}
\Exps\big[\vecprod{\MarDif{\newtime_1 +
      1}}{\errvec{\newtime_1}}\vecprod{\MarDif{\newtime_2 +
      1}}{\errvec{\newtime_2}} \big] \; = & \;
\Exps\big[\Exps\big[\vecprod{\MarDif{\newtime_1 +
        1}}{\errvec{\newtime_1}}\vecprod{\MarDif{\newtime_2 +
        1}}{\errvec{\newtime_2}} \mid \Filt{\newtime_1}\big]\big]\\ \;
= & \; \Exps\big[\vecprod{\MarDif{\newtime_2 +
      1}}{\errvec{\newtime_2}} \: \Exps\big[\vecprod{\MarDif{\newtime_1 +
        1}}{\errvec{\newtime_1}} \mid \Filt{\newtime_1}\big]\big] \; = \; 0,
\end{align*}
for all $\newtime_1 > \newtime_2$, the cross product term $\secterm_2$
vanishes. On the other hand, the martingale difference sequence is
bounded. This can be shown as follows: from part (b) we know
$\vnorm{\VF(\Mess^{\newtime}, \indx^{\newtime + 1})}{2} /
\vnorm{\Mesast}{2} \le \sqrt{\prefac(\Pot)}$; also using the fact that
$\vnorm{\cdot}{2}$ is convex, Jensen's inequality yields $\vnorm{\MVF
  (\Mess^{\newtime})} {2} / \vnorm{\Mesast}{2} \le
\sqrt{\prefac(\Pot)}$; therefore, we have
\begin{align*}
\vnorm{\MarDif{\newtime+1}}{2} \; \le \;
\frac{\vnorm{\VF(\Mess^{\newtime},
    \indx^{\newtime+1})}{2}}{\vnorm{\Mesast}{2}} \, + \,
\frac{\vnorm{\MVF(\Mess^{\newtime})}{2}}{\vnorm{\Mesast}{2}} \; \le \;
2\:\sqrt{\prefac(\Pot)}.
\end{align*}
Moving on to the first term $\secterm_1$, we exploit the Cauchy
Schwartz inequality in conjunction with the fact that the martingale
difference sequence is bounded to obtain
\begin{align*}
\Exps\big[ \vecprod{\MarDif{\newtime + 1}}{\errvec{\newtime}}^2\big]
\; \le \; \Exps\big[ \vnorm{\MarDif{\newtime + 1}}{2}^2 \:
  \vnorm{\errvec{\newtime}}{2}^2 \big] \; \le \; 4\:\prefac(\Pot) \:
\Expt{\Error{\newtime}}.
\end{align*}
Taking the expectation from both sides of the
inequality~\eqref{EqnBoundActualError} yields
\mbox{$\Expt{\Error{\newtime}} \le (\prefac(\Pot)/\slack^2) \:
  (1+\log{\newtime})/\newtime$}; and hence we have
\begin{align*}
\Exps\big[ \vecprod{\MarDif{\newtime + 1}}{\errvec{\newtime}}^2\big]
\; \le \; \frac{4\:\prefac(\Pot)^2}{\slack^2} \, \frac{1 +
  \log{\newtime}}{\newtime},
\end{align*}
for all $\newtime \ge 1$. Moreover, since 
\begin{align*}
\frac{\vnorm{\Mess^{0}} {2}} {\vnorm{\Mesast}{2}} \; & \le \;
\left(\frac {\sum _{(\Fnode \to \Snode) \in \DirSet} \big( \max_{i\in
    \Alphabet} \UBound{\Fnode \Snode} {i} {0} \big)} {\sum _{(\Fnode
    \to \Snode) \in \DirSet} \big( \min_{i\in \Alphabet}
  \LBound{\Fnode \Snode} {i} {0} \big)}\right)^{\frac{1}{2}} \; = \;
\sqrt{\frac{\prefac(\Pot)}{4}},
\end{align*}
the initial term $\Exps\big[ \vecprod{\MarDif{1}}{\errvec{0}}^2\big]
\le 4\:\prefac(\Pot) \: \Expt{\Error{0}} $ is upper bounded by
$4\:\prefac(\Pot)^2$. Finally, putting all the pieces together, we
obtain
\begin{align*}
\var \bigg(\frac{1}{\Time+1} \sum_{\newtime = 0}^{\Time} \:
\vecprod{\MarDif{\newtime + 1}}{\errvec{\newtime}} \bigg) \; \le & \;
\frac{4\:\prefac(\Pot)^2}{\slack^2 \: (\Time + 1)^2} \:
\sum_{\newtime=1}^{\Time} \frac{1 + \log{\newtime}} {\newtime} \, + \,
\frac{4\:\prefac(\Pot)^2}{(\Time + 1)^2}\\ \; \stackrel{(\mathrm{i})}{\le} & \;
\frac{4\:\prefac(\Pot)^2}{\slack^2} \, \frac{(1 + \log(\Time + 1))^2 +
  4}{(\Time+1)^2},
\end{align*}
where inequality (i) follows from the facts $\sum_{\newtime=1}^{\Time}
(1 + \log{\newtime}) /\newtime \le (1+\log{\Time})^2$, and
\mbox{$\slack < 2$}. Consequently, we may apply Chebyshev's inequality
to control the stochastic deviation \mbox{$\sum_{\newtime = 1}^{\Time
    + 1} \: \vecprod{\MarDif{\newtime + 1}}{\errvec{\newtime}} /
  (\Time+1)$}. More specifically, for $\gamma > 0$ (to be specified)
we have
\begin{align}
\label{EqnChebyshev}
\mProb \bigg( \big|\frac{2}{\slack \: (\Time + 1)} \sum_{\newtime =
  0}^{\Time} \: \vecprod{\MarDif{\newtime + 1}}{\errvec{\newtime}}
\big| \; > \; \gamma \bigg) \; \le \;
\frac{16\:\prefac(\Pot)^2}{\slack^4 \: \gamma^2} \,
\frac{(1+\log(\Time + 1))^2 + 4}{(\Time+1)^2}.
\end{align}
We now combine our earlier bound~\eqref{EqnBoundActualError} with the
tail bound~\eqref{EqnChebyshev}, making the specific choice
\begin{align*}
\gamma \; = \; \frac{4\:\prefac(\Pot)}{\slack^2 \: \sqrt{\epsilon}} \,
\frac{\sqrt{(1 + \log(\Time+1))^2 + 4}}{\Time + 1},
\end{align*}
for a fixed $0 < \epsilon < 1$, thereby concluding that
\begin{align*}
 \Error{\Time + 1} \; & \leq \; \frac{\prefac(\Pot)}{\slack^2} \frac{1
   + \log(\Time + 1)}{\Time + 1} \, + \,
 \frac{4\:\prefac(\Pot)}{\slack^2 \: \sqrt{\epsilon}} \,
 \frac{\sqrt{(1 + \log(\Time+1))^2 + 4}}{\Time + 1},
\end{align*}
with probability at least $1 - \epsilon$. Simplifying the last bound,
we obtain
\begin{align*}
 \Error{\Time + 1} \; & \leq \; \frac{\prefac(\Pot)}{\slack^2} \bigg(1
 + \frac{8}{\sqrt{\epsilon}}\bigg) \, \frac{1 + \log(\Time + 1)}{\Time +
   1},
\end{align*}
for all $\Time \ge 1$, with probability at least $1 - \epsilon$.


\subsection{Proof of Proposition~\ref{PropUpFunLipschitz}}
\label{SecProofProp}

Recall the definition~\eqref{EqnProbMass} of the probability mass
function $\{\pbmas{\Fnode \Snode}{j}\}_{j \in \Alphabet}$ used in the
update of directed edge $(\Fnode \to \Snode)$.  This probability
depends on the current value of the message, so we can view it as
being generated by a function $q_{\Fnode \Snode}: \real^\Dimn
\rightarrow \real^\dimn$ that performs the mapping $\Mess \mapsto
\{\pbmas{\Fnode \Snode}{j}\}_{j \in \Alphabet}$.  In terms of this
function, we can rewrite the BP message
update~\eqref{EqnBPUpdateEntry} on directed edge $(\Fnode \to \Snode)$
as $\UpFun (\Mess) = \tBPMat_{\Fnode \Snode} \: q_{\Fnode
  \Snode}(\Mess)$, where the renormalized compatibility matrix
$\tBPMat_{\Fnode \Snode}$ was defined
previously~\eqref{EqnNormalizedCol}.  We now define the $\Dimn \times
\Dimn$ block diagonal matrix $\tBPMat \defn \blkdiag
\{\tBPMat_{\Fnode\Snode} \}_{(\Fnode \to \Snode) \in \DirEdge}$, as
well as the function $q:\real^\Dimn \rightarrow \real^\Dimn$ obtained
by concatenating all of the functions $q_{\Fnode \Snode}$, one for
each directed edge.  In terms of these quantities, we rewrite the
global BP message update in the compact form $\UpFunc(\Mess) = \tBPMat
\: q(\Mess)$.

With these preliminaries in place, we now bound the Lipschitz constant
of the mapping $\UpFunc: \real^\Dimn \rightarrow \real^\Dimn$.  Given
an arbitrary pair of messages $\Mess, \Mess' \in \Ball$, we have
\begin{align}
\label{EqnEarlyRep}
\|\UpFunc(\Mess) - \UpFunc(\Mess')\|^2_2 \;  & = \; \| \tBPMat \, \big(
q(\Mess) - q(\Mess') \big)\|^2_2 \;\; = \!\! \sum_{(\Fnode \to \Snode) \in
  \DirEdge} \| \tBPMat_{\Fnode \Snode} \big(q_{\Fnode \Snode}(\Mess) -
q_{\Fnode \Snode}(\Mess') \big) \|_2^2.
\end{align} 
By the Perron-Frobenius theorem~\cite{Horn85}, we know that
$\tBPMatrix{\Fnode \Snode}$ has a unique maximal eigenvalue of $1$,
achieved for the left eigenvector $\onevec \in \real^{\dimn}$, where
$\onevec$ denotes the vector of all ones.  Since the
$\dimn$-dimensional vectors $q_{\Fnode \Snode}(\Mess)$ and $q_{\Fnode
  \Snode}(\Mess')$ are both probability distributions, we have
$\inprod{\onevec}{q_{\Fnode \Snode}(\Mess) - q_{\Fnode
    \Snode}(\Mess')} = 0$. Therefore, we conclude that
\begin{align*}
\tBPMat_{\Fnode \Snode} \big( q_{\Fnode \Snode}(\Mess) - q_{\Fnode
  \Snode}(\Mess') \big)\; & = \; \big(\tBPMatrix{\Fnode \Snode} -
\frac{ \Leigvec{\Fnode\Snode} \onevec^T}{\onevec^T
  \Leigvec{\Fnode\Snode} }\big) \big ( q_{\Fnode \Snode}(\Mess) -
q_{\Fnode \Snode}(\Mess') \big),
\end{align*}
where $\Leigvec{\Fnode\Snode}$ denotes the right eigenvector of
$\tBPMatrix{\Fnode \Snode}$ corresponding to the eigenvalue one.
Combining this equality with the representation~\eqref{EqnEarlyRep},
we find that
\begin{align}
\|\UpFunc(\Mess) - \UpFunc(\Mess')\|^2_2 \; & = \!\! \sum_{(\Fnode \to
  \Snode) \in \DirEdge} \| \big(\tBPMatrix{\Fnode \Snode} -  \frac{ \Leigvec{\Fnode\Snode} \onevec^T}{\onevec^T  \Leigvec{\Fnode\Snode} }\big) \big (
q_{\Fnode \Snode}(\Mess) - q_{\Fnode \Snode}(\Mess') \big) \|_2^2
\nonumber \\
\label{EqnFirstStep}
\; & \leq \max_{(\Fnode \to \Snode) \in \DirEdge}
\matsnorm{\tBPMatrix{\Fnode \Snode} -  \frac{ \Leigvec{\Fnode\Snode} \onevec^T}{\onevec^T  \Leigvec{\Fnode\Snode} } }{2}^2 \; \; \|q(\Mess) -
q(\Mess')\|_2^2.
\end{align}
It remains to upper bound the Lipschitz constant of the mapping $q:
\real^\Dimn \rightarrow \real^\Dimn$ previously defined.
\begin{lemma}
\label{LemOpNormJacob}
For all $\Mess \neq \Mess'$, we have
\begin{align}
\label{EqnOpNormJacob}
\frac{\| q(\Mess) - q(\Mess')\|_2}{\|\Mess - \Mess'\|_2} \; & \leq \;
2 \, \max_{ (\Fnode \to \Snode) \in \DirSet}\PHIFUN(\Fnode, \Snode) \;
\max_{ (\Tnode \to \Fnode) \in \DirSet} \PHIFUNTWO(\Tnode, \Fnode),
\end{align}
where the quantities $\PHIFUN(\Fnode, \Snode)$, and
$\PHIFUNTWO(\Tnode, \Fnode)$ were previously defined
in~\eqref{EqnDefnBigPhi} and~\eqref{EqnDefnBigPhiTwo} .
\end{lemma}
\noindent As this proof is somewhat technical, we defer it to
Appendix~\ref{AppLemOpNormJacob}.  Combining the upper
bound~\eqref{EqnOpNormJacob} with the earlier
bound~\eqref{EqnFirstStep} completes the proof of the proposition.


\section{Experimental Results} 
\label{SecSimulations}

In this section, we present a variety of experimental results that
confirm the theoretical predictions, and show that SBP is a practical
algorithm.  We provide results both for simulated graphical models,
and real-world applications to image denoising and disparity
computation.

\subsection{Simulations on Synthetic Problems}

We start by performing some simulations for the Potts model, in which
the edge potentials are specified by a parameter $\pott \in (0,1]$, as
  discussed in Example~\ref{ExaPotts}.  The node potentials are
  generated randomly, on the basis of fixed parameters $\mean \ge \std
  > 0$ satisfying $\mean + \std < 1$, as follows: for each $\Fnode \in
  \node$ and label $i \neq 1$, we generate an independent random
  variable $Z_{\Fnode; i}$ uniformly distributed on the interval $(-1,
  +1)$, and then set
\begin{align*}
\NPot (i) & = \begin{cases} 1 &  i = 1
  \\ \mean + \std Z_{\Fnode; i} &  i \ge 2
\end{cases}.
\end{align*}
\begin{figure}[h]
\begin{center}
\begin{tabular}{cc}
\widgraph{.48\textwidth}{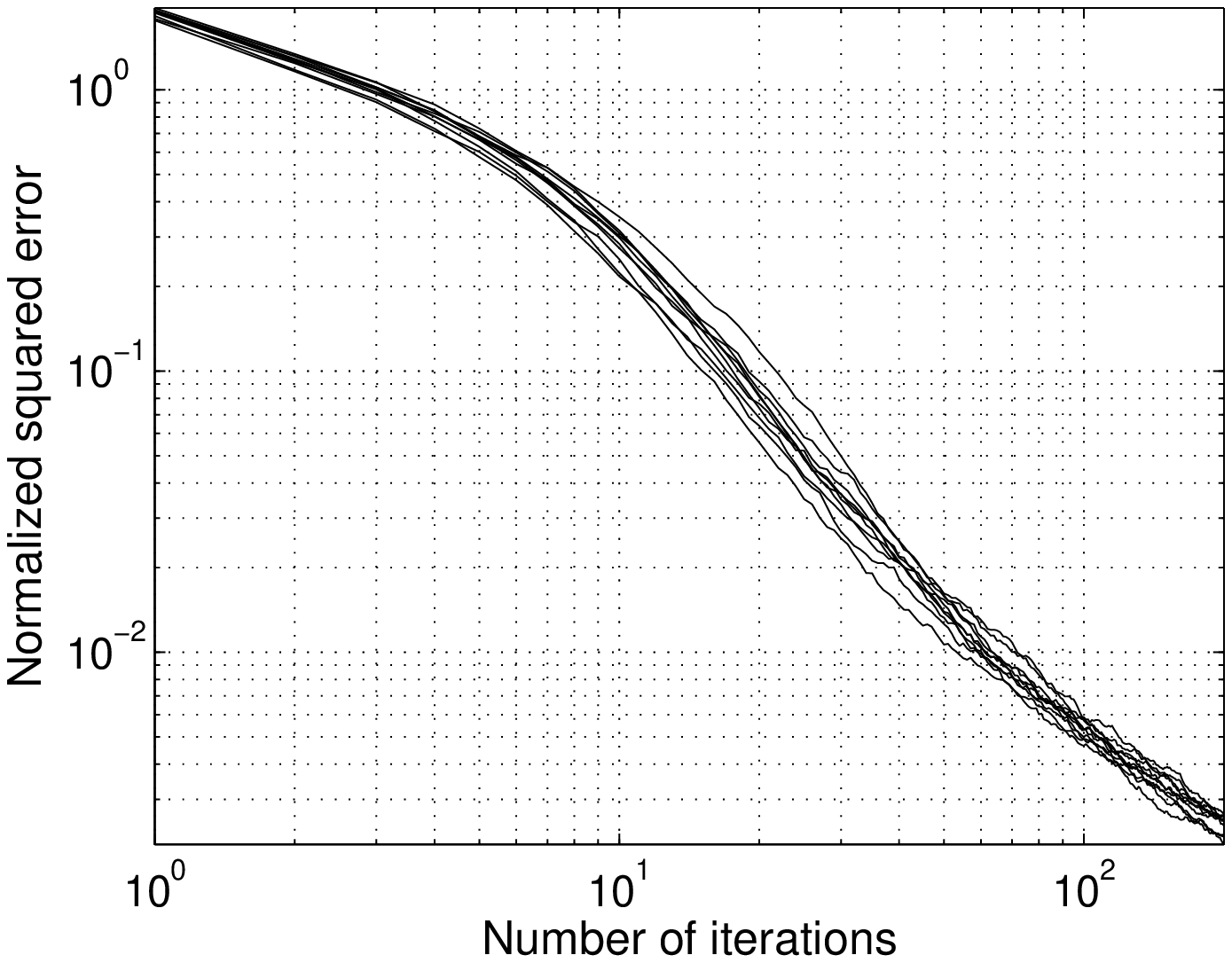} &
\widgraph{.48\textwidth}{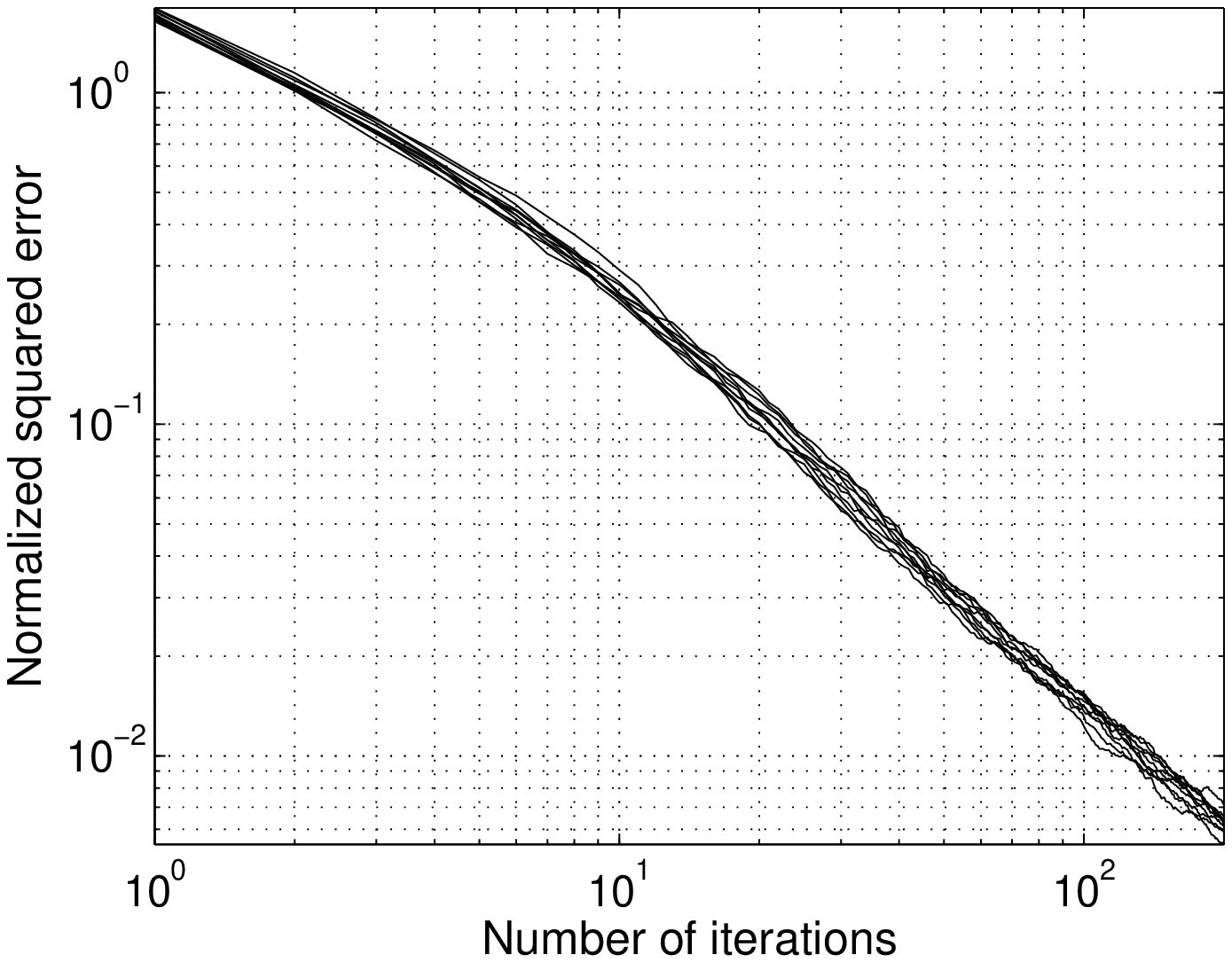}\\ (a) & (b)
\end{tabular}
\end{center}
\caption{Panels illustrate the normalized squared-error
  $\vnorm{\Mess^{\Time} - \Mesast}{2}^2 / \vnorm{\Mesast}{2} ^2$
  versus the number of iterations $\Time$ for a chain of size
  $\numnode = 100$ and state dimension $\dimn = 64$. Each plot
  contains 10 different sample paths.  Panel (a) corresponds to the
  coupling parameter $\pott = 0.02$ whereas panel (b) corresponds to
  $\pott = 0.05$.  In all cases, the SBP algorithm was implemented
  with step size $\step = 2/(\Time + 1)$, and the node potentials were
  generated with parameters $(\mean, \std) = (0.1, 0.1)$.}
\label{FigConsistency}
\end{figure}

For a fixed graph topology and collection of node/edge potentials, we
first run BP to compute the fixed point $\Mesast$.\footnote{We stop
  the BP iterations when $\vnorm{\Mess^{\Time + 1} -
    \Mess^{\Time}}{2}$ becomes less than $10^{-4}$.}  We then run SBP
algorithm to find the sequence of messages $\{ \Mess^\Time \}_{\Time =
  0}^{\infty}$ and compute the normalized squared error
$\vnorm{\Mess^{\Time} -\Mesast}{2}^2 / \vnorm{\Mesast}{2} ^2$.  In
cases where the mean squared error is reported, we computed it by
averaging over $20$ different runs of the algorithm.  (Note that the
runs are different, since the SBP algorithm is randomized.)

\begin{figure}
\begin{center}
\begin{tabular}{cc}
  \widgraph{.48\textwidth}{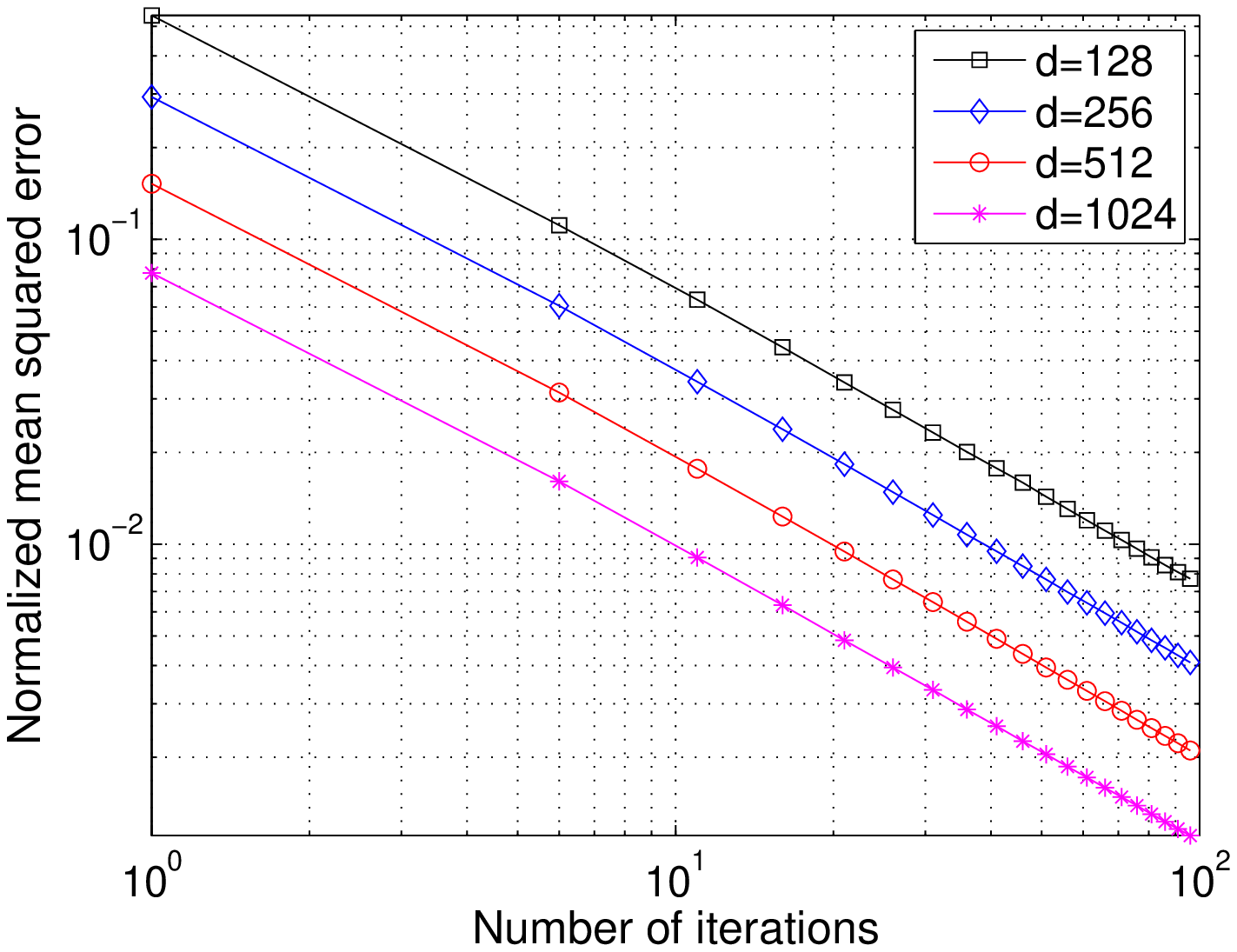} & 
\widgraph{.48\textwidth}{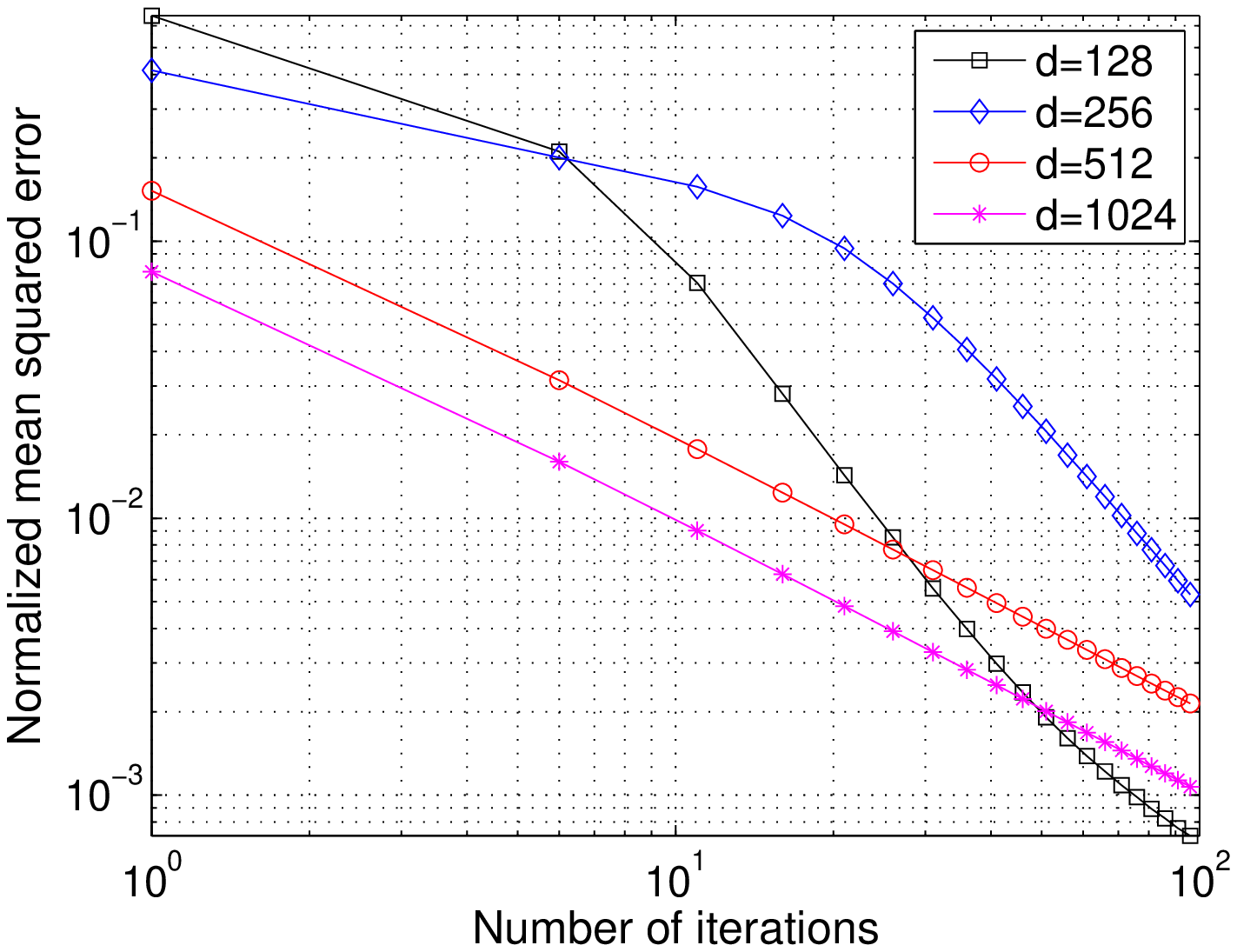} \\
(a) &  (b)
\end{tabular}
\end{center}
\caption{Effect of increasing state dimension on convergence rates.
  Plots of the normalized mean squared-error
  $\Expt{\vnorm{\Mess^{\Time} - \Mesast}{2}^2} / \vnorm{\Mesast}{2}
  ^2$ versus the number of iterations for two different graphs: (a)
  chain with $\numnode = 100$ nodes, and (b) two-dimensional square
  grid with $\numnode = 100$ nodes.  In both panels, each curve
  corresponds different state dimension $\dimn \in \{128, 256, 512,
  1024\}$.  All simulations were performed with step sizes $\step =
  2/(\Time + 1)$, and the node/edge parameters were generated with
  parameters $(\mean, \std) = (0.1, 0.1)$ and $\pott = 0.1$
  respectively.}
\label{FigDiffDimension}
\end{figure}

In our first set of experiments, we examine the consistency of the SBP
on a chain-structured graph, as illustrated in
Figure~\ref{FigGraphicalModels}(b), representing a particular instance
of a tree.  We implemented the SBP algorithm with step size $\step =
2/(\Time + 1)$, and performed simulations for a chain with $\numnode =
100$ nodes, state dimension $\dimn = 64$, node potential parameters
$(\mean, \std) = (0.1, 0.1)$, and for two different choices of edge
potential $\pott \in \{0.02, 0.05\}$.  The resulting traces of the
normalized squared error versus iteration number are plotted in
Figure~\ref{FigConsistency}; each panel contains $10$ different sample
paths. These plots confirm the prediction of strong consistency given
in Theorem~\ref{ThmTree}(a)---in particular, the error in each sample
path converges to zero.  We also observe that the typical performance
is highly concentrated around its average, as can be observed from the
small amount of variance in the sample paths.

Our next set of simulations are designed to study the effect of
increasing of the state dimension $\dimn$ on convergence rates.  We
performed simulations both for the chain with $\numnode = 100$ nodes,
as well as a two-dimensional square grid with $\numnode = 100$
nodes. In all cases, we implemented the SBP algorithm with step sizes
$\step = 2 / (\Time + 1)$, and generated the node/edge potentials with
parameters $(\mean, \std) = (0.1, 0.1)$ and $\pott = 0.1$
respectively.  In Figure~\ref{FigDiffDimension}, we plot the
normalized mean-squared error (estimated by averaging over $20$
trials) versus the number of iterations for the chain in panel (a),
and the grid in panel (b).  Each panel contains four different curves,
each corresponding to a choice of state dimension $\dimn \in \{128,
256, 512, 1024\}$.  For the given step size, Theorem~\ref{ThmMain}
guarantees that the convergence rate should be $1/\Time^\alpha$
($\alpha \le 1$) with the number of iterations $\Time$.  In the
log-log domain plot, this convergence rate manifests itself as a
straight line with slope $-\alpha$.  For the chain simulations shown
in panel (a), all four curves exhibit exactly this behavior, with the
only difference with increasing dimension being a vertical shift (no
change in slope).  For the grid simulations in panel (b), problems
with smaller state dimension exhibit somewhat faster convergence rate
than predicted by theory, whereas the larger problems ($\dimn \in
\{512, 1024\}$) exhibit linear convergence on the log-log scale.

\begin{table}
\begin{center}
\begin{tabular}{cc|c|c|c|c|}
\cline{3-6} & & $\dimn = 128$ & $\dimn = 256$ & $\dimn = 512$ & $\dimn
= 1024$ \\ \cline{1-6}
\multicolumn{1}{|c|}{\multirow{4}{*}{Chain}}&
\multicolumn{1}{|c|}{BP (per iteration)} & 0.0700 & 0.2844 & 2.83 & 18.0774 \\
\cline{2-6} 
\multicolumn{1}{|c|}{} & \multicolumn{1}{|c|}{SBP (per iteration)} & 0.0036 & 0.0068 & 0.0145 &  0.0280 \\
\cline{2-6}
\multicolumn{1}{|c|}{} & \multicolumn{1}{|c|}{BP (total)} & 0.14 & 0.57 & 5.66  &  36.15  \\
\cline{2-6}
\multicolumn{1}{|c|}{} & \multicolumn{1}{|c|}{SBP (total)} & 0.26 & 0.27 & 0.29 &  0.28 \\ \hline \hline
\multicolumn{1}{|c|}{\multirow{4}{*}{Grid}} &
\multicolumn{1}{|c|}{BP (per iteration)} & 0.1300 & 0.5231 & 5.3125 & 32.5050 \\
\cline{2-6} 
\multicolumn{1}{|c|}{} & \multicolumn{1}{|c|}{SBP (per iteration)} & 0.0095 & 0.0172 & 0.0325 & 0.0620 \\
\cline{2-6} 
\multicolumn{1}{|c|}{} & \multicolumn{1}{|c|}{BP (total)} & 0.65  & 3.66  & 10.63 & 65.01 \\
\cline{2-6} 
\multicolumn{1}{|c|}{} & \multicolumn{1}{|c|}{SBP (total)} & 0.21 & 1.31 & 0.65 & 0.62 \\ 
\cline{1-6}
\end{tabular}
\vspace{0.2in}
\caption{Comparison of BP and SBP computational cost for two different
  graphs each with $\numnode = 100$ nodes.  For each graph type, the
  top two rows show per iteration running time (in seconds) of the BP
  and SBP algorithms for different state dimensions.  The bottom two
  rows show total running time (in seconds) to compute the message
  fixed point to $\delta = 0.01$ accuracy.}
\end{center}
\label{TabRunningTime}
\end{table}

As discussed previously, the SBP message updates are less expensive by
a factor of $\dimn$.  The top two rows of Table~\ref{TabRunningTime}
show the per iteration running time of both BP and SBP algorithms, for
different state dimensions as indicated.  As predicted by theory, the
SBP running time per iteration is significantly lower than BP, scaling
linearly in $\dimn$ in contrast to the quadratic scaling of BP.  To be
fair in our comparison, we also measured the total computation time
required for either BP or SBP to converge to the fixed point up to a
$\delta$-tolerance, with $\delta = 0.01$.  This comparison allows for
the fact that BP may take many fewer iterations than SBP to converge
to an approximate fixed point.  Nonetheless, as shown in the bottom
two rows of Table~\ref{TabRunningTime}, in all cases except one (chain
graph with dimension $\dimn = 128$), we still see significant
speed-ups from SBP in this overall running time.  This gain becomes
especially pronounced for larger dimensions, where these types of
savings are more important.

\subsection{Applications in Image Processing and Computer Vision}

\newcommand{\NIMSIZE}{.38\textwidth}
\begin{figure}[h!]
\begin{center}
\begin{tabular}{ccc}
  \widgraph{\NIMSIZE}{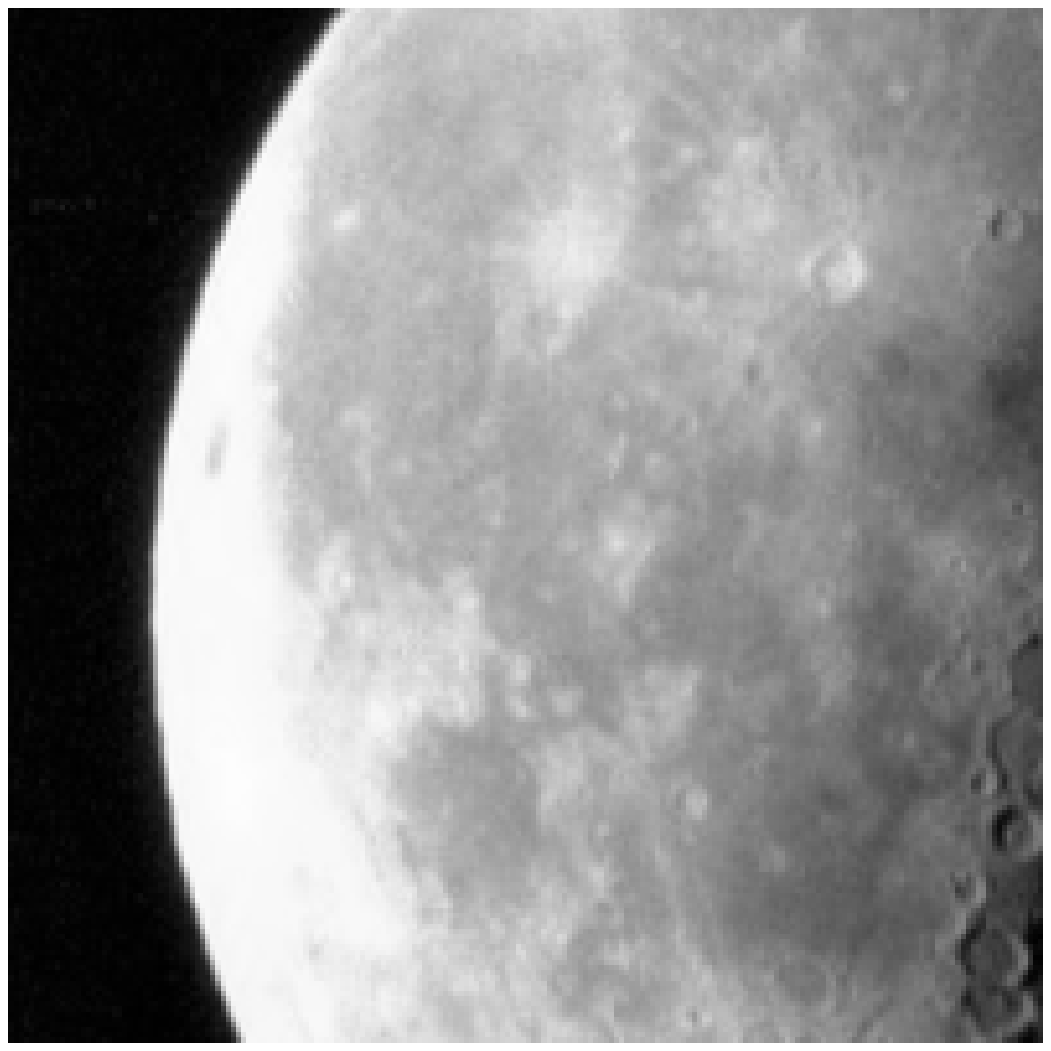} & &
  \widgraph{\NIMSIZE}{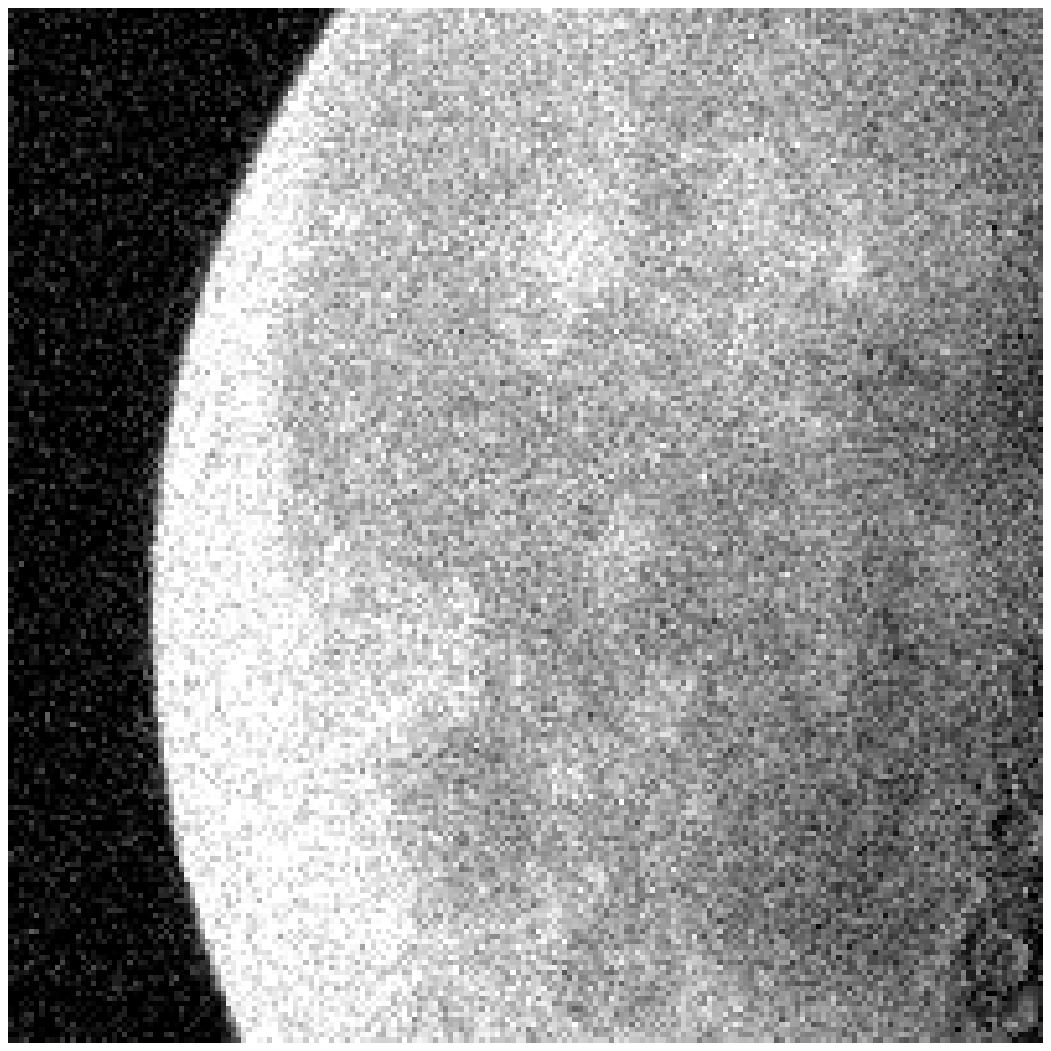} \\ (a) & & (b) \\ 
  \widgraph{\NIMSIZE}{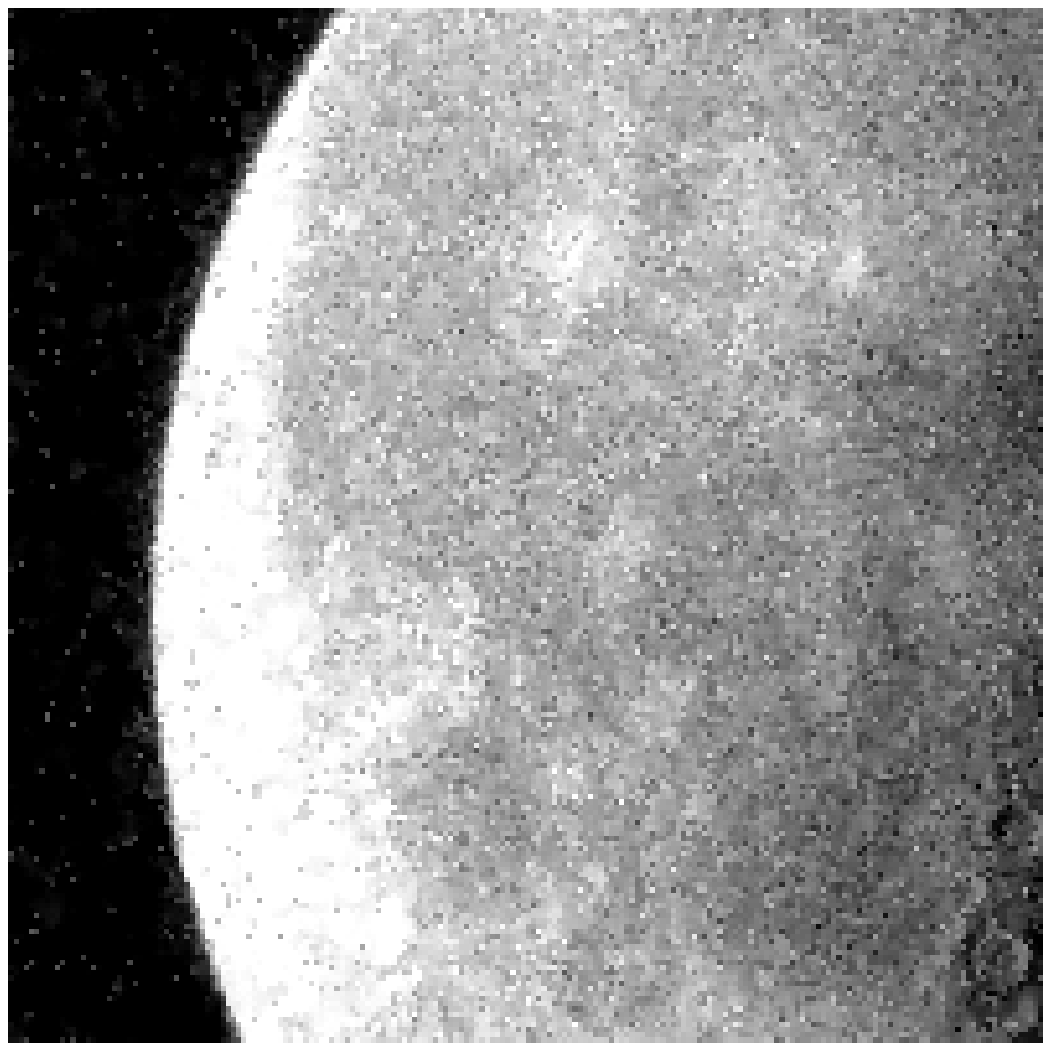} & &
  \widgraph{\NIMSIZE}{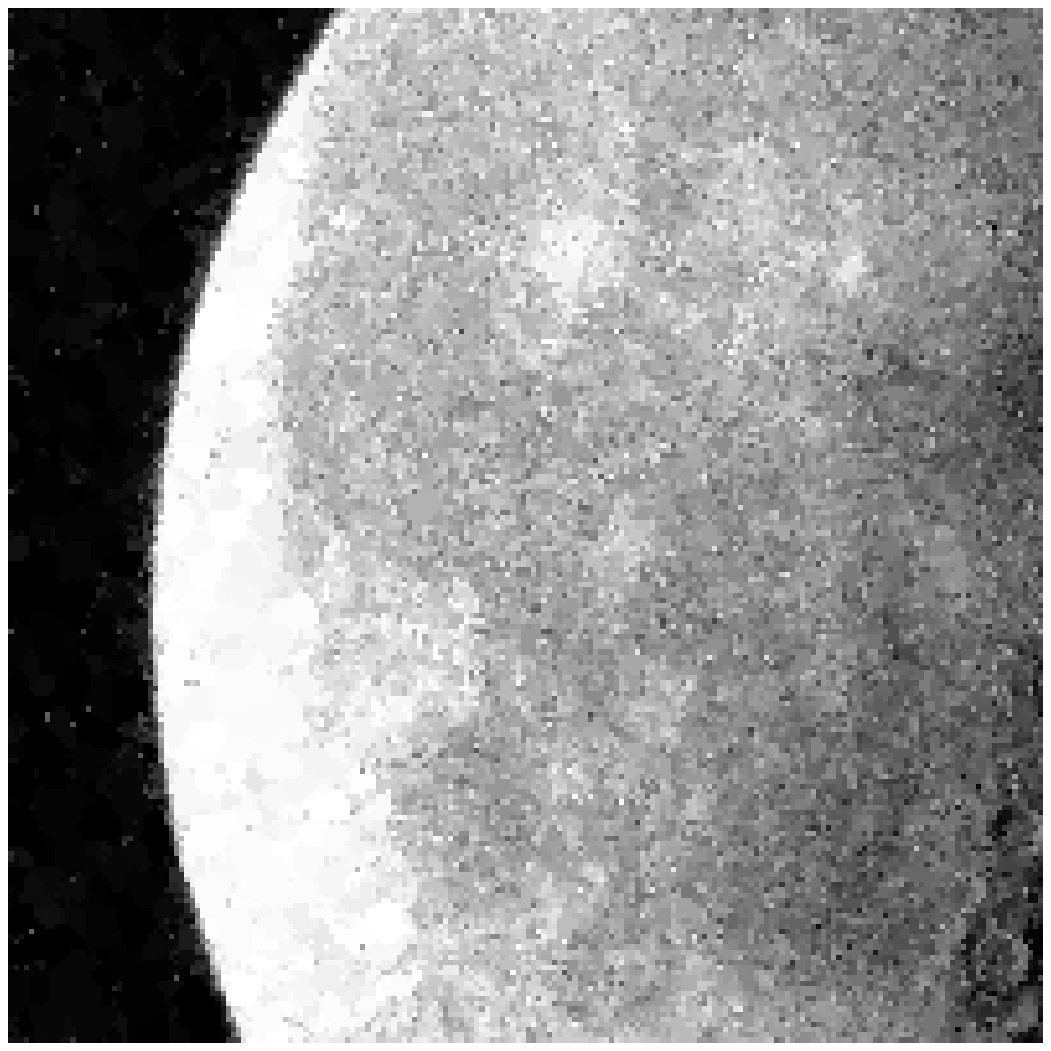} \\ (c) & & (d)
\end{tabular}
\end{center}
\caption{Image denoising application, (a) original image, (b) noisy
  image, (c) refined image obtained from BP after $\Time = 5$
  iterations, and (d) refined image obtained from SBP after $\Time =
  100$ iterations. The image is $200 \times 200$ with $\dimn = 256$
  gray-scale levels. The SBP step size, the Potts model parameter, and
  noise standard deviation are set to $\step = 1/(\Time + 1)$, $\pott
  = 0.05$ and $\sigma = 0.1$ respectively.}
\label{FigDenoisingImages}
\end{figure}

\begin{figure}[h]
\begin{center}
  \widgraph{.65\textwidth}{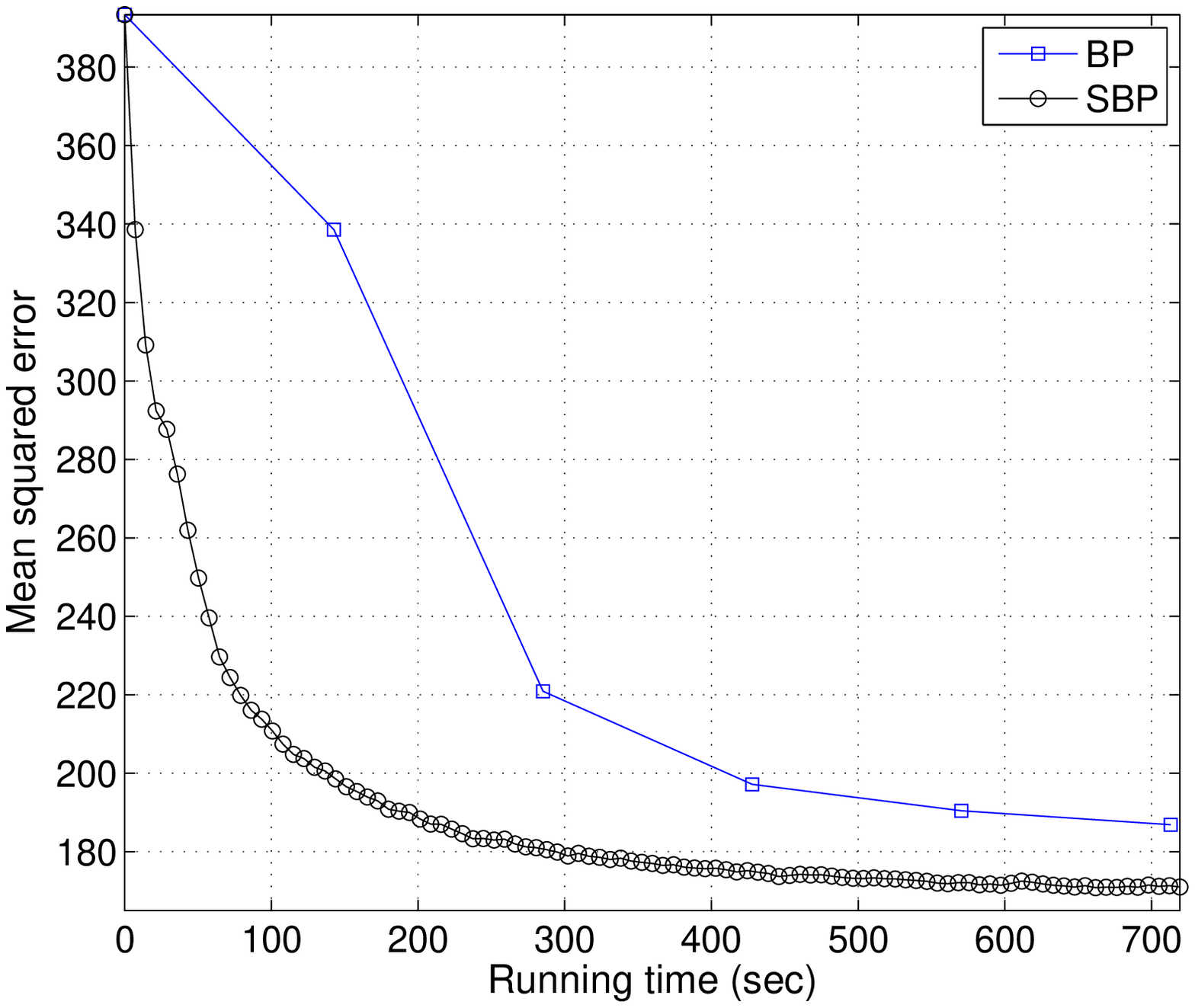}
\end{center}
\caption{Mean squared error versus the running time (in seconds) for
  both BP and SBP image denoising. The simulations are performed with
  the step size $\step = 1/(\Time+1)$, and the Potts model parameter
  $\pott = 0.05$ on a $200 \times 200$ image with $\dimn = 256$
  gray-scale levels. The noise is assumed to be additive, independent
  Gaussian random variables with standard deviation $\sigma = 0.1$.}
\label{FigDenoiseRunTime}
\end{figure}

In our next set of experiments, we study the SBP on some larger scale
graphs and more challenging problem instances, with applications to
image processing and computer vision.  Message-passing algorithms can
be used for image denoising, in particular, on a two dimensional
square grid where every node corresponds to a pixel. Running the BP
algorithm on the graph, one can obtain (approximations to) the most
likely value of every pixel based on the noisy observations. In this
experiment, we consider a $200 \times 200$ image with $\dimn = 256$
gray-scale levels, as showin in Figure~\ref{FigDenoisingImages}(a). We
then contaminate every pixel with an independent Gaussian random
variable with standard deviation $\sigma = 0.1$, as shown in
Figure~\ref{FigDenoisingImages}(b). Enforcing the Potts model with
smoothness parameter $\pott = 0.05$ as the edge potential, we run BP
and SBP for the total of $\Time = 5$ and $\Time = 100$ iterations
respectively to obtain the refined images (see panels (c) and (d),
respectively, in Figure~\ref{FigDenoisingImages}).
Figure~\ref{FigDenoiseRunTime} illustrates the mean squared error
versus the running time for both BP and SBP denoising. As one can
observe, despite smaller jumps in the error reduction, the
per-iteration running time of SBP is substantially lower than
BP. Overall, SBP has done a marginally better job than BP in a
substantially shorter amount of time in this instance.\footnote{Note that
  the purpose of this experiment is not to analyze the potential of
  SBP (or for that matter BP) in image denoising, but to rather
  observe their relative performances and computational complexities.}

Finally, in our last experiment, we apply SBP to a computer vision
problem. Graphical models and message-passing algorithms are popular
in application to the stereo vision
problem~\cite{SunEtal03,KlaEtal06}, in which the goal is to estimate
objects depth based on the pixel dissimilarities in two (left and
right view) images. Adopting the original model in Sun et
al.~\cite{SunEtal03}, we again use a form of the Potts model in order
to enforce a smoothness prior, and also use the form of the
observation potentials given in the Sun et al. paper. We then run BP
and SBP (with step size $3/(\Time+2)$) for a total of $\Time = 10$ and
$\Time = 50$ iterations respectively in order to estimate the pixel
dissimilarities. The results for the test image ``map'' are presented
in Figure~\ref{FigStereo}. Here, the maximum pixel dissimilarity is
$\dimn = 32$, which makes stereo vision a relatively low-dimensional
problem. In this particular application, the SBP is faster by about a
factor of $3-4$ times per iteration; however, the need to run more
iterations makes it comparable to BP.  This is to be expected since
the state dimension $\dimn = 32$ is relatively small, and the relative
advantage of SBP becomes more significant for larger state dimensions
$\dimn$.

\newcommand{\NIMSIZEB}{.38\textwidth}
\begin{figure}[h]
\begin{center}
\begin{tabular}{cc}
  \widgraph{\NIMSIZEB}{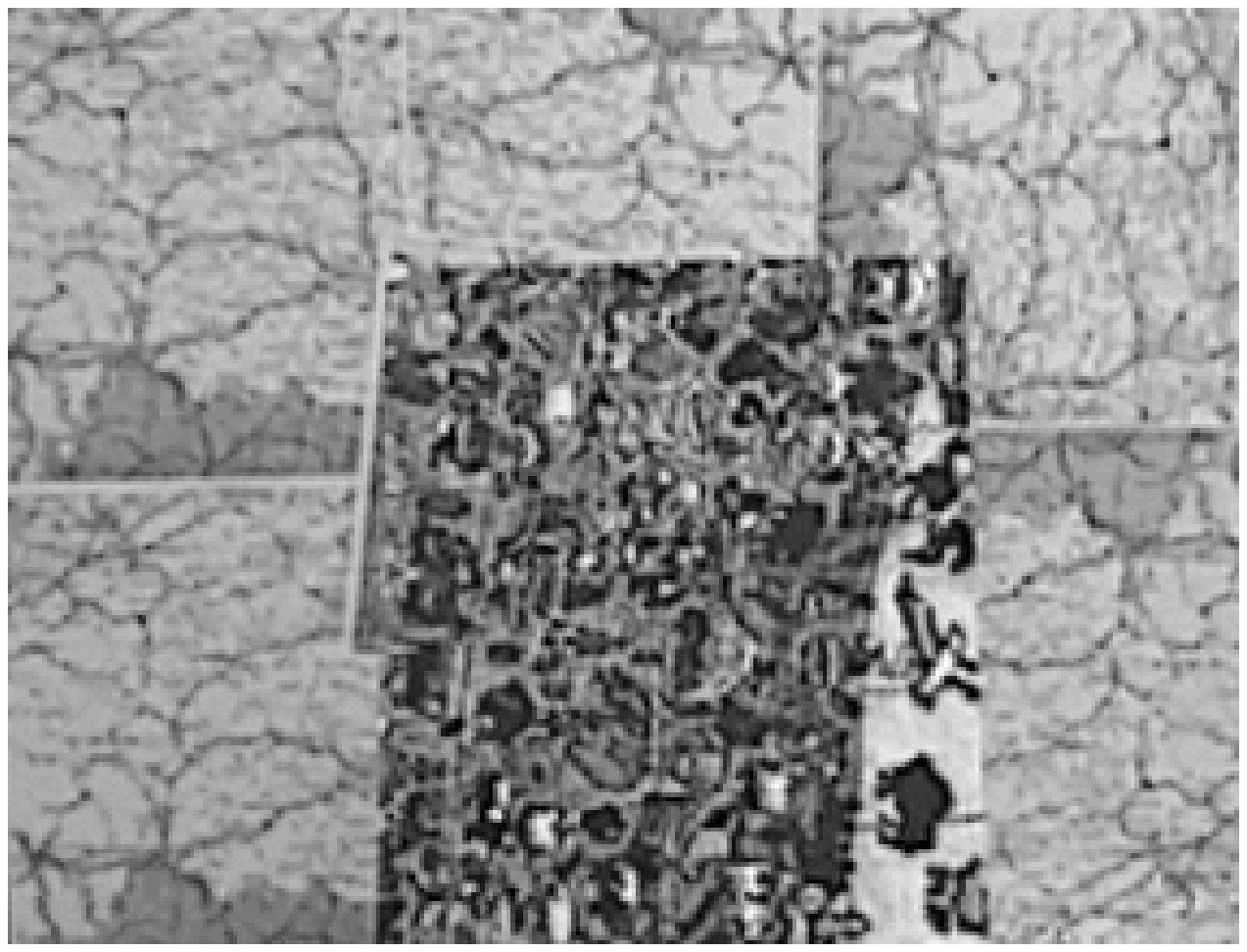} & 
  \widgraph{\NIMSIZEB}{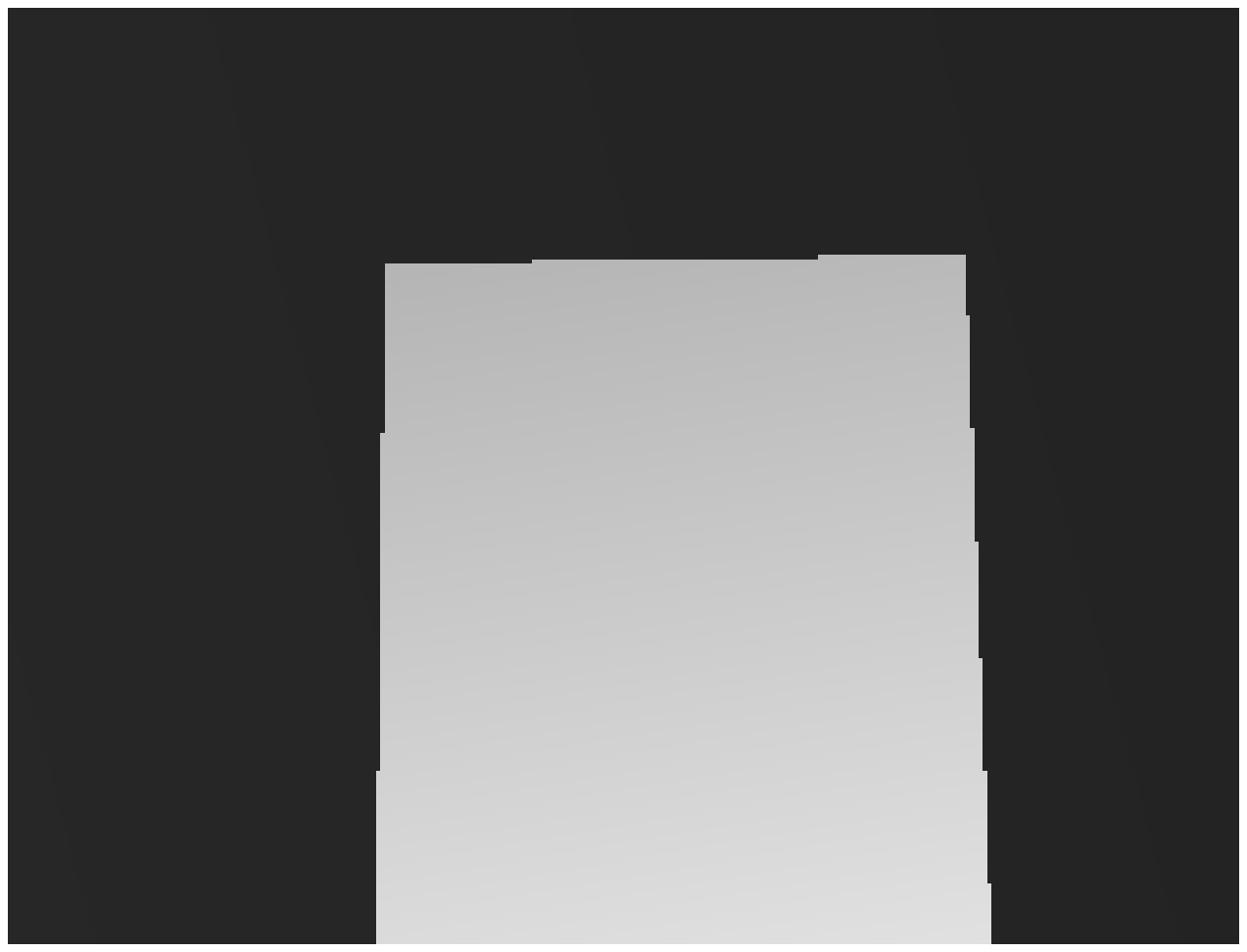} \\ (a) & (b)\\ 
  \widgraph{\NIMSIZEB}{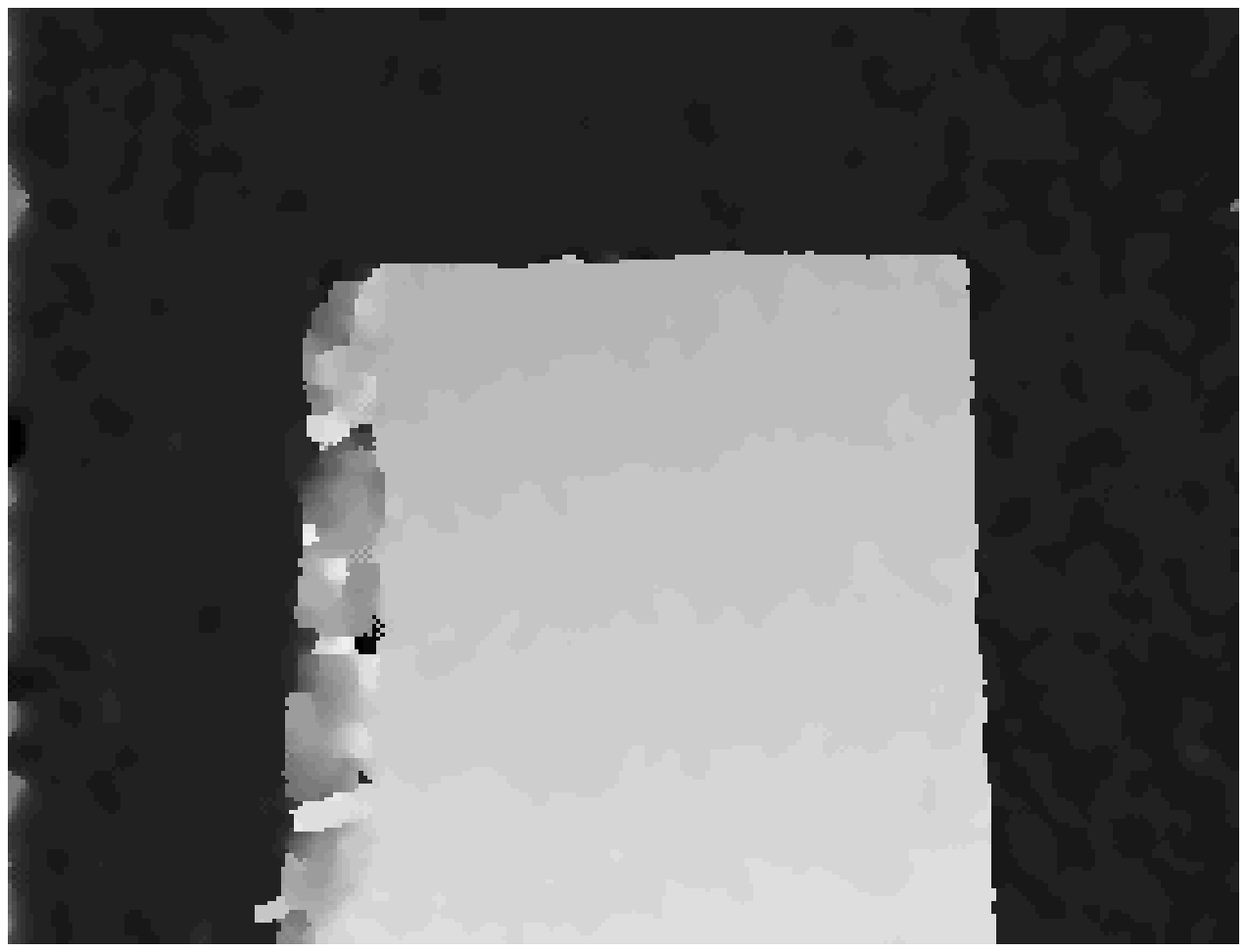} & 
  \widgraph{\NIMSIZEB}{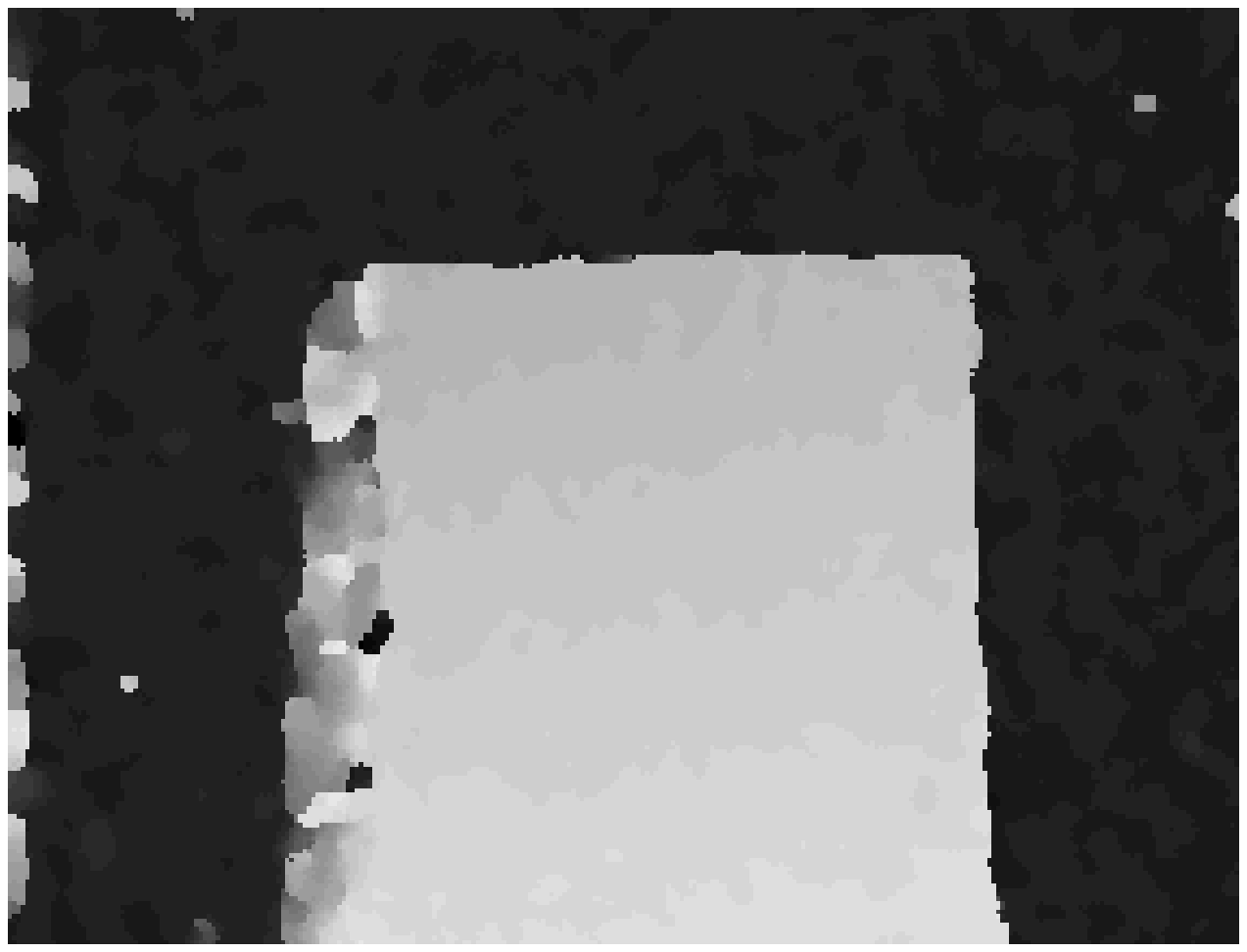} \\ (c) & (d)
\end{tabular}
\end{center}
\caption{Stereo vision, depth recognition, application, (a) reference
  image, (b) ground truth, (c) BP estimate after $\Time = 10$
  iterations, and (d) SBP estimate after $\Time = 50$ iterations. The
  algorithms are applied to the standard ``map'' image with maximum
  pixel dissimilarity $\dimn = 32$. The SBP step size is set to $\step
  = 3/(\Time + 2)$.}
\label{FigStereo}
\end{figure}


\section{Discussion}
\label{SecDiscussion}

In this paper, we have developed and analyzed a new and low-complexity
alternative to BP message-passing.  The SBP algorithm has per
iteration computational complexity that scales linearly in the state
dimension $\dimn$, as opposed to the quadratic dependence of BP, and a
communication cost of $\log \dimn$ bits per edge and iteration, as
opposed to $\dimn-1$ real numbers for standard BP message updates.
Stochastic belief propagation is also easy to implement, requiring
only random number generation and the usual distributed updates of a
message-passing algorithm.  Our main contribution was to prove a
number of theoretical guarantees for the SBP message updates,
including convergence for any tree-structured problem, as well as for
general graphs for which the ordinary BP message update satisfies a
suitable contraction condition.  In addition, we provided
non-asymptotic upper bounds on the SBP error, both in expectation and
in high probability.

The results described here suggest a number of directions for future
research. First, the ideas exploited here have natural generalizations
to problems involving continuous random variables and also other
algorithms that operate over the sum-product semi-ring, including the
generalized belief propagation algorithm~\cite{Yedidia05} as well as
reweighted sum-product algorithms~\cite{Wainwright05}. More generally,
the BP can be seen as optimizing the dual of the Bethe free energy
function~\cite{Yedidia05}, and it would be interesting to see if SBP
can be interpreted as a stochastic version of this Bethe free energy
minimization. It is also natural to consider whether similar ideas can
be applied to analyze stochastic forms of message-passing over other
semi-rings, such as the max-product algebra that underlies the
computation of maximum a posteriori (MAP) configurations in graphical
models.  In this paper, we have developed SBP for applications to
Markov random fields with pairwise interactions.  In principle, any
undirected graphical model with discrete variables can be reduced to
this form~\cite{Yedidia05,WaiJorBook08}; however, in certain
applications, such as decoding of LDPC codes over non-binary state
spaces, this could be cumbersome.  For such cases, it would be useful
to derive a variant of SBP that applies directly to factor graphs with
higher-order interactions.  Finally, our analysis for general graphs
has been done under a contractivity condition, but it is likely that
this requirement could be loosened.  Indeed, the SBP algorithm works
well for many problems where this condition need not be satisfied.

\subsection*{Acknowledgements}
Both authors were partially supported by MURI grant N00014-11-1-0688
to MJW.  Both authors would like to thank Alekh Agarwal for helpful
discussions on stochastic approximation and optimization at the
initial phases of this research; the anonymous reviewers for their
helpful feedback, as well as Associate Editor Pascal Vontobel for his
careful reading and detailed suggestions that helped to improve the
paper.


\appendix


\section{Details of Example~\ref{ExaPotts}}
\label{AppPotts}

In this appendix, we verify the sufficient condition for
contractivity~\eqref{EqnSuffPotts}.  Recall the
definition~\eqref{EqnZeroBound} of the zero'th order bounds.  By
construction, we have the relations
\begin{align*}
\LBnd{\Fnode \Snode}{i} \, & = \, \LBound {\Fnode \Snode}{i}{0} \, =
\, \frac{\pott} {1 + (\dimn - 1) \pott}, \quad \mbox{and} \\
\UBnd{\Fnode \Snode}{i} \, & = \, \UBound {\Fnode \Snode}{i}{0} \, =
\, \frac{1} {1 + (\dimn - 1) \pott} \quad \mbox{for all $i \in
  \Alphabet$ and $(\Fnode \to \Snode) \in \DirEdge$.}
\end{align*}
Substituting these bounds into the definitions~\eqref{EqnDefnPhi}
and~\eqref{EqnDefnChi} and doing some simple algebra yields the upper
bounds
\begin{align*}
\TermPhi{\Fnode \Snode} {\Tnode \Fnode} \, & \le \, \max_{j \in
  \Alphabet} \: \bigg\{ \frac {\ColSum{j} \prod_{\Ftnode \in
    \Neig(\Fnode) \setminus \{ \Snode, \Tnode\} } \UBnd {\Ftnode
    \Fnode} {j}} {\sum_{\ell=1} ^{\dimn} \ColSum{\ell} \prod_{\Ftnode
    \in \Neig(\Fnode) \setminus \Snode} \LBnd {\Ftnode \Fnode}{\ell}}
\bigg\} \; = \; \, \frac{1 + (\dimn - 1)\pott} {\pott ^{\degr_\Fnode -
    1}} \: \max_{j \in \Alphabet} \bigg\{\frac{\NPot(j)}{\sum_{\ell =
    1} ^{\dimn} \NPot(\ell)}\bigg\}, \quad \mbox{and} \\
\TermChi {\Fnode \Snode} {\Tnode \Fnode} \, & \le \, \max_{j \in
  \Alphabet} \bigg\{ \frac {\ColSum{j} \prod_{\Ftnode \in \Neig(\Fnode)
    \setminus \Snode } \UBnd {\Ftnode \Fnode} {j}} { \sum_{\ell=1}
  ^{\dimn} \ColSum{\ell} \prod_{\Ftnode \in \Neig(\Fnode) \setminus
    \Snode} \LBnd {\Ftnode \Fnode}{\ell}} \bigg\} \max_{j \in
  \Alphabet} \bigg\{ \frac {1} {\LBnd {\Tnode \Fnode} {j}} \bigg\} \; =
\; \, \frac{1 + (\dimn - 1)\pott} {\pott ^{\degr_\Fnode }} \: \max_{j
  \in \Alphabet} \bigg\{\frac{\NPot(j)}{\sum_{\ell = 1} ^{\dimn}
  \NPot(\ell)}\bigg\},
\end{align*}
where we have denoted the degree of the node $\Fnode$ by
$\degr_{\Fnode}$.  Substituting these inequalities into
expression~\eqref{EqnUpFunLipsConst} and noting that $\pott \le 1$, we
find that the global update function has Lipschitz constant at most
\begin{align*}
\Lips \, & \le \, 4 \: (1 - \pott) (1 + (\dimn - 1)\pott) \:
\max_{\Fnode \in \node} \bigg\{ \frac {(\degr_{\Fnode} - 1)^2} {\pott ^{2
    \degr_{\Fnode}} } \, \max_{j \in \Alphabet} \bigg\{ \frac
    {\NPot(j)} {\sum_{\ell} \NPot(\ell)} \bigg\} ^2 \bigg\},
\end{align*}
as claimed.



\section{Proof of Lemma~\ref{LemNilpotent}}
\label{AppLemNilpotent}

By construction, for each directed edge $(\Fnode \to \Snode)$, the
message vector $\Mess_{\Fnode \Snode}$ belongs to the probability
simplex---that is, $\sum_{i\in\Alphabet} \Mess_{\Fnode \Snode} (i) =
1$, and $\Mess_{\Fnode \Snode} \conegeq \zerovec$.  From
equation~\eqref{EqnBPExpUpdate}, the vector $\Mess_{\Fnode \Snode}$ is
a convex combination of the columns of the matrix $\tBPMat$.
Recalling bounds~\eqref{EqnZeroBound}, we conclude that the message
vector must belong to the set $\Ball$, as defined in
equation~\eqref{EqnDefnBall}, in particular with $\LBnd{\Fnode
  \Snode}{i} = \LBound{\Fnode \Snode}{i}{0}$ and $\UBnd{\Fnode
  \Snode}{i} = \UBound{\Fnode \Snode}{i}{0}$. Note that the set
$\Ball$ is compact, and any member of it has strictly positive
elements under our assumptions.

For directed edges $(\Fnode \to \Snode)$ and $(\Tnode \to \Ftnode)$,
let $\parderv \in \real^{\dimn \times \dimn}$ denote the Jacobian
matrix obtained from taking the partial derivative of the update
function $\UpFunc_{\Fnode \Snode}$ with respect to the message vector
$\Mess_{\Tnode \Ftnode}$.  By inspection, the function $\UpFun$ is
continuously differentiable; consequently, the function
$\frac{\partial \UpFun(i; \Mess)} {\partial \Mess_{\Tnode \Ftnode}
  (j)}$ is continuous, and hence must achieve its supremum
over the compact set $\Ball$.  Consequently, we may use these
Jacobian matrices to define a matrix $\blkMat{\Fnode\Snode}{\Tnode
  \Ftnode} \in \real^{\dimn \times \dimn}$ with entries
\begin{align*}
\blkMat{\Fnode\Snode}{\Tnode \Ftnode}(i, j) & \defn \max_{\Mess \in
  \Ball} \bigg| \frac{\partial \UpFun(i; \Mess)} {\partial
  \Mess_{\Tnode \Ftnode} (j)} \bigg|, \quad \mbox{for $i, j = 1,
  \ldots, \dimn$.}
\end{align*}
We then use these matrices to define a larger matrix $\nilpot \in
\real^{\Dimn \times \Dimn}$, consisting of $2 |\edge| \times 2
|\edge|$ sub-blocks each of size $\dimn \times \dimn$, with the
sub-blocks indexed by pairs of directed edges $(\Fnode \to \Snode) \in
\DirEdge$. In particular, the matrix $\blkMat{\Fnode\Snode}{\Tnode
  \Ftnode}$ occupies the sub-block indexed by the edge pair $(\Fnode
\to \Snode)$ and $(\Tnode \to \Ftnode)$.  Note that by the structure
of the update function $\UpFunc$, the matrix
$\blkMat{\Fnode\Snode}{\Tnode \Ftnode}$ can be non-zero only if
$\Ftnode = \Fnode$ and $\CompNeig$.  \\

Now let $\grad \UpFunc \in \real^{\Dimn \times \Dimn}$ denote the
Jacobian matrix of the update function $\UpFunc$. By the integral form
of the mean value theorem, we have the representation
\begin{align*}
\UpFunc (\Mess) - \UpFunc(\MessP) \, & = \, \bigg [\int_{0}^{1} \grad
  \UpFunc (\MessP + \tau (\Mess - \MessP)) \: d\tau \bigg] \:
(\Mess - \MessP).
\end{align*}
Applying triangle inequality separately to each component of this
$\Dimn$-vector and then using the definition of $\nilpot$, we obtain
the elementwise upper bound
\begin{align*}
|\UpFunc(\Mess) - \UpFunc(\MessP)| \, & \coneleq \, \nilpot \: |\Mess -
\MessP|.
\end{align*}

It remains to show that $\nilpot$ is nilpotent: more precisely, we
show that $\nilpot^r$ is the all-zero matrix, where $r = \diameter
(\graph)$ denotes the diameter of the graph $\graph$.  In order
to do so, we first let $\blkInd \in \real ^{2|\edge| \times 2
  |\edge|}$ be the ``block indicator'' matrix---that is, its entries
are given by
\begin{align*}
\blkInd (\diredge{\Fnode}{\Snode}, \diredge{\Tnode}{\Ftnode}) \; = \;
\begin{cases}
1 & \mbox{if $\blkMat {\Fnode \Snode}{\Tnode \Ftnode} \neq 0$} \\ 0 &
\mbox{otherwise.}
\end{cases}
\end{align*}
Based on this definition, it is straightforward to verify that if
$\blkInd ^{\diam} = 0$ for some positive integer $r$, then we also
have $\nilpot^{r} = 0$.  Consequently, it suffices to show that
$\blkInd^r = 0$ for $r = \diameter(\graph)$. \\

Fix a pair of directed edges $(\Fnode \to \Snode)$ and
$(\diredge{\Tnode}{\Ftnode})$, and some integer $\ell \geq 1$.  We first
claim that the matrix entry $\blkInd ^{\ell}(\diredge{\Fnode}{\Snode},
\diredge{\Tnode}{\Ftnode})$ is non-zero only if there exists a
\emph{directed path} of length $\ell + 1$ from $\Tnode$ to $\Snode$
that includes both $\Ftnode$ and $\Fnode$, meaning that there exist
nodes $\Ftnode_1, \Ftnode_2, \ldots, \Ftnode_{\ell-2}$ such that
\begin{equation*}
\Tnode \in \Neig(\Ftnode) \setminus \Ftnode_1, \quad \Ftnode_1 \in
\Neig(\Ftnode_2) \setminus \Ftnode_3, \ldots, \quad \mbox{and} \quad
\Ftnode_{\ell-2} \in \Neig(\Fnode) \setminus \Snode.
\end{equation*}
We prove this claim via induction.  The base case $\ell = 1$ is true
by construction.  Now supposing that the
claim holds at order $\ell$, we show that it must hold at order $\ell
+ 1$.  By definition of matrix multiplication, we have
\begin{align*}
\blkInd ^{\ell+1} (\diredge{\Fnode}{\Snode},
\diredge{\Tnode}{\Ftnode}) \; = \; \sum _{(\diredge{x}{y}) \in
  \DirSet} \blkInd ^{\ell} (\diredge{\Fnode}{\Snode}, \diredge{x}{y})
\: \blkInd (\diredge{x}{y}, \diredge{\Tnode}{\Ftnode}).
\end{align*}
In order for this entry to be non-zero, there must exist a directed
edge $(\diredge{x}{y})$ that forms a $(\ell+1)$-directed path to
$(\diredge{\Fnode}{\Snode})$, and moreover, we must have $\Ftnode =
x$, and $\Tnode \in \Neig(x) \setminus y$. These conditions are
equivalent of having a directed path of length $\ell + 2$ from
$\Tnode$ to $\Snode$, with $\Ftnode$ and $\Fnode$ as intermediate
nodes, thereby completing the proof of our intermediate claim. \\

Finally, we observe that in a tree-structured graph, there can be no
directed path of length greater than $r = \diameter(\graph)$.
Consequently, our intermediate claim implies that $\blkInd ^{\diam} =
0$ for any tree-structured graph, which completes the proof.


\section{Proof of Lemma~\ref{LemAux}}
\label{AppLemAux}

Noting that it is equivalent to bound the logarithm, we have
\begin{align}\label{EqnAuxEqOne}
\log \prod _{\ell = i+1} ^{\Time + 2} \bigg( 1 - \frac {\Coef} {\ell}
\bigg) \, & = \, \sum _{\ell = i+1} ^{\Time + 2} \log \bigg( 1 -
\frac{\Coef} {\ell} \bigg)
\, \le  \, - \Coef \: \sum _{\ell = i+1} ^{\Time + 2} \frac {1}
{\ell},
\end{align}
where we used the fact that $\log (1 - x) \le - x$ for $x \in (0,
1)$. Since the function $1/x$ is decreasing, we have
\begin{align}
\label{EqnAuxEqTwo}
\sum _{\ell = i+1} ^{\Time + 2} \frac{1} {\ell} \, \ge \, \int _{i+1}
^{\Time + 3} \frac {1}{x} \: dx \, = \, \log (\Time + 3) \: - \: \log
(i + 1).
\end{align}
Substituting inequality \eqref{EqnAuxEqTwo} into \eqref{EqnAuxEqOne}
yields $\log \prod _{\ell = i+1} ^{\Time + 2} \big( 1 - \frac{\Coef}
{\ell} \big) \, \leq \, \Coef \: \big( \log (i + 1) \: - \: \log
(\Time + 3 ) \big)$, from which the claim stated in the lemma
follows.


\section{Proof of Lemma~\ref{LemOpNormJacob}}
\label{AppLemOpNormJacob}

Let $\JacMat(\Mess) \in \real^{\Dimn \times \Dimn}$ denote the
Jacobian matrix of the function $q: \real^\Dimn \rightarrow
\real^\Dimn$ evaluated at $\Mess$.  Since $q$ is differentiable, we
can apply the integral form of the mean value theorem to write
\mbox{$q(\Mess) - q(\MessP) = \big[\int _{0} ^{1} \JacMat (\MessP +
    \tau (\Mess - \MessP)) \, d\tau \big] \: (\Mess - \MessP)$.}
From this representation, we obtain the upper bound
\begin{align*}
\|q(\Mess) - q(\MessP)\|_2 \, \leq \, \bigg[\int _{0}^{1} \matsnorm{\JacMat
  (\MessP + \lambda (\Mess - \MessP))}{2} \, d\lambda \bigg]
  \|(\Mess - \MessP)\|_2 \, \leq \, 
\sup_{\Mess \in \Ball}
  \matsnorm{\JacMat(\Mess)}{2} \, \|\Mess - \MessP\|_2,
\end{align*}
showing that it suffices to control the quantity $\sup_{\Mess \in
  \Ball} \matsnorm{\JacMat(\Mess)}{2}$. \\

Let $\LocalJacob{\Fnode \Snode}{\Tnode \Ftnode}$ be the $\dimn \times
\dimn$ matrix of partial derivatives of the function $q_{\Fnode
  \Snode}: \real^\Dimn \rightarrow \real^\dimn$ obtained from
taking the partial derivatives with respect to the message vector
$\Mess_{\Tnode \Ftnode} \in \real^\dimn$.  We then define a $2|\edge|
\times 2|\edge|$-dimensional matrix $\SpMat$ with the entries
\begin{align} 
\label{EqnDefnSpMat}
\SpMat (\diredge{\Fnode} {\Snode} , \diredge {\Tnode} {\Ftnode}) &
\defn \; 
\begin{cases} 
\sup_{\Mess \in \Ball} \matsnorm{\LocalJacob
    {\Fnode \Snode} {\Tnode \Ftnode}}{2} & \mbox{if $\Ftnode =
    \Fnode$, and $\CompNeig$} \\
0 & \mbox{otherwise.}
\end{cases}
\end{align}
Our next step is to show that $\sup_{\Mess \in \Ball} \: \matsnorm
{\JacMat(\Mess)}{2} \; \le \; \matsnorm {\SpMat} {2}$.  Let $\vecone =
\{\vecone_{\Fnode\Snode}\}_{(\Fnode \to \Snode) \in \DirSet}$ be an
arbitrary $\Dimn$-dimensional vector, where each sub-vector
$\vecone_{\Fnode\Snode}$ is an element of $\real^{\dimn}$.  By
exploiting the structure of $\JacMat(\Mess)$ and $\vecone$, we have
\begin{align*}
\vnorm{\JacMat (\Mess) \: \vecone}{2}^ 2 \; & = \sum_{(\Fnode \to
  \Snode) \in \DirSet} \vnorm { \sum _ {\CompNeig} \LocalJacob {\Fnode
    \Snode} {\Tnode \Fnode} \: \vecone _ {\Tnode \Fnode} }{2}^2 \\
\; & \stackrel{\mathrm{(i)}}{\leq} \sum_{(\Fnode \to \Snode) \in \DirSet} \bigg(
\sum _ {\CompNeig} \vnorm {\LocalJacob {\Fnode \Snode} {\Tnode \Fnode}
  \: \vecone _ {\Tnode \Fnode}} {2} \bigg)^2 \\
\; & \stackrel{\mathrm{(ii)}}{\leq} \sum_{(\Fnode \to \Snode) \in \DirSet} \bigg(
\sum _ {\CompNeig} \matsnorm { \LocalJacob {\Fnode \Snode} {\Tnode
    \Fnode} } {2} \vnorm {\vecone _ {\Tnode \Fnode}} {2} \bigg)^2 \\
\; & \stackrel{\mathrm{(iii)}}{\leq} \sum_{(\Fnode \to \Snode) \in \DirSet} \bigg(
\sum _ {\CompNeig}
\SpMatEntry{\diredge{\Fnode}{\Snode}}{\diredge{\Tnode}{\Fnode}} \vnorm
           {\vecone _ {\Tnode \Fnode}} {2} \bigg)^2 ,
\end{align*}
where the bound (i) follows by triangle inequality; the bound (ii)
follows from definition of the operator norm; and the final inequality
(iii) follows by definition of $A$.

Defining the vector $\vectwo \in \real^{2 |\edge|}$ with the entries
$\vectwo_{\Tnode \Fnode} = \vnorm {\vecone _ {\Tnode \Fnode}} {2}$, we
have established the upper bound \mbox{$\vnorm{\JacMat (\Mess) \:
    \vecone}{2}^2 \leq \|A \vectwo\|_2^2$,} and hence that
\begin{align*}
\vnorm{\JacMat (\Mess) \: \vecone}{2}^2 \, & \leq \, \matsnorm{A}{2}^2 \,
\|\vectwo\|_2^2 \, = \, \matsnorm{A}{2}^2 \, \|\vecone\|_2^2,
\end{align*}
where the final equality uses the fact that $\vnorm{\vecone}{2}^{2} =
\vnorm{\vectwo}{2}^2$ by construction.  Since both the message $\Mess$
and vector $\vecone$ were arbitrary, we have shown that $\sup_{\Mess
  \in \Ball} \: \matsnorm {\JacMat (\Mess)}{2} \leq \matsnorm
{\SpMat}{2}$, as claimed. \\

Our final step is to control the quantities $\sup_{\Mess \in \Ball}
\matsnorm{\LocalJacob {\Fnode \Snode} {\Tnode \Ftnode}}{2}$ that
define the entries of $A$.  In this argument, we make repeated use of
the elementary matrix inequality~\cite{Horn85}
\begin{align}
\label{EqnHornInq}
\matsnorm{B}{2}^2 \, & \leq \,\underbrace{\bigg( \max_{i=1, \ldots, n}
  \sum_{j=1}^n |B_{ij}| \bigg)}_{\matsnorm{B}{\infty}} \;
\underbrace{\bigg( \max_{j=1, \ldots, n} \sum_{i=1}^n |B_{ij}| \bigg)
  }_{\matsnorm{B}{1}},
\end{align}
valid for any $n \times n$ matrix.

Recall the definition of the probability
distribution~\eqref{EqnProbMass} that defines the function
\mbox{$q_{\Fnode \Snode}: \real^\Dimn \rightarrow \real^\dimn$,} as
well as our shorthand notation $\AgMes{}{\Fnode}{\Snode} (\Fvar) =
\prod_{\CompNeig} \Mes{}{\Tnode}{\Fnode} (\Fvar)$.  Taking the
derivatives and performing some algebra yields
\begin{align*}
\frac {\ro q_{\Fnode \Snode} (i \: ; \: \Mess)} {\ro \Mess_{\Tnode
    \Fnode} (j)} \, & = \,  \sum_{k = 1}^{\dimn} \frac {\ro q_{\Fnode \Snode}
  (i \: ; \: \Mess)} {\ro \AgMess (k)} \, \frac{\ro \AgMess (k)} {\ro
  \Mess_{\Tnode \Fnode} (j)} \\
 \, & = \, \frac {\ro q_{\Fnode \Snode} (i \: ; \: \Mess)} {\ro \AgMess
  (j)} \: \frac {\AgMess(j)} {\Mess_{\Tnode \Fnode} (j)} \\
\, & = \, \frac {- \ColSum{i} \: \AgMess (i) \: \ColSum{j}} {\big(
\sum _{k = 1} ^{\dimn} \ColSum{k} \AgMess (k) \big) ^{2}} \: \frac
{\AgMess(j)} {\Mess_{\Tnode \Fnode} (j)},
\end{align*}
for $i \neq j$, and $\CompNeig$.  For $i = j$, we have
\begin{align*}
\frac {\ro q_{\Fnode \Snode} (i \: ; \: \Mess)} {\ro \Mess_{\Tnode
    \Fnode} (i)}
\, & = \, \frac {\ro q_{\Fnode \Snode} (i \: ; \: \Mess)} {\ro \AgMess
  (i)} \: \frac {\AgMess(i)} {\Mess_{\Tnode \Fnode} (i)} \\
\, & = \, \bigg[\frac {\ColSum{i}} {\sum _{k = 1} ^{\dimn} \ColSum{k}
    \AgMess (k)} \, - \, \frac {\ColSum{i} ^{2} \: \AgMess (i)} {\big(
    \sum _{k = 1} ^{\dimn} \ColSum{k} \AgMess (k) \big) ^{2}} \bigg] \:
\frac {\AgMess(i)} {\Mess_{\Tnode \Fnode} (i)}.
\end{align*}
Putting together the pieces leads to the upper bounds

\begin{align*}
\matsnorm {\LocalJacob{\Fnode \Snode}{\Tnode \Fnode}}{1} & \leq \; 2
\: \max_{j \in \Alphabet} \bigg\{ \frac{\ColSum {j} \: \AgMess (j) } {\sum _{k
    =1} ^{\dimn} \ColSum {k} \: \AgMess (k) } \: \frac {1}
   {\Mess_{\Tnode \Fnode} (j)} \bigg\}, \quad \mbox{and} \\
%
\matsnorm {\LocalJacob{\Fnode \Snode}{\Tnode \Fnode}} {\infty} & \le
\; \max_{i \in \Alphabet} \bigg\{ \frac {\ColSum {i} \: \AgMess (i)}
   {\sum_{k=1} ^{\dimn} \ColSum {k} \: \AgMess (k) } \: \frac {1}
   {\Mess_{\Tnode \Fnode} (i)} \, + \, \frac {\ColSum {i} \: \AgMess
     (i)} {\big(\sum_{k=1} ^{\dimn} \ColSum {k} \: \AgMess (k) \big)
     ^{2}} \: \sum_{j=1} ^{\dimn} \frac{\ColSum{j} \: \AgMess(j)}
   {\Mess_{\Tnode \Fnode (j)}}\bigg\}.
\end{align*}
Recalling the definitions~\eqref{EqnDefnPhi} and~\eqref{EqnDefnChi} of
$\TermPhi{\Fnode \Snode}{\Tnode \Fnode}$ and $\TermChi{\Fnode
  \Snode}{\Tnode \Fnode}$ respectively, we find that
\begin{align*}
\matsnorm {\LocalJacob{\Fnode \Snode}{\Tnode \Fnode}} {1} \, \le \, 2
\: \TermPhi{\Fnode \Snode} {\Tnode \Fnode}, \quad \mbox{and} \quad
\matsnorm{\LocalJacob{\Fnode \Snode}{\Tnode \Fnode}} {\infty} \, \le
\, \TermPhi{\Fnode \Snode} {\Tnode \Fnode} + \TermChi{\Fnode
  \Snode}{\Tnode \Fnode}.
\end{align*}
Thus, by applying inequality~\eqref{EqnHornInq} with $B =
\LocalJacob{\Fnode \Snode}{\Tnode \Fnode}$, we conclude that
\begin{align*}
\matsnorm{\LocalJacob{\Fnode \Snode}{\Tnode \Fnode}} {2} ^{2} \; \le
\; 2 \: \TermPhi{\Fnode \Snode} {\Tnode \Fnode} \: (\TermPhi{\Fnode
  \Snode} {\Tnode \Fnode} + \TermChi{\Fnode \Snode} {\Tnode \Fnode}).
\end{align*}
Since this bound holds for any message $\Mess \in \Ball$, we conclude
that each of the matrix entries
\mbox{$\SpMatEntry{\diredge{\Fnode}{\Snode}}{\diredge{\Tnode}{\Fnode}}$}
satisfies the same inequality.  Again applying the basic matrix
inequality~\eqref{EqnHornInq}, this time with $B = A$, we conclude
that $\matsnorm{\SpMat}{2}$ is upper bounded by
\begin{align*}
 2 \max_{(\Fnode \to \Snode) \in \DirSet} \sum_{\CompNeig}
 \big(\TermPhi{\Fnode \Snode}{\Tnode \Fnode} \: (\TermPhi{\Fnode \Snode}
         {\Tnode \Fnode} + \TermChi{\Fnode \Snode} {\Tnode \Fnode})
         \big) ^{\frac{1}{2}}
    \max_ {(\Tnode \to \Fnode) \in
           \DirSet} \sum_ {\Snode \in \Neig (\Fnode) \setminus \Tnode}
       \big( \TermPhi{\Fnode \Snode}{\Tnode \Fnode} \: (\TermPhi{\Fnode
           \Snode} {\Tnode \Fnode} + \TermChi{\Fnode \Snode} {\Tnode
           \Fnode}) \big) ^{\frac{1}{2}},
\end{align*}
which concludes the proof.


\bibliographystyle{plain}
\bibliography{NewSBP_bibfile}

\end{document}